\documentclass[10pt]{aastex}
\usepackage{emulateapj5}
\usepackage{apjfonts}
\usepackage{natbib}
\slugcomment{{\sc Submitted to ApJ:} July 26, 2003\\{\sc Accepted: September 14, 2003}
}

\newcommand{\begit}{\begin{itemize}}
\newcommand{\enit}{\end{itemize}}
\newcommand{\beq}{\begin{equation}} 
\newcommand{\eeq}{\end{equation}} 
\newcommand{\beqa}{\begin{eqnarray}} 
\newcommand{\eeqa}{\end{eqnarray}} 

\def\undertext#1{$\underline{\smash{\hbox{#1}}}$}

\newcommand{\modot}{M$_\odot$\hspace*{0.05cm}}
\def\sgreat{\lower3pt\hbox{$\buildrel {\scriptstyle >}
   \over {\scriptstyle\sim}$}}

\newcommand{\numberofmodels}{72\hspace{.12cm}}

\newcommand{\bari}{{ \scriptstyle \hspace{-0.45em} \stackrel{--}{}}} 

\def\sles{\lower2pt\hbox{$\buildrel {\scriptstyle <}
   \over {\scriptstyle\sim}$}}
\def\sgreater{\lower2pt\hbox{$\buildrel {\scriptstyle >}
   \over {\scriptstyle\sim}$}}

\def\sgreat{\lower2pt\hbox{$\buildrel {\scriptstyle >}
   \over {\scriptstyle\sim}$}}

\begin{document} 

\title{Gravitational Waves from Axisymmetric, Rotating Stellar Core Collapse}

\author{Christian D. Ott}
\affil{Institut f\"{u}r Theoretische Astrophysik, Universit\"{a}t Heidelberg,\\
Heidelberg, Germany; cott@ita.uni-heidelberg.de}
\author{Adam Burrows}
\affil{Steward Observatory, 
The University of Arizona, Tucson, AZ 85721;
burrows@zenith.as.arizona.edu}
\author{Eli Livne}
\affil{Hebrew University, Jerusalem, Israel;
eli@frodo.fiz.huji.ac.il
}
\author{Rolf Walder}
\affil{Steward Observatory, 
The University of Arizona, Tucson, AZ 85721;
rwalder@as.arizona.edu}

\vspace{.5cm}

\begin{abstract}

We have carried out an extensive set of two-dimensional, axisymmetric, purely-hydrodynamic calculations
of rotational stellar core collapse with a realistic, finite-temperature
nuclear equation of state and realistic massive star progenitor models.   
For each of the total number of \numberofmodels different simulations we performed, 
the gravitational wave signature was extracted via the quadrupole formula in the 
slow-motion, weak-field approximation.
We investigate the consequences of variation in the initial ratio of 
rotational kinetic energy to gravitational potential energy and in the initial degree of
differential rotation. Furthermore, we include in our model suite progenitors from 
recent evolutionary calculations that take into account the effects of rotation
and magnetic torques.  For each model, we calculate gravitational 
radiation wave forms, characteristic wave strain spectra,
energy spectra, final rotational profiles, and total radiated energy.
In addition, we compare our model signals with the anticipated sensitivities
of the 1st- and 2nd-generation LIGO detectors coming on line.  We find that
most of our models are detectable by LIGO from anywhere in the Milky Way.

\end{abstract}


\section{Introduction}

The classical constraints on core-collapse supernova theory are nucleosynthetic
yields, residue neutron star or black hole mass, explosion energy, neutrino signal, pulsar
fields, and pulsar kicks.  Any viable theory of supernova explosions
must in the long run reproduce these data.  However, we are still a long
way from this situation and much work remains before the roles of neutrinos,
multi-dimensional hydrodynamics, rotation, convection, and magnetic fields in the mechanism
of core-collapse supernovae are fully elucidated (Bethe and Wilson 1985; Herant et al. 1994;
Burrows, Hayes, and Fryxell 1995; Fryer et al. 1999; Fryer and Heger 2000;
Rampp and Janka 2000,2002; Liebend\"orfer et al. 2001a,b; Buras et al. 2003;
Thompson, Burrows, and Pinto 2003; Akiyama et al. 2003).  

There is, however, another quite dramatic potential constraint
on core-collapse supernovae: their gravitational radiation signatures.
Massive stars (ZAMS mass $\sgreat$ 8 \modot) develop degenerate  
cores in the final stages of nuclear burning and achieve the Chandrasekhar 
mass.  Gravitational collapse ensues, leading to dynamical compression to nuclear
densities, subsequent core bounce, and hydrodynamical shock wave generation.
These phenomena involve large masses at high velocities ($\sim c/4$) and great
accelerations.  Such dynamics, if only slightly aspherical, will lead
to copious gravitational wave emission and, arguably, to one of the most
distinctive features of core-collapse supernovae. The gravitational waveforms and 
associated spectra bear the direct stamp of the hydrodynamics and rotation 
of the core and speak volumes about internal supernova evolution.
Furthermore, they provide data that complement (temporally 
and spectrally) those from the neutrino pulse (which also originates from the core), enhancing
the diagnostic potential of each.

As the current generation of gravitational wave detectors comes on line, 
gravitational wave astronomy might soon be able to shed light on 
the supernova phenomenon.  Gravitational waves couple so weakly to matter that they 
propagate almost undistorted from their source in the ultra-dense collapsing 
and rebounding stellar core to detectors like LIGO (\citealt{gust:99}), VIRGO
 (\citealt{punturo:03}), GEO600 (\citealt{wilke:02}), and TAMA (\citealt{ando:01}) on Earth.  
No other physical signal, apart from neutrinos, can give comparable ``live'' dynamical data of 
a star's death. However, there is a major caveat: in order for gravitational waves 
to be emitted the collapsing core must have a sizable and rapidly varying asphericity, 
since gravitational radiation is of quadrupole nature (Misner, Thorne, and Wheeler 1973).

Fortunately, as astronomical observations have shown, most stars rotate (\citealt{fukuda:82}; 
\citealt{spruit:03}). Rotation can result in large asphericity at and around bounce
and, hence, provides hope that the emission of gravitational radiation from stellar
core collapse can be significant.  Furthermore, Rayleigh-Taylor-like convection in the protoneutron
star, the aspherical emission of neutrinos and post-bounce triaxial rotational instabilities 
are also potential sources of gravitational radiation.
Together these phenomena, with their characteristic spectral and temporal signatures,
make core-collapse supernovae promising and interesting generators of gravitational radiation.   

Early studies of gravitational wave emission from stellar core collapse 
used either spherically symmetric models and applied perturbation techniques to analyze 
the gravitational radiation (\citealt{turnerwagoner:79}; \citealt{seidelmoore:87}; 
\citealt{seideletal:88}) or applied semi-analytic methods to the aspherical Newtonian 
collapse of homogeneous and inhomogeneous, uniformly-rotating, and degenerate ellipsoids 
(Shapiro 1977; Saenz and Shapiro 1978,1979,1981; Moncrief 1979; Ipser and Managan 1984).
Subsequent studies were based on 2-D Newtonian hydrodynamic core-collapse
calculations with rotation and various simple treatments of the microphysics. \cite{muellerhille:81} 
and M\"uller (1982) performed a limited set of simulations with a 
finite-temperature equation of state (EOS) using initial 
models from stellar evolution calculations,  
and did not treat the neutrino physics. 
\cite{finnevans:90}, \cite{bonazzola:93}, \cite{yamadasato:95} 
and \cite{zm:97} studied the gravitational wave signature from collapsing 
$n=3$ polytropes with a simplified equation of state, consisting of a polytropic 
and a thermal part. In all of the above
studies, rotation was imposed upon spherically
symmetric initial models by an artificial rotation law. \cite{zm:97} performed the  
most comprehensive sweep through values of the rotation parameter $\beta$,  
defined by
\begin{equation}
\beta = \frac{E_{rot}}{|E_{grav}|}\, ,
\end{equation}
where $E_{rot}$ is the total rotational kinetic energy and $E_{grav}$
is the total gravitational energy.
They also varied the degree of initial differential rotation and  
used self-consistent initial models in rotational equilibrium, calculated 
using the method of Eriguchi and M\"uller (1985). Dimmelmeier et al. (2001a,b) 
extended the aforementioned study to general-relativistic gravity in the conformally-flat limit.

\cite{mm:91} accounted for electron capture on protons 
and employed an approximate neutrino leakage scheme.  Their 
limited set of detailed 2-D calculations used rotationally non-equilibrium 
initial models from stellar evolution calculations and a finite-temperature 
EOS (Hillebrandt and Wolff 1985). 
Using a smooth particle hydrodynamics code with a 
neutrino diffusion scheme and the  
Lattimer-Swesty EOS (Lattimer and Swesty 1991), Fryer and Heger (2000)
performed 2-D rotating collapse simulations.  \cite{fhh:02} followed this up by
computing the corresponding gravitational wave signature.  

In a recent Newtonian study, \cite{kotake:03} used an EOS based on the relativistic
mean field theory (\citealt{shen:98}). They took electron capture into account
 and made use of a leakage scheme for simplified neutrino transport. They performed
a limited set of calculations that employed realistic 15 \modot ZAMS progenitor
models of Woosley and Weaver (1995) onto which they imposed rotation by rotation
laws similar to those of \cite{mm:91}.

While most of the above studies were primarily concerned with the gravitational 
wave signatures due to core bounce itself, others have 
highlighted the gravitational wave signatures from later stages of supernova 
evolution and due to other phenomena.  Epstein (1978), Burrows and Hayes (1996),
and M\"uller and Janka (1997) explored the gravitational wave signature of anisotropic neutrino emission.
Burrows and Hayes (1996) and M\"uller and Janka (1997) studied  
aspherical convective motion behind the shock 
and in the protoneutron star.   Rampp, M\"uller, and Ruffert (1998), Brown (2001),
Centrella et al.\ (2001), and Shibata, Karino and Eriguchi (2002,2003) 
investigated triaxial instabilities in the 
quickly spinning neutron star remnant (New 2003).  

In this paper, we present the results from our 2-D axisymmetric purely hydrodynamical 
simulations of rotating stellar core collapse, performed with the  
code VULCAN/2D (\citealt{livne:93}).  For this study, we used  
the 11, 15 and 20 M$_{\odot}$ progenitor models of Woosley and Weaver (1995).
We imposed rotation using the same prescriptions employed by 
\cite{zm:97} and \cite{harry:02b}.  In addition, we used the recent rotating progenitor models 
of \cite{heger:00} and \cite{spruit:03}.  The latter are 1-D calculations 
that employ a prescription for angular momentum 
transport and mass loss in the evolving massive star,
but do not include the back-reaction of centrifugal 
effects on the dynamics after carbon burning.
All of our calculations, except for the comparison studies 
to previous work, have made use of the Lattimer-Swesty EOS in tabular form
(Thompson, Burrows, and Pinto 2003), as well as realistic 
progenitor structures. We have not included any approximate treatment of
electron capture and neutrino transport since these schemes (e.g., the leakage
scheme used in \citealt{kotake:03}) only crudely approximate
full neutrino transport.

In \S\ref{section:prog}, we review the progenitor model suite we 
have relied upon.  Section \ref{section:rotation} summarizes
the initial model rotational law that we used to set models into rotation.
In \S\ref{section:nummeth}, we provide an overview of our implementation of VULCAN/2D and 
discuss the two different equations of state, realistic and polytropic, 
that we have used. Section \ref{section:extract} 
deals with our method for the extraction of the gravitational wave signature from the 
hydrodynamic data. Section \ref{validation} covers our calculations with
polytropic models and polytropic equations of state.  These are provided
and compared with previous work in order to validate the methods and codes we have employed. 
In \S\ref{realistic}, our results with more realistic 
progenitors and a more detailed equation of state are presented 
and compared with those of previous studies.  We discuss the hydrodynamics
of rotating collapse, the generation of vortices, the damping effect
of a stalled shock wave, and the excitation of $l=2$ pulsations, as well
as the signatures of these hydrodynamic features in the gravitational wave pulse.  
Importantly, we provide model gravitational wave spectra
and estimate the detectability of these signals in the LIGO detector (\citealt{gust:99}).
We verify the two major types of wave forms and post-bounce behaviors
originally identified by Zwerger and M\"uller (1997) that 
depend upon the initial rotational energy and the degree of differential rotation.
Moreover, we derive the progenitor model dependence of the gravitiational wave signatures.
In \S\ref{section:omegaev}, we discuss the angular velocity profiles
associated with a subset of the models we have studied, 
identifying in particular the maximum spin rate and the magnitude and position of the maximum rotational 
shear that result from collapse. Section \ref{section:detection}
provides both a context in which to discuss the detection of the gravitational wave signatures
of collapse and estimates of their detectability in the galaxy.
In \S\ref{summary}, we summarize our conclusions
concerning the generic features of the gravitational wave signature of core-collapse
supernovae.  A major goal of our study is the illumination of the importance
of using realistic progenitors and equations of state when deriving
and analyzing the gravitational wave forms and spectra from supernova explosions.   
Signal templates derived using polytropes and a polytropic EOS can
deviate significantly from those obtained using more realistic assumptions
and starting points.  A further goal of our study is the derivation
of the systematic variation with the degree of initial rotation of
a supernova's gravitational wave form and its spectrum.

\section{Progenitor Models}
\label{section:prog}

Since this study is concerned with the gravitational wave signal originating from the 
highly dynamical aspherical bulk mass motions at core collapse and bounce and does not 
aim at the solution of the full supernova problem, it is sensible to restrict our 
simulations to the modeling of the central rotating iron core 
whose hydrodynamical collapse and bounce is believed to yield the dominant contribution to the 
gravitational wave energy emitted during the supernova phenomenon.
A simple approach for constructing progenitor data used in many 
previous studies is to approximate the iron core as an 
ultra-relativistic, electron-degenerate Chandrasekhar core with a central density 
of $10^{10}$~g~cm$^{-3}$ and an electron number fraction ($Y_e$) of 0.5. The equation
of state for such a core is then polytropic with a $\Gamma$ of  4/3 ($n=3$) and  
a polytropic constant $K$ given by eq. (\ref{eq:polyk}) 
(\citealt{shapteu:83}). We have made use of this approximation in conjunction 
with the hybrid equation of state described in \S\ref{section:hybrideos} to 
show that our numerical model yields results that match those of \cite{zm:97}.

\begin{figure}
\epsscale{1.0}
\plotone{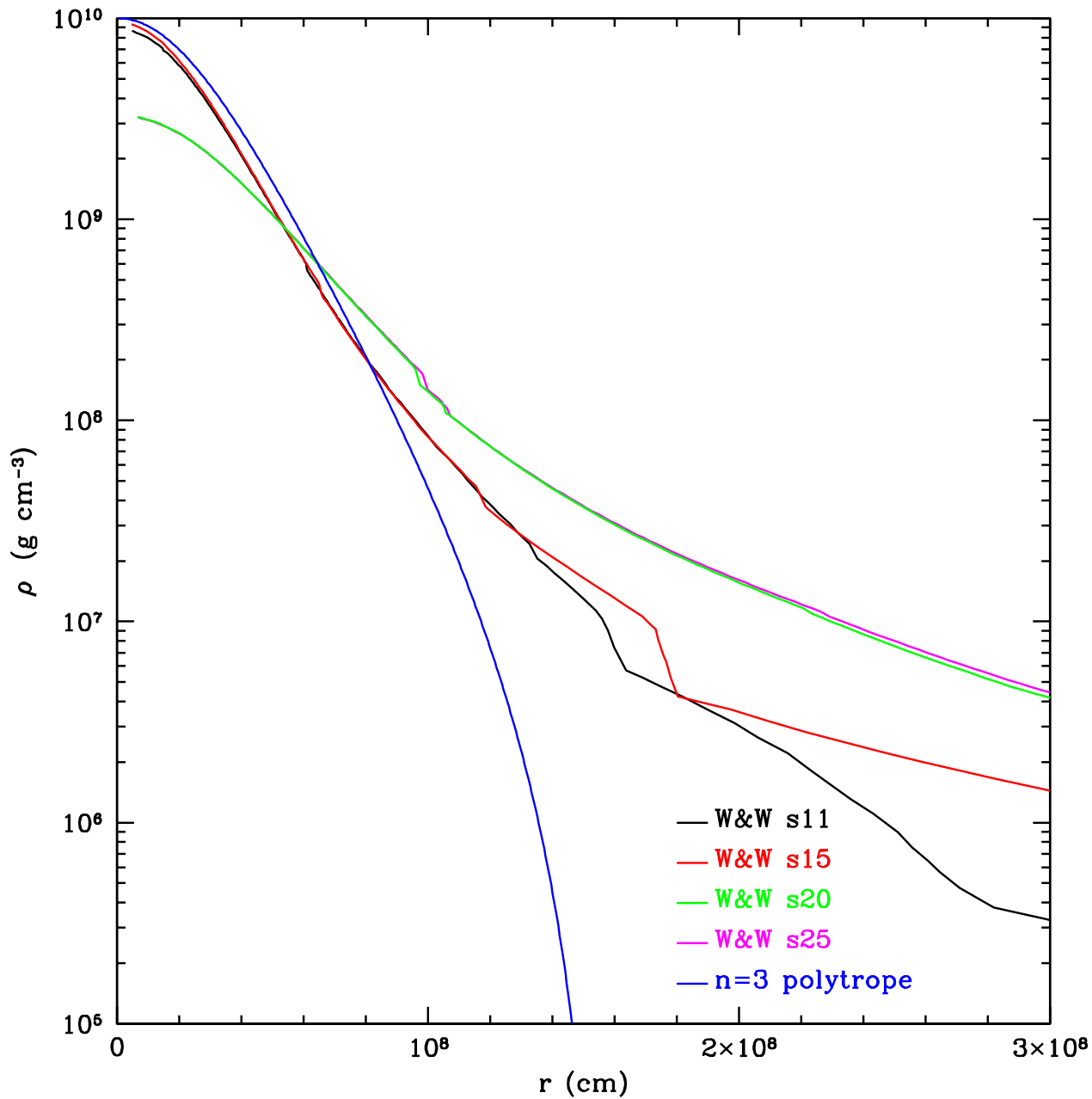}
\caption{Mass density as a function of radius of the
Woosley and Weaver (1995) progenitor models and the $n=3$ 
polytropic progenitor of Zwerger and M\"uller (1997). 
Note the good correspondence in the central regions 
between the less massive iron cores and the 
polytrope, but the general disagreement at 
larger radii. The polytrope has a sharper edge. \label{fig:initialrho}}

\end{figure}

\begin{figure}
\epsscale{1.0}
\plotone{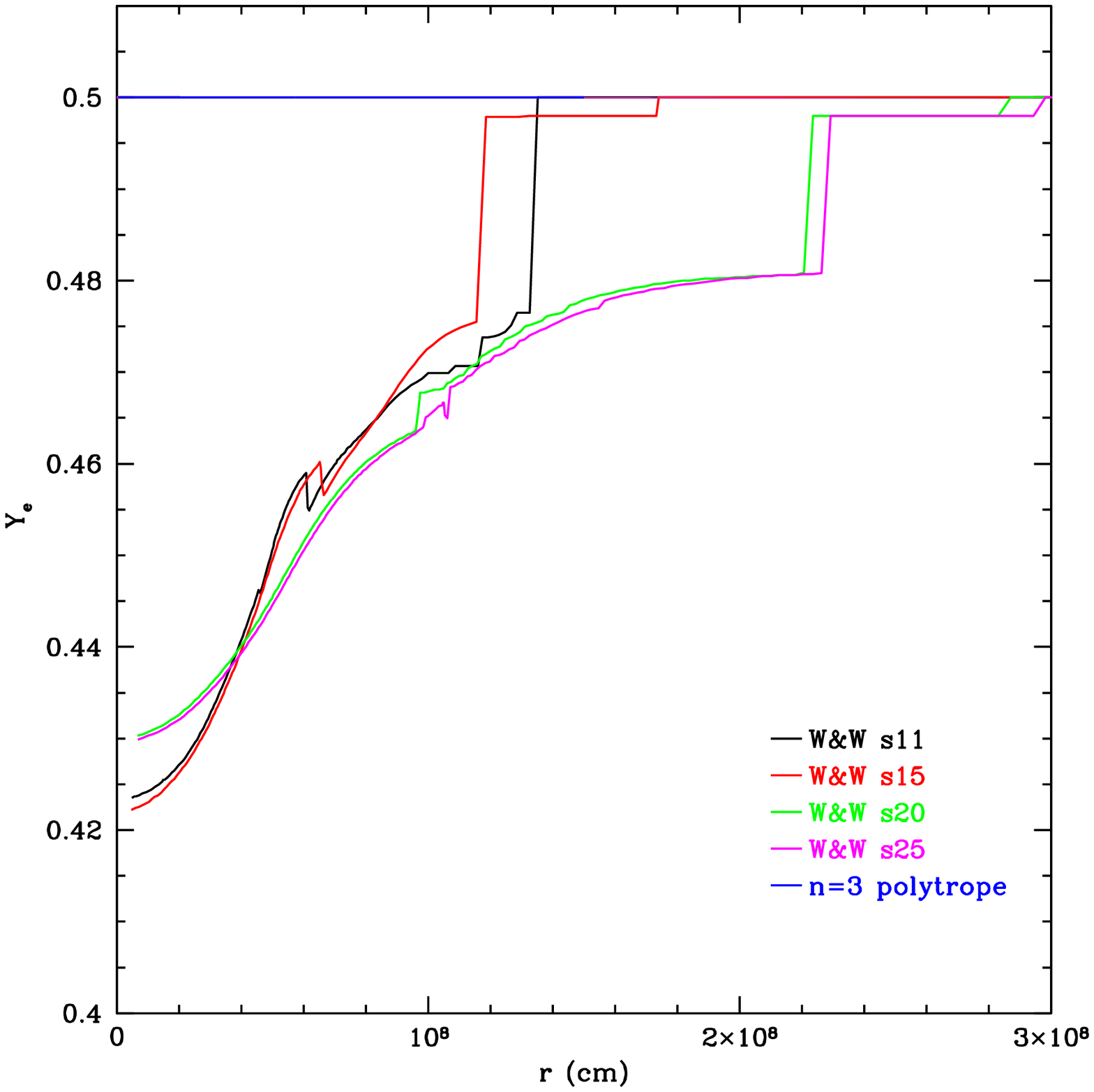}

\caption{Electron number fraction $Y_e$ as a function 
of radius for the Woosley and Weaver (1995) progenitor models. The central 
iron core has been neutronized due to electron capture during 
core silicon burning (\citealt{wwz:78}). For the polytrope, $Y_e$ 
has been assumed to be 0.5 (\citealt{zm:97}). The discontinuity 
in $Y_e$ exhibited in the Woosley and Weaver models indicates the 
approximate boundary between the central iron core and the surrounding  
shells. \label{fig:initialye}
}
\end{figure}

However, nature is not this simple and detailed stellar evolution calculations 
(\citealt{wwh:02}) have shown 
that the iron cores of evolved massive stars are not perfect polytropes. 
Rather, they have a complicated thermodynamic and compositional structure. Fig. 
\ref{fig:initialrho} depicts the differences between the density 
profiles of progenitor models from \cite{ww:95} and the density profile of the 
polytrope used by \cite{zm:97}. In Fig. \ref{fig:initialye}, we show the profiles of the electron 
fraction ($Y_e$) of the \cite{ww:95} 
progenitors, and contrast them with the flat $Y_e=0.5$ profile of the Zwerger and M\"uller 
polytrope. 

For our study we have used 11, 15, and 20 \modot presupernova 
models from \cite{ww:95}. As Figure \ref{fig:initialrho} demonstrates, 
the pre-collapse core density and $Y_e$ profiles 
of the Woosley and Weaver (1995) 25 and 20 \modot models are
very similar.  The temperature profiles also match.  Thus, one 
would expect very similar collapse dynamics and gravitational wave signatures. A test 
calculation with the 25 \modot progenitor bears out this expectation 
(Table \ref{table:resultss20}). 
\section{Rotation}
\label{section:rotation}

Rotation is a key agent in the development of asphericity 
during core collapse and, hence, for the emission of gravitational radiation. Depending 
on the initial rotation rate and the angular momentum distribution, conservation of angular 
momentum may lead to very rapidly rotating compact remnants, which are unstable on secular, 
or even dynamical, timescales. In the approximation of MacLaurin spheroids (incompressible, 
uniform density, rigidly rotating equilibrium configurations), triaxial instabilities may grow if $\beta \ge 0.14$
and $\beta \ge 0.27$ for secular and dynamical instabilities, respectively  
(\citealt{tassoul:78}). Recently, \cite{cent:01} have shown that models with off-center 
density maxima (i.e. with a toroidal structure) can already become dynamically unstable 
at values of $\beta \ge 0.14$. However, the critical $\beta$s required for either secular or dynamical
instability were derived in the past using either constant-density models or very compact cores
whose deviation from uniform density was modest. Recently, Shibata, Karino and Eriguchi (2003) 
have shown
that rotating polytropes with centrally located density maxima 
can become dynamically unstable even for $\beta$ on the order of 1\% if they are strongly
differentially rotating. In this paper, we calculate the total $\beta$s
for an entire realistic iron core with a large dynamic range of densities ($\sim$eight
orders of magnitude).  This is not the $\beta$ for only the inner, more uniform, core.
The outer regions of realistic iron cores do not move much
during the crucial dynamical phases of the inner core important for
the estimation of the gravitational wave signature.  Importantly, the actual critical
$\beta$s necessary for triaxial deformation have yet to be determined
for such structures.
To our knowledge, all investigations to date of the growth 
of triaxial instabilities have lacked either realistic initial models or sophisticated
equations of state. These limitations should be kept in mind when assessing previous work.

The centrifugal forces connected to rotation do not only cause asphericity, but slow down 
the core collapse and may, provided the configuration has the right angular momentum distribution, 
stop  the collapse before nuclear matter density is reached (``subnuclear bounce''). 
A critical condition for the stabilizing effect of rotation 
on (pseudo-) radial modes of stars in the 
Newtonian regime is
\begin{equation}
\label{eq:gammacrit}
\Gamma > \Gamma_{crit} = \frac{2}{3} \frac{(2-5\beta)}{(1-2\beta)},
\end{equation}
where $\Gamma$ is the effective adiabatic index that describes the change of 
pressure along a collapse trajectory of a given mass element:
\begin{equation}
\Gamma = \frac{\partial \ln P}{\partial \ln \rho}\bigg|_M
\end{equation}
(Ledoux 1945; \citealt{tohline:84}; \citealt{mm:91}). 
Hence, for a given progenitor structure and
equation of state with an effective adiabatic index $\Gamma$, there is a critical
value of the rotation parameter above which the configuration is stable against collapse:
\begin{equation}
\label{eq:betacrit}
\beta > \beta_{crit} = \frac{1}{2}\frac{(4-3\Gamma)}{(5-3\Gamma)} \, .
\end{equation}

It is known that massive stars on the main sequence rotate rapidly, 
with typical equatorial rotational velocities of $\sim$200~km~s$^{-1}$ (\citealt{fukuda:82}). 
This is a significant fraction of their breakup velocity. Unfortunately, since  
observations of the stellar surface tell us little about the  
angular momentum of the stellar interior (or its evolution), 
one has to rely on parameter dependent, semi-phenomenological 
prescriptions to follow these quantities numerically. 
\cite{heger:00} have built upon the one-dimensional calculations of \cite{ww:95} and
incorporate a prescription for angular momentum transport. We include
in our progenitor model suite their ``rotating''
15 and 20 \modot progenitor models. \cite{spruit:03} have extended the work of \cite{heger:02}
by the inclusion of the effects of magnetic torques on rotational
evolution and provide corresponding cores for 15, 20
and 25 \modot models. Table \ref{initial.tab} lists all the presupernova 
models we have employed in this study.
Since the gravitational wave signature is
sensitive to the distribution of angular momentum 
throughout the iron core, gravitational waves
may eventually be used to learn about the interior rotational
structure of massive stars.

We have used two different approaches to include rotation in our calculations. First, we 
follow \cite{zm:97} in forcing the one-dimensional initial models to rotate with constant
angular velocity on cylinders on our axisymmetric grid according to the rotation law
\begin{equation}
\label{eq:rotlaw}
\Omega(r) = \Omega_0 \, \bigg[ 1 + \bigg(\frac{r}{{\rm A}}\bigg)^2 \bigg]^{-1}\, ,
\end{equation}
where $\Omega(r)$ is the angular velocity, $r$ is the distance from 
the rotation axis, and $\Omega_0$ and A are free parameters that 
determine the rotational speed/energy of the model and the 
distribution of angular momentum.
Large values of A lead to very rigid rotation, small values 
to strongly differential rotation. 
Our parameter studies are performed over a range of $\beta_i$ and A,
where $\beta_i$ is the initial $\beta$ of the model.  The choices for  
$\beta_i$ and A were based upon current knowledge of iron core rotation, but 
exclude strongly differential
rotation (\citealt{spruit:03}; \citealt{heger:00}). Note that there has 
been some confusion in the
literature concerning the meaning of $r$ in eq. \ref{eq:rotlaw}. \cite{mm:91} and 
\cite{kotake:03} interpreted
$r$ as radial distance from the origin, whereas \cite{zm:97} and \cite{harry:02b} understood it
as distance from the rotation axis. We follow the latter definition of $r$ 
as it accords with the
Poincar\'{e}-Wavre theorem which predicts that the specific angular momentum is constant on cylinders
for degenerate rotating objects (for a review see Tassoul 1978).

We name our runs according to the following convention: 
[initial model name]{\rm A}[in km]$\beta_i$[in \%]. For example, s11A1000$\beta$0.3 is a Woosley
and Weaver (1995) 11 \modot model with A=1000 km and an initial $\beta_i$ of 0.3\%.

In contrast to \cite{zm:97}, we do not use rotational equilibrium configurations, since these
can only be found consistently for models with constant entropy and $Y_e$ (\citealt{hach:86}). 
For direct comparisons
with \cite{zm:97} we used models with small initial rotation 
rates ($\beta_i$) in which the initial deformation due to rotation 
would be negligible. \cite{zm:97} found a maximum 
difference of 10\% in the central densities at bounce for a strongly rotating model 
evolved with and one evolved without an initial rotational equilibrium configuration. 
Since the progenitor models with $\beta_i$ \sgreater$\,$ 1\% tend not to collapse 
($\beta_i$ > $\beta_{crit}$, if nothing artificial is done to alter their structure),
we have limited our study to models with $\beta_i \le$ 1\%.  Hence, we expect the error that 
is introduced by the non-equilibrium 
rotational configuration at the onset of collapse
to be very small and, in the worst case, to be on the order of a few percent. 
In addition to this, we point out that \cite{zm:97} argue that the use of 
non-equilibrium models is justified if the stellar core collapses slowly enough 
to allow for the adjustment to the appropriate angular 
density stratification for its rate of rotation. This is certainly the case for 
our models, which all collapse on a timescale on the order of 100-500 milliseconds (ms).

Secondly, we have made use of the recent presupernova models of \cite{heger:00} and 
\cite{spruit:03} that, though they are intrinsically one-dimensional, take into 
account the effects of centrifugal 
forces on the stellar structure before carbon burning ends. 
Furthermore, redistribution of angular momentum and chemical species 
were modeled using a set of prescriptions and assumptions for mixing and transport processes. In 
particular, all torques were assumed to lead to rigid rotation on some physical  
timescale (Fryer and Heger 2000). The ``magnetic'' models of \cite{spruit:03} 
assume a magnetic dynamo process that generates fields which inhibit differential 
rotation and lead to slower core rotation at collapse. 

\begin{figure}
\plotone{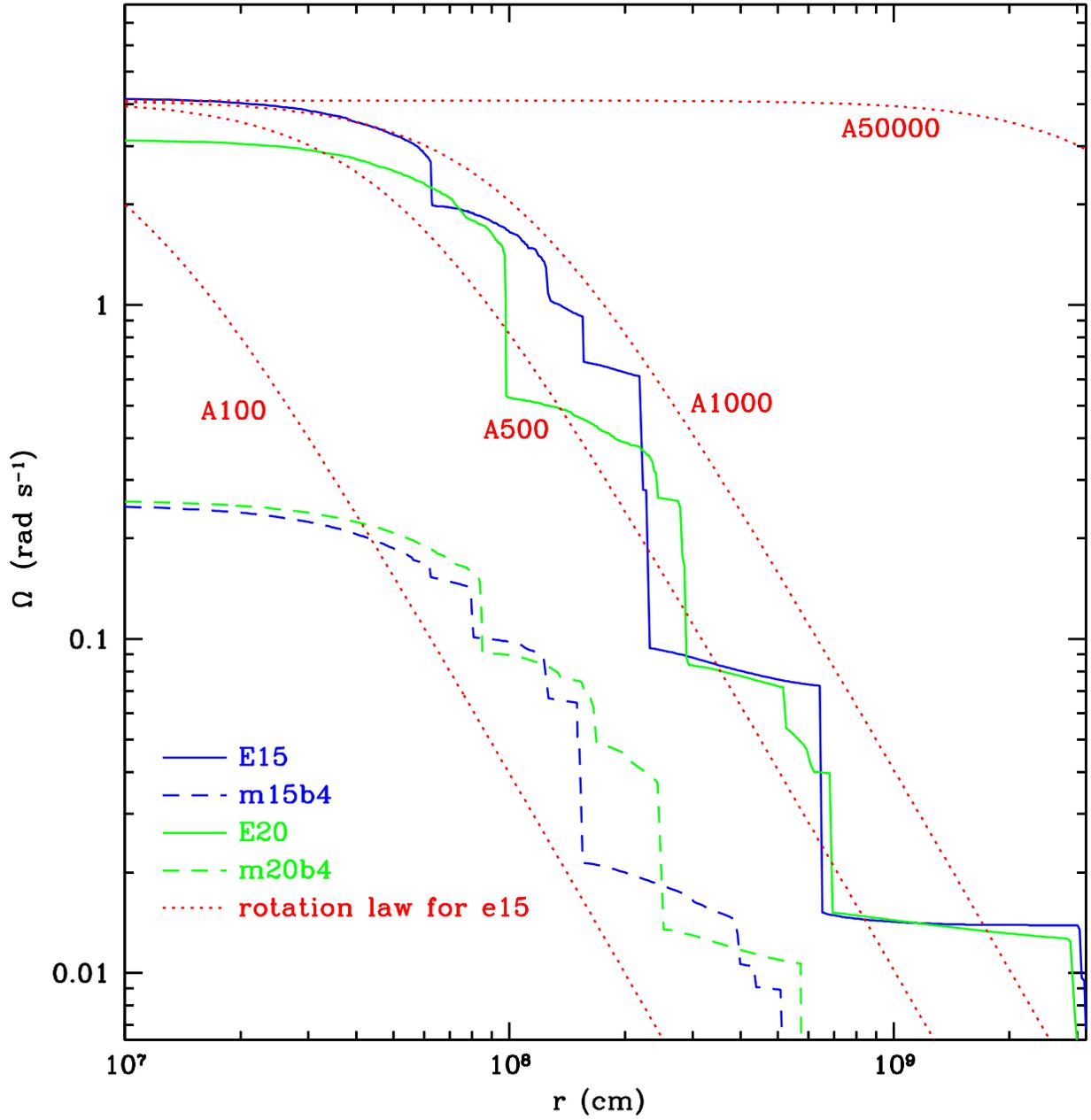}
\caption{\label{fig:initialomega} Initial angular velocity profiles of 
the rotating 15 (blue) and 20 (green) \modot progenitor models 
(see Table \ref{table:initialmodels} for model 
parameters). The dotted red profiles were generated with the rotation
law of eq. (\ref{eq:rotlaw}) using the central $\Omega$ of model e15 for $\Omega_0$.
All realistic presupernova models exhibit near rigid rotation inside $\simeq$ 1000 km. 
Note the much smaller angular velocities exhibited by models m15b4 and m20b4, 
which were evolved with the inclusion of magnetic fields. }
\end{figure}

In Fig. \ref{fig:initialomega}, the profiles for selected models of the initial 
angular velocity versus radius are shown. Note that
the differences due to different progenitor masses are negligible compared with the order-of-magnitude
differences introduced by the inclusion of magnetic-field effects during stellar evolution.  
One should be cautious, however, in accepting these results since 
research on stellar evolution with rotation is still 
in its infancy.

\section{Numerical Techniques}
\label{section:nummeth}

\subsection{Equations of State}
\label{section:hybrideos}
\label{section:lseos}

For all our calculations involving realistic progenitor models we
have made use of the equation of state of \cite{lseos:91} (the LSEOS). It is based on the 
finite-temperature liquid drop model of nuclei developed in \cite{lattimeretal:85}. Our particular
implementation is the one presented in Thompson, Burrows, 
and Pinto (2003) that uses a three-dimensional table in
temperature ($T$), density ($\rho$), and $Y_e$. At each point in the table 
the specific internal energy, the pressure ($P$), the entropy per baryon ($s$), 
and compositional information are stored. Using integer arithmetic to find nearest neighbor
points for a given set of $\rho,T,Y_e$, the need for time-consuming search algorithms
has been eliminated. Given $\rho, T$ and $Y_e$, the code performs three six-point 
bivariant interpolations
in the $T\,-\,\rho$ planes nearest to and bracketing the given $Y_e$ point. 
A quadratic interpolation
is then executed between $Y_e$ points to obtain the desired 
thermodynamic quantity. Since our hydrodynamic
routine updates specific internal energy, we employ a Newton-Raphson/bisection 
scheme which iterates 
on temperature at a fixed internal energy until the root is found to within a part in $10^8$.

The LSEOS extends down to only $\sim 5 \times 10^6$ g cm$^{-3}$ and its validity 
in this density regime is guaranteed only for fairly high temperatures, where the 
assumption of nuclear statistical 
equilibrium (NSE) still holds. For calculations involving lower densities, Thompson, Burrows, 
and Pinto (2003) have coupled the LSEOS 
to the Helmholtz EOS (\citealt{timmesarnett:99}; \citealt{timmesswesty:00}), 
which contains electrons
and positrons at arbitrary degeneracy and relativity, photons, nuclei and 
nucleons as non-relativistic
ideal gases, and Coulomb corrections.

To facilitate the comparison of our results with those of \cite{zm:97} and \cite{harry:02b}, we 
have implemented the ``hybrid'' equation of state used in those studies of collapsing 
$\Gamma = 4/3$  ($n=3$) polytropes (Janka, Zwerger, and M\"onchmeyer 1993). It consists 
of a polytropic part $P_p$ and a thermal contribution $P_{th}$
\begin{equation}
\label{eq:zmpress}
P = P_p + P_{th}\,.
\end{equation}
The thermal part accounts for the thermal pressure of the high-entropy material heated
by the bounce shock and is given by
\begin{equation}
\label{eq:P_th}
P_{th} = (\Gamma_{th} - 1)\, u_{th}\, ,
\end{equation}
where $\Gamma_{th}$ is set to 1.5. The thermal energy density $u_{th}$ is given by 
the total energy density $u$ through the relation
\begin{equation}
\label{eq:u}
u = u_p + u_{th}\, ,
\end{equation}
where $u_p$ is the energy density of the degenerate electron gas.

The polytropic part 
\begin{equation}
\label{eq:P_p}
P_p = K\rho^{\Gamma}
\end{equation}
reflects the pressure contributions due to the degenerate and relativistic electrons and 
(in the regime of nuclear density) the repulsive action of nuclear forces. The polytropic
 constant is initially set to
\begin{equation}
\label{eq:polyk}
K = \frac{3}{4} \bigg(\frac{\pi}{3}\bigg)^{2/3} \hbar c \bigg(\frac{Y_e}{m_B}\bigg)^{4/3} 
= 1.2435 \times 10^{15}\, Y_e^{4/3} cgs\, ,
\end{equation}
where $Y_e$ is the electron number fraction and the other quantities have their 
usual meaning (\citealt{shapteu:83}). $\Gamma$ is chosen to be 1.32 to
initiate the collapse of the $n=3$ polytrope. $Y_e$ is set to $0.5$. To mimic the stiffening
of the equation of state at nuclear density 
(set to $\rho_{nuc} = 2.0 \cdot 10^{14}$ g cm$^{-3}$),
$\Gamma$ is for $\rho \ge \rho_{nuc}$ set to $2.5$ and $K$ is modified by the requirement of
continuity of the thermodynamic variables at $\rho_{nuc}$ (Janka, Zwerger, 
and M\"onchmeyer 1993).  

\subsection{Hydrodynamics - VULCAN/2D}

Our simulations were performed with the Newtonian two-dimensional finite-volume hydrodynamic
code VULCAN/2D developed by Eli Livne (\citealt{livne:93}).  VULCAN/2D uses a scalar von
Neumann-Richtmyer artificial viscosity scheme for shock handling. 
The hydrodynamic equations are solved in the Lagrangian formulation and the hydrodynamical
data are remapped after each time step onto a fixed Eulerian grid.
VULCAN/2D can be run in implicit or explicit
time integration mode. Since we are dealing with supersonic flows, we use VULCAN/2D in
explicit mode. VULCAN/2D is second-order accurate in time and space and has been rigorously 
tested and compared with one-dimensional Lagrangian and Eulerian hydrodynamic 
codes (\citealt{livne:93}).

\begin{figure}
\epsscale{1.0}
\plotone{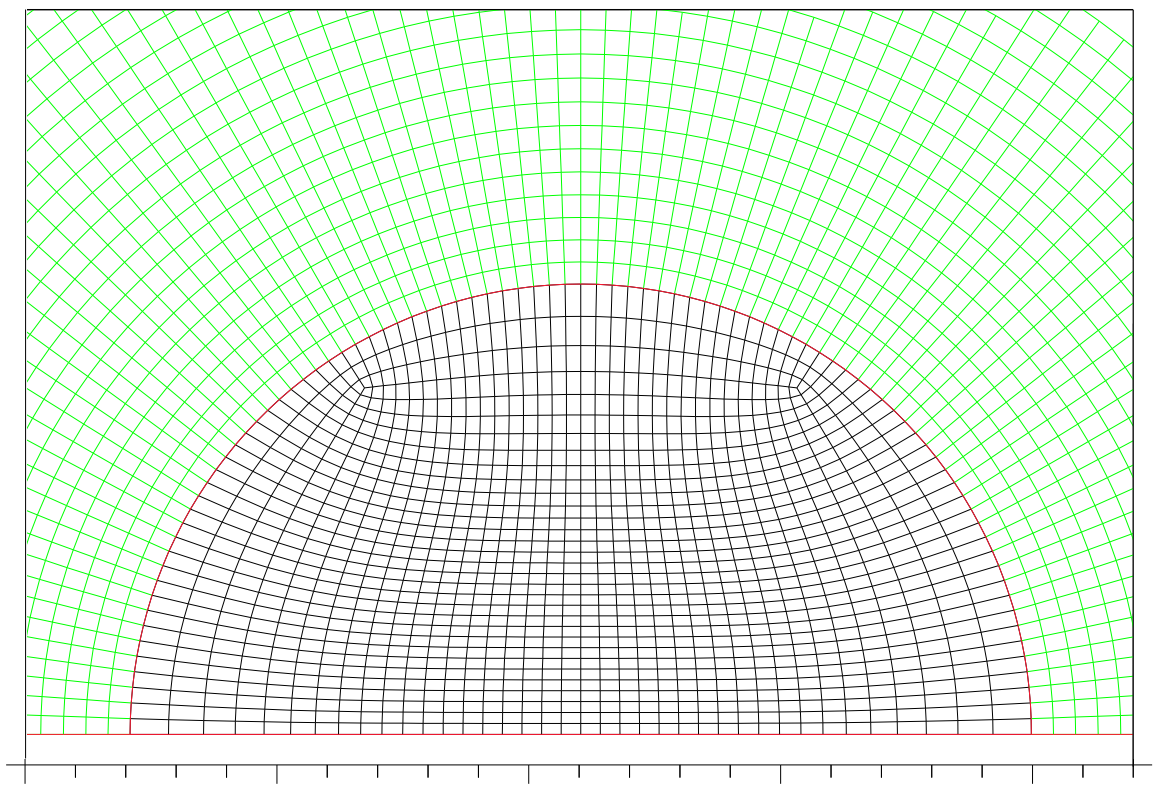}
\caption{Central region of the VULCAN/2D computational grid. Shown in black 
is the inner 10 km, where the capability of VULCAN/2D to work with arbitrary 
grid configurations has been used to perform
a smooth transition from the outer radial grid to a Cartesian grid 
in the innermost region.\label{fig:grid}}
\end{figure}

A feature of VULCAN/2D is its ability to deal with arbitrarily shaped grids while using
cylindrical coordinates. For the problem of stellar collapse we have chosen a polar grid
with logarithmic spacing for the region outside the inner 10 km and an inner Cartesian
grid (see Fig. \ref{fig:grid}).  In this way we circumvent severe time step constraints that
would be imposed in the angular direction in the central region of a regular polar grid. 
The price we pay for this is the ``horns'' seen
in Fig. \ref{fig:grid} in the central region which are 
a consequence of the demand for continuity at the boundary
between the central and the outer grid.

For our production runs we use 412 radial and 91 angular zones (including the central
region), encompassing 1500 to 3000 km in radial extent 
(depending on the initial model) and covering the full 180 degrees of the
symmetry domain.  In our comparisons with \cite{zm:97}, this resolution has 
been shown to reproduce their results to better than 10\%.

\subsection{Gravitational Wave Signature Extraction and Waveforms}
\label{section:extract}

We have calculated the gravitational wave field in the 
slow-motion, weak-field quadrupole approximation (Misner, Thorne, and Wheeler 1973). 
The dimensionless gravitational wave strain $h$ is 
\begin{equation}
\label{eq:sqf}
h^{TT}_{ij}(\mathbf{\vec{D}},t) = \frac{2G}{Dc^4}\,
\ddot{I}{\bari}\,\,\,^{TT}_{ij}(t - \frac{D}{c})\, ,
\end{equation}
where $D=|\mathbf{\vec{D}}|$ is the distance between the observer and the source and 
\begin{equation}
\label{eq:itensor}
I{\bari}\,^{TT}_{ij} = P_{ijkl}(\mathbf{\vec{N}}) \int d^3\mathbf{x}\,\, \rho\,\, 
\bigg[\, x_k x_l
 - \frac{1}{3}\delta_{ij}\,x_mx^m\, \bigg]\end{equation}
is the transverse-traceless part of the reduced Cartesian mass-quadrupole tensor. 
$P_{ijkl}(\mathbf{\vec{N}})$ (with $\mathbf{\vec{N}} = \mathbf{\vec{D}}/D$) is the transverse-traceless (TT) 
projection operator onto the plane orthogonal to the outgoing wave direction 
$\mathbf{N}$ and is of the form:
\beqa
P_{ijkl}(\mathbf{\vec{N}}) = (\delta_{ik} - N_i N_k)(\delta_{jl} - N_jN_l) \nonumber \\
- \frac{1}{2}(\delta_{ij} - N_iN_j)(\delta_{kl} - N_k N_l)\, .
\eeqa

Direct application of eq. (\ref{eq:sqf}) (known in the literature as the 
``standard quadrupole formula'' (SQF)) in a numerical fluid dynamics calculation
is problematic, since numerically troublesome second time derivatives of the 
quadrupole moment are involved and the moment arm emphasizes  
contributions of low-density material far from the central regions (Finn and Evans 1990). 

Using the Euler equations of inviscid hydrodynamics, Finn and Evans (1990), Nakamura and Oohara (1989),
and Blanchet, Damour, and Sch\"afer (1990) derived formulations of  
the quadrupole formula involving either only one time derivative and easier, 
more tractable, spatial derivatives or spatial derivatives of the 
hydrodynamic observables exclusively. We use the formulation of Nakamura and Oohara (1989)
and Blanchet, Damour, and Sch\"afer (1990):   
\beqa
\label{eq:blanchet}
h^{TT}_{ij}(\mathbf{\vec{D}},t) = \frac{2G}{Dc^4}\,P_{ijkl}(\mathbf{\vec{N}}) \times \nonumber \\
\int\,d^3 \mathbf{x}\,\,\rho\,\,\bigg[\,2v^kv^l - x^k\, \partial_l \Phi - x^l\, \partial_k \Phi\,\bigg]\, ,
\eeqa
where $\Phi$ is the Newtonian gravitational potential, $\rho$ is the 
mass-density, and $v$ the velocity. 

For our 2-dimensional axisymmetric calculations, it is useful to rewrite
the full gravitational radiation field in 
terms of the ``pure-spin tensor harmonics'' $T_{ij}^{E2,lm}$ and 
$T_{ij}^{B2,lm}$ (\citealt{thorne:80}; \citealt{mm:91}):
\beqa
\label{eq:tensorharmonics}
\tilde{h}^{TT}_{ij}(\mathbf{\vec{D}},t) = \frac{1}{D} \sum^{\infty}_{l=2} 
\sum^{l}_{m=-l} \bigg[ \,\, A_{lm}^{E2} (t-\frac{D}{c}) T_{ij}^{E2,lm}
(\theta,\phi) + \nonumber \\
A_{lm}^{M2} (t-\frac{D}{c}) T_{ij}^{M2,lm}(\theta,\phi) \bigg] \, .
\eeqa

The coefficients $A_{lm}^{E2}$ and $A_{lm}^{M2}$ represent the mass quadrupole 
and the mass-current quadrupole contributions, respectively. In the quadrupole approximation,
higher-order as well as mass-current contributions are neglected and due to the 
assumption of axisymmetry only one non-vanishing term remains in eq. 
(\ref{eq:tensorharmonics}), namely $A_{20}^{E2}$. By comparing eq. 
(\ref{eq:blanchet}) with the lowest-order term of eq. (\ref{eq:tensorharmonics}),
M\"onchmeyer et al. (1991) write $A_{20}^{E2}$ in terms of the hydrodynamic variables:
\beqa
\label{eq:MM}
A_{20}^{E2} =   \frac{16\, \pi^{3/2}}{\sqrt{15}}\frac{G}{c^4}  \int_{-1}^1\!\! 
\int_0^{\infty}r^2 d\mu \, dr \, \cdot \rho \, \cdot \, \nonumber \\
\Big[ v_r^2(3 \mu^2 -1)+v_{\theta}^2(2-3\mu^2)-v_{\phi}^2 
-6\, v_r v_{\theta} \mu \sqrt{1-\mu^2} \nonumber \\
- r\, \partial_r \Phi\, (3 \mu^2 - 1) + 3\, \partial_{\theta}
\Phi\, \mu \sqrt{1 - \mu^2} \Big]\, ,
\eeqa
where $\mu = \cos{\theta}$ and $v_r$, $v_\theta$, and $v_\phi$ are the components
of the velocity vector in the $r$, $\theta$, and $\phi$ 
directions. Furthermore, $\partial_r = \partial /\partial r$ and 
$\partial_\theta=\partial/\partial\theta$. The components of the approximate 
gravitational wave field $h^{TT}$ are then given by (\citealt{thorne:80}; \citealt{mm:91}):
\begin{equation}
\label{eq:MM2}
h_{\theta \theta}^{TT} = \frac{1}{8}\sqrt{\frac{15}{\pi}}\sin^2 \alpha\,\, \frac{A_{20}^{E2}}{D} \equiv\, h_{+}\, ,
\end{equation}
where $\alpha$ is the angle between the symmetry axis and the line of sight 
of the observer. The only other nonzero component is 
$h_{\phi\phi}^{TT} = - h_{\theta\theta}^{TT} = - h_+$.   $h_\times$ equals zero,
due to the assumption of axisymmetry. $h_+$ and $h_\times$ are the dimensionless wave 
strains corresponding to the two independent polarizations of the 
gravitational radiation field (Misner, Thorne, and Wheeler 1973).

The total gravitational energy radiated over time is given by
\begin{equation}
\label{eq:egw}
E_{GW} = \frac{c^3}{32\pi G} \int_{-\infty}^{\infty}\,
\bigg|\frac{dA^{E2}_{20}}{dt}\bigg|^2 dt\,\,.
\end{equation}

As an alternative to eqs. (\ref{eq:MM}) and (\ref{eq:MM2}), we also 
implemented the ``first moment of momentum divergence'' formula of 
Finn and Evans (1990) (eq. 38 of their paper):
\beqa
\label{eq:finnevans1}
{d I\bari_{zz} \over dt} \ = \ {4\pi \over 3} \int_{-1}^{1} d\mu
\int_{0}^{\infty} dr\,\, r^{3}\, \rho \, \times \nonumber \\ 
\left[ P_{2}(\mu)v_{r} + {1 \over 2}
{\partial P_{2}(\mu ) \over \partial\theta } v_{\theta} \right] \, ,
\eeqa
where $P_2(\mu)$ is the second Legendre polynomial in $\mu$ and 
$I \bari _{zz}$ is the zz-component of the reduced mass-quadrupole 
moment tensor. The gravitational wave strain is then obtained through
\begin{equation}
\label{eq:finnevans2}
h_{\theta \theta}^{TT} = \frac{6G}{Dc^4}\, 
\sin^2 \alpha \frac{d^2}{dt^2}\, I\bari_{zz}\,\,. 
\end{equation}

\subsubsection{Energy Spectra}
\label{spectra}

Writing $A_{20}^{E2}$ in eq. (\ref{eq:egw}) in terms of the inverse Fourier transform
\begin{equation}
{A}_{20}^{E2}(t) = \int_{-\infty}^{\infty}\,\tilde{A}_{20}^{E2}(f) e^{- 2\pi i f t} dt
\end{equation}
and after several algebraic steps, we obtain:
\begin{equation}
\label{eq:dEdnu}
\frac{dE(f)}{df} =  \frac{c^3}{G}\,\frac{(2\pi f)^2}{16\pi} 
\bigg|\tilde{A}_{20}^{E2}(f)\bigg|^2\\
\end{equation}
in terms of the Fourier transform
\begin{equation}
\tilde{A}_{20}^{E2}(f) = \int_{-\infty}^{\infty}\,A_{20}^{E2}(t) e^{2\pi if t} df \,\,.
\end{equation}
The total radiated energy is then obtained from the integral over the energy spectrum
\begin{equation}
\label{eq:dEdnuE}
E_{GW} = \int_{0}^{\infty} \frac{dE(f)}{df} df\, ,
\end{equation}
which should be identical to the result obtained from eq.~(\ref{eq:egw}).

We have implemented eq. (\ref{eq:dEdnu}) using the Fast 
Fourier Transform (FFT) technique (Press et al. 1992). 
Since the wave amplitude is calculated at unequal time intervals 
due to variations in the time step, 
we first interpolate the data onto an evenly-spaced temporal grid
before applying the FFT.  We have verified that the value of $E_{GW}$ 
obtained using eq. (\ref{eq:egw}) is always within a few percent of that 
obtained using eq. (\ref{eq:dEdnuE}).

\section{Method Validation - Polytropes}
\label{validation}

Table \ref{table:polytropes} summarizes the results of our comparisons with the work of 
\cite{zm:97}, who used simple $n=3$ polytropes and the hybrid
EOS discussed in \S\ref{section:hybrideos}.    
Our results match those of \cite{zm:97} in density at bounce ($\rho_{max}$), 
maximum gravitational wave amplitude ($A^{E2}_{20}\,_{max}$) and total gravitational
wave energy ($E_{GW}$), in most cases to better than 10\%. The largest difference in 
maximum density is 22\% and is found in model A500$\beta$0.9 
(Zwerger and M\"uller's A3B3G2). However, \cite{harry:02b} performed a similar
comparison study with their code and obtained for this model a density at bounce much closer
to ours. As to $E_{GW}$, our models generally yield larger values
than those of \cite{zm:97}, since ours were evolved for longer periods of time after 
bounce, thus tracking a greater part of the post-bounce aspherical motion of the
compact remnant.

Using the publicly available data of the Newtonian runs of \cite{harry:02b},
we find that our waveforms match perfectly up to bounce and then start to exhibit a
slight shift in post-bounce maxima, minima and periods. We attribute
these differences in part to the fact that our artificial viscosity scheme for hydrodynamic shock
capture leads to slightly greater artificial damping than the piecewise parabolic method (PPM) used in the
cited work and to our simulation on the full 180$^{\circ}$ domain.

\section{Results Using Realistic Progenitor Models}
\label{realistic}

Collapse dynamics of rotating supernova progenitors is governed by 
three major forces: gravity, pressure gradients, and centrifugal forces.
In the canonical model of non-rotating core collapse,
photo-dissociation of iron peak nuclei and electron capture
on nuclei and free protons initiate core collapse.
As core collapse progresses, an almost homologously ($v$ $\propto$ r) 
collapsing central region, the inner core, forms, while the outer core collapses  
supersonically. With increasing density, electron capture
rates grow until a density of about 
3 $\times$ 10$^{12}$ g cm$^{-3}$ is reached, at which time the matter becomes 
opaque to the electron capture neutrinos which are then trapped in 
the core. As nuclear density is approached, nuclear repulsive forces lead to a sudden
stiffening of the equation of state, initiating the bounce of the inner core and the
subsequent outward propagation of the bounce shock.

When rotation is included, centrifugal forces counteracting gravity's pull change
the dynamics of collapse. Depending on the total amount of angular momentum and
its distribution throughout the collapsing core, the effects can be either minor, leading
only to small deformation (oblateness) of the core and bounce at slightly lower
maximum densities or - if there is a great amount of angular momentum - major, resulting 
in ``fizzlers'' with large deformations and slower bounces at subnuclear 
densities (\citealt{shapirolightman:76}; \citealt{tohline:84}). \cite{mm:91} give a detailed 
description of the hydrodynamics of rotational core collapse. Since this paper is 
primarily concerned with the gravitational wave signature of core collapse, in the following 
we provide only brief descriptions of the most salient features of the 
hydrodynamics.

\subsection{Hydrodynamical Evolution and \\ Gravitational Wave Signature}
\label{section:mainresults}

We separate our models into two types. Type I 
encompasses those models that experience core bounce predominantly due to 
the stiffening of the nuclear equation of state at or above nuclear densities 
and type II comprises models that bounce
due to centrifugal forces. We refer to those models 
that experience significant centrifugal forces, but still bounce
at or close to nuclear densities, as type I/II.  We show that types I and II have
distinctive and characteristic gravitational wave signatures,  while
the wave signature of an type I/II exhibits a mixture of 
type I and type II features. \cite{zm:97} introduced a similar 
classification scheme. We, however, do not see the behavior
which they call type III since its occurence is connected with very low
effective $\Gamma$s and extremely
rapid collapsei, which none of our models shows.

We first present the systematics with $\beta_i$ that we have found for the 
s15 model from \cite{ww:95} with the parameter A of the rotation law
(eq. \ref{eq:rotlaw}) set to 1000 km.   $\Omega_0$ is adjusted
to yield the wanted value of $\beta_i$. Recall that our $\beta$
is calculated for the entire iron core with its many decades of density,
and not just for the inner or homologously collapsing region.  In 
\S\ref{section:intramodel} and \S\ref{section:intermodel} we discuss
the differences introduced by different choices for A and initial model.
Finally, in \S\ref{section:hegermodels}, we present the results
we obtained from core collapse calculations with the rotating progenitor
models from \cite{heger:00} and \cite{spruit:03}.

\bf Type I\rm: Models of type I rotate so slowly that centrifugal forces
are not able to stop collapse before nuclear densities are reached. The central
region of the stellar core plunges deeply into the potential well and quickly and 
significantly overshoots to supranuclear densities before its infall velocities
are reversed on a timescale of less than a millisecond by the solid-wall-like 
action of the repulsive nuclear forces.  As in nonrotating models, a strong shock 
forms at the boundary between the subsonically collapsing inner core
and the supersonically infalling outer mantle. 

Due to angular momentum conservation and, hence, the growing
influence of centrifugal forces with increasing compactness, 
the core bounce is not spherically symmetric, but happens first and most
strongly at the poles. Hence, the strong bounce shock is aspherical, 
propagating faster along the poles. 

\begin{figure}
\epsscale{1.0}
\begin{center}
\includegraphics[width=10cm]{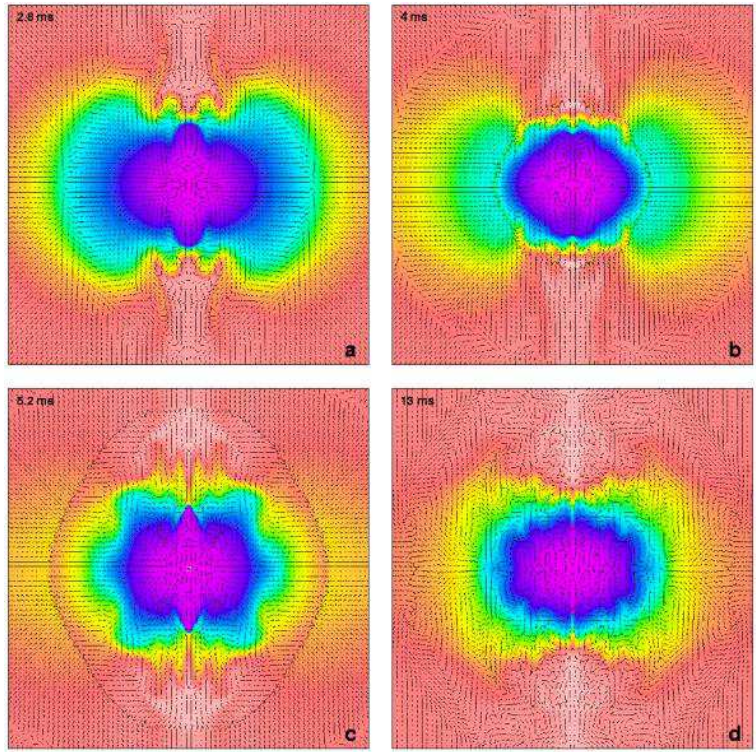}
\end{center}
\caption{2-D plots of the specific entropy of the inner 200$\times$200 km$^2$ of
the s15A1000$\beta$0.2 model depicting the ring-down epoch of the 
evolution of this generic type I model. The color map goes from
light red for high specific entropy ($s$=16 k$_B$) to magenta
for low entropy ($s$=0.7 k$_B$).  The times given in the top 
left-hand corners are the snapshot times after bounce.
Velocity vectors for the $r$ and $\theta$ motions (not the $\phi$ motion
in the angular direction) are superposed.
As the core executes successive radial and non-radial
post-bounce oscillations, it generates strong sound waves (weak shock waves)
that are seen in panels (b) and (c) as contours of velocity
discontinuity. At the time of panel (a), the actual bounce shock
has already left the frame. Panel (d) shows
the compact remnant at 13 ms after bounce when it has already
lost most of its excess pulsational energy. The velocities at that time
are dominated by incoherent vortical motions. 
(This figure is available in high-quality format from
http://www.ita.uni-heidelberg.de/\~{}cott/gwpaper~.)
\label{fig:2ds15}}
\end{figure}

In type I models, the initial oblateness of the core at bounce
becomes an oblate-prolate ($l=2$) oscillation, accompanied by higher-order modes, in addition
to the already present post-bounce radial ``ringing'' of the compact remnant.
This is seen in Fig. \ref{fig:2ds15}, which shows the innermost
200 km by 200 km of the collapsed core of model s15A1000$\beta$0.2 (type I).
Shortly after bounce and after the shock wave is launched, the core oscillates
wildly in a superposition of
modes that are damped on timescales of milliseconds, since
oscillation leads to dissipation by the periodic emission 
of strong sound waves.  As Fig.~\ref{fig:2ds15} indicates,
the amplitude of the $l=2$ deformation has decayed considerably 13 ms after bounce,
at which time predominantly vortical and incoherent motions have emerged
to dominate.

\begin{figure}
\epsscale{1.0}
\plotone{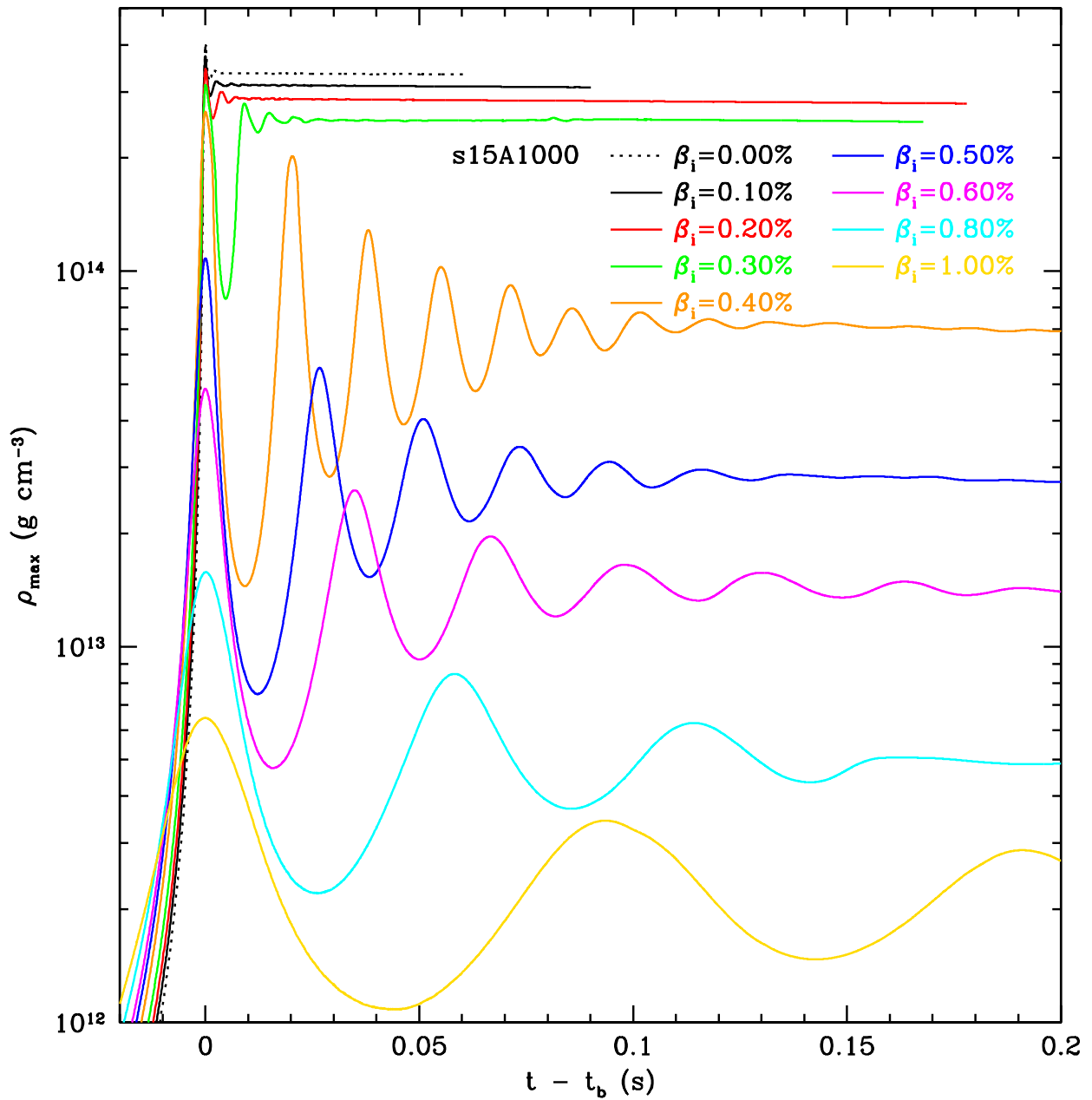}
\caption{Evolution of the maximum density of the s15A1000 model series, which rotates almost 
rigidly inside 1000 km, for different initial rotation parameters $\beta_i$.  For comparision, the dashed line
shows the evolution of the central density of a nonrotating model. The
models on this graph undergo core bounce between 280 and 500 ms after the start
of the evolution. The time to bounce increases
with initial rotation rate since centrifugal forces, acting as additional pressure support,
slow down collapse. All times are relative to the time of bounce ($t_b$) for each 
individual model. One can clearly see the transition in collapse dynamics that takes 
place in the interval in $\beta$-space between 0.20\% and 0.40\% which leads from type I
to type II behavior (see text for details).   
\label{fig:s15A1000_rho}
}
\end{figure}

Figure \ref{fig:s15A1000_rho} portrays the evolution of the maximum density of the 
s15A1000 model series with A=1000 km.   
Plotted with a dashed line is the evolution of the maximum density of a nonrotating 
model. The models with $\beta_i$=0.10\% and $\beta_i$=0.20\% bounce at or 
very close to nuclear density and exhibit 
type I model behavior. The $\beta_i$=0.30\% model, however, though
experiencing bounce at about nuclear density, 
exhibits one cycle of pronounced large-scale coherent radial  
expansion and re-contraction, followed by a second, but weaker bounce
after which the volume mode is quickly damped due to the formation of 
a second shock wave. We refer in the following to this cycle of expansion, re-contraction
and secondary bounce as the ``expansion-collapse-bounce cycle.'' It marks
the beginning of the transition from type I to type II behavior
with increasing $\beta_i$.

The evolution of the rotation parameter $\beta=E_{rot}/|E_{grav}|$, 
an integral quantity and a global measure of the system,
is shown in Fig.~\ref{fig:s15A1000_beta}. For
models that show clean type I behavior, namely those with $\beta_i$ less than
0.30\%, $\beta$(t) manifests only one notable post-bounce minimum, directly 
paralleling the evolution of the maxmium density. The  
first transition model with $\beta_i$=0.30\% reaches a $\beta$ that is 30\% higher 
at bounce and exhibits multiple
post-bounce minima and maxima before reaching its final value of 6.37\%, 
the largest in the s15A1000 model series. 

\begin{figure}
\epsscale{1.0}
\plotone{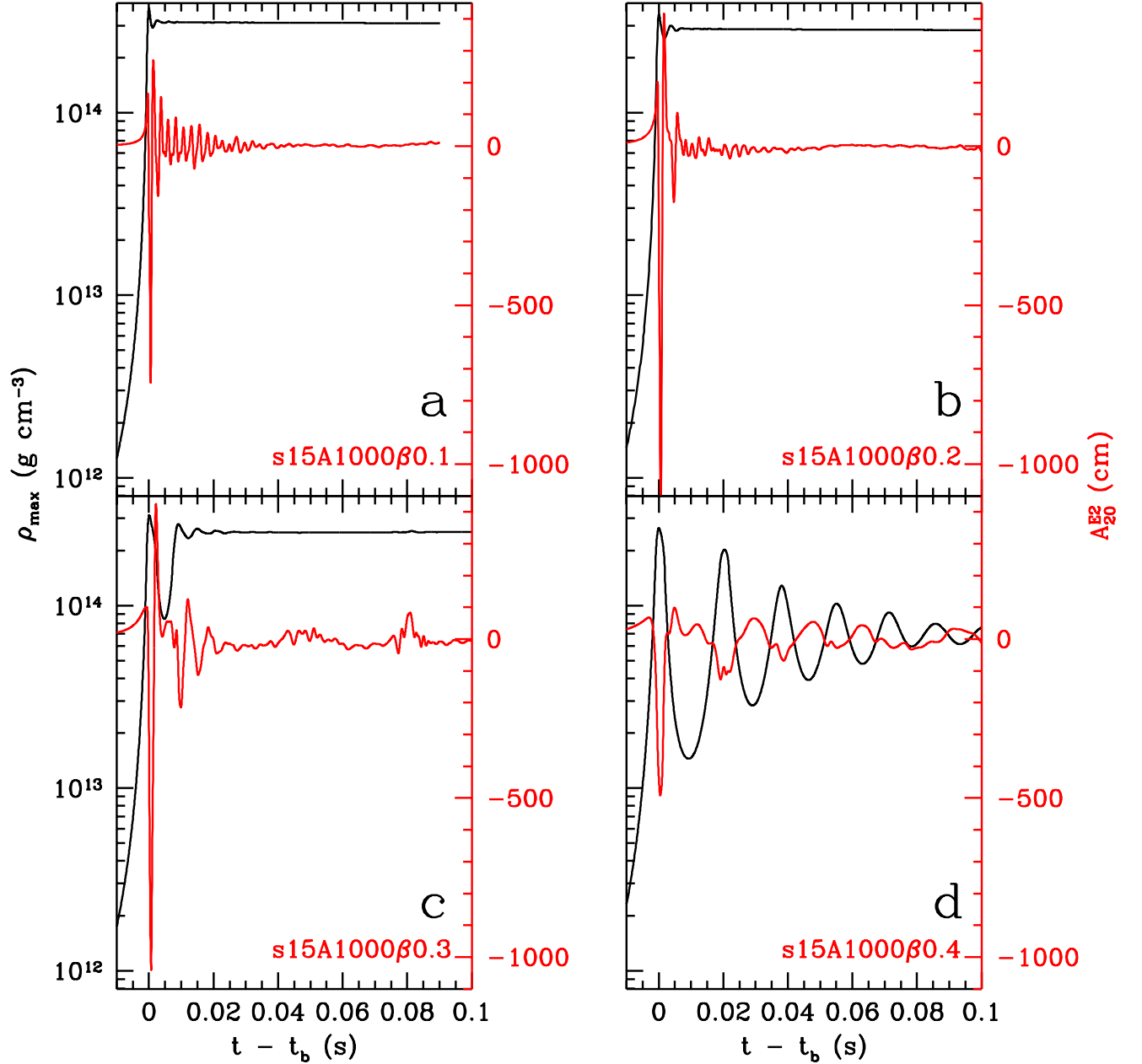}
\caption{
Maximum density ($\rho_{max}$, black) and gravitational wave amplitude ($A^{E2}_{20}$,
red) as functions of time for the \cite{ww:95} s15 model and A = 1000 km.  
For each model, the zero of the time axis is set
to the time of core bounce (t$_b$). \undertext{{\bf a (upper left):}}
$\beta_i$=0.10\%, bounce at supranuclear density with negligible influence of centrifugal
forces (type I). 
Note the subsequent ring-down waveform, which is associated with both  
radial- and non-radial modes of the compact remnant. 
\undertext{{\bf b (upper right):}} $\beta_i$=0.20\%, similar to 
the previous model, but with
greater asphericity leading to larger amplitudes. \undertext{{\bf c (lower left):}} 
Transitional model with $\beta_i$=0.30\%, which is beyond the $\beta_i$ for the largest 
gravitational wave amplitudes. The typical oscillation period of the waveform is significantly
larger than those of the previous models and the hydrodynamic data exhibit at 
least one additional coherent large scale expansion-collapse-bounce cycle caused
by the growing influence of centrifugal forces. \undertext{{\bf d (lower right):}} 
$\beta_i$=0.40\%. The hydrodynamic evolution and the associated waveform of this
model are already largely influenced by centrifugal forces. The core exhibits
subsequent multiple coherent bounces that are quasi-exponentially damped (type II).
\label{fig:s15A1000_1}
}
\end{figure}

We now focus on the characteristics of the gravitational wave signature
of a type I model and see how the signature changes with increasing $\beta_i$.  
In the quadrupole approximation that we made in
\S\ref{section:extract}, the amplitude of the gravitational waves is directly connected
to the second time derivative of the reduced mass-quadrupole moment ($I\bari_{zz}$; eq. \ref{eq:itensor}). 
Figures \ref{fig:s15A1000_1}a and \ref{fig:s15A1000_1}b
show the evolution of the maximum density and the quadrupole gravitational wave
amplitude for times shortly before and after bounce of the typical type I models
s15A1000$\beta$0.1 and s15A1000$\beta$0.2.
From the way $I\bari_{zz}$ evolves, we can divide the evolution of the wave signal into 3 
phases (\citealt{mm:91}).

Phase 1 (the ``infall phase'') coincides with the hydrodynamic collapse phase, 
and is marked by the accelerated increase (in absolute value) of the quadrupole moment 
as angular momentum conservation forces the collapsing core to deviate more and 
more from spherical symmetry.
When the core approaches bounce, the increase of $|I \bari _{zz}|$ decelerates (i.e., the wave signal
becomes negative) and eventually is reversed into a decrease of $|\, I \bari _{zz}|$,  
the smaller core size counteracting the effect of the increasing density on the quadrupole moment.  
This marks the transition to phase 2 (the ``bounce phase''). The wave signal reaches its
absolute minimum a fraction of a millisecond after core bounce and then increases,
reaching positive values during the first post-bounce local minimum of the
central density. As said, right before and during core bounce, 
the increase of $|I \bari _{zz}|$ started in phase 1 is dramatically reversed. 
Nevertheless, the first time derivative of $I\bari _{zz}$ reaches positive values for only a short period
and $I \bari_{zz}$ itself remains negative throughout the entire evolution, increasing again 
in absolute value after bounce. 
Finally,  phase 3 (the ``ring-down phase'') is
characterized by small amplitude oscillations of the wave signal that reflect the
core ring-down occurring in type I models.

The three phases that we described
in the preceding paragraph are easily discernible in Figs. 
\ref{fig:s15A1000_1}a and \ref{fig:s15A1000_1}b.
As centrifugal forces become increasingly significant with larger $\beta_i$, the evolution
of the maximum density and of the waveform itself change.  Figure 
\ref{fig:s15A1000_1}c, depicting model s15A1000$\beta$0.3, shows this well. This model undergoes one large
post-bounce oscillation, an  ``expansion-collapse-bounce cycle.''  
A second, less strong, bounce occurs and is 
reflected in a second pronounced spike in the waveform. Increasing $\beta_i$ 
to 0.40\% changes the dynamics drastically to what we classify
as type II behavior.

\bf Type II\rm: Models of type II experience bounces at subnuclear densities caused
by centrifugal forces. Whereas core bounce due to repulsive nuclear forces
results in an abrupt reversal of the quasi-homologously collapsing inner core 
on a timescale of less than 1 ms, a bounce due to the action of
centrifugal forces occurs on longer timescales and with smaller accelerations. This
can be understood from the fact that rotation acts approximately like a gas
with a $\Gamma$ of 5/3 (\citealt{tassoul:78}; \citealt{mm:91}).
Fig. \ref{fig:s15A1000_rho} depicts the evolution of the maximum density of 
models with $\beta_i$s from 0.0\% to 1.00\%. 
For $\beta_i \sgreat 0.4\%$ , the models exhibit type II behavior.

With increasing $\beta_i$, the density at bounce decreases, the 
maxima become much wider, and the accelerations
grow smaller. The rotation-dominated, highly aspherical
bounce of type II models happens at larger radii and creates, 
since it is much less abrupt, less entropy and much
weaker bounce shocks than in type I models. After core bounce, the core
expands coherently, leading to an almost order of magnitude drop in the maximum
density. This expansion is reversed when gravitational forces
again begin to dominate over pressure gradients and centrifugal forces. In this way, the
quickly spinning core undergoes several damped expansion-collapse-bounce cycles until
it settles into an equilibrium configuration. In fact, it acts much like a
damped harmonic oscillator with a Hamiltonian consisting of radial kinetic,
rotational, internal, and gravitational energies (\citealt{mm:91}).

\begin{figure}
\epsscale{1.0}
\plotone{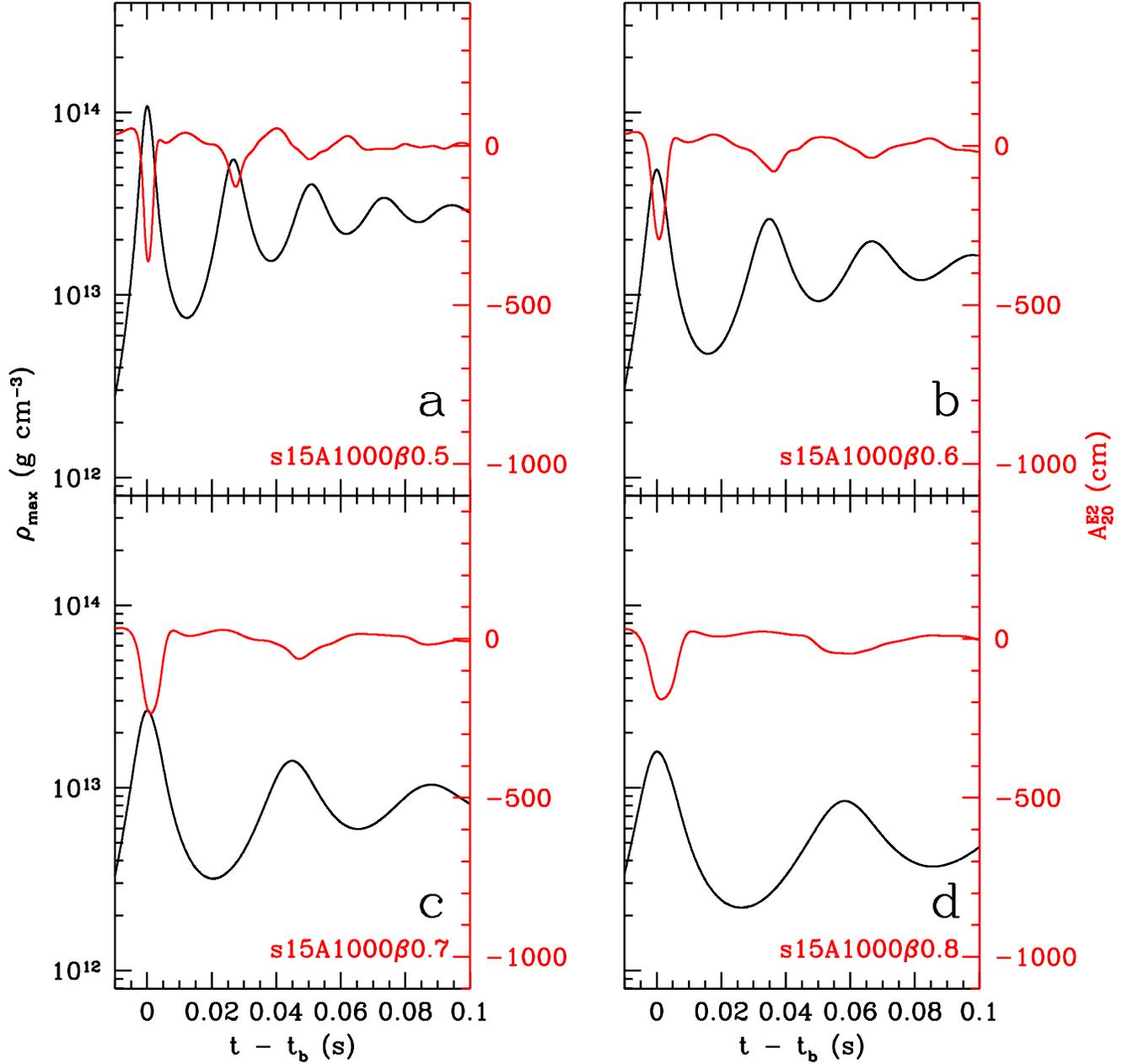}
\caption{Same as Figure \ref{fig:s15A1000_1}, but for higher values of $\beta_i$.
$\rho_{max}(t)$ (black) is the maximum density and $A^{E2}_{20} (t)$ (red) is the gravitational 
wave amplitude for s15A1000. \undertext{{\bf a (upper left):}} 
This model with $\beta_i$=0.50\% undergoes a bounce at subnuclear density 
when centrifugal forces overcome gravitational attraction. Note the significantly
larger dynamical timescales compared with models with smaller $\beta_i$.  
\undertext{{\bf Panels b, c and d:}} As $\beta_i$ increases from 0.60\%
to 0.80\%, bounce occurs more slowly and at progressingly lower densities.   
The waveforms are dominated by the subsequent
expansion-collapse-bounce cycles of the quickly spinning deformed stellar core (type II).
\label{fig:s15A1000_2}
}
\end{figure}

Figures \ref{fig:s15A1000_1}d and \ref{fig:s15A1000_2}a through \ref{fig:s15A1000_2}d
depict the time evolution of the maximum density and of the quadrupole gravitational
wave amplitude for the type II models of the s15A1000 model series. For these models,
the ``ring-down'' signature seen in type I models is replaced by the signature
of coherent expansion-collapse-bounce cycles, manifest by 
wide peaks in the wave signal. Model s15A1000$\beta$0.4 (Fig. 
\ref{fig:s15A1000_1}d) still shows shorter period substructure, originating
from additional nonspherical pulsation modes, but such substructure  
fades away with larger $\beta_i$.  The
absolute values of the gravitational wave amplitude peaks are smaller than those
of type I models, since deceleration and acceleration occur over longer
timescales. This is also reflected in the longer characteristic periods and in the  
lower characteristic frequencies of the gravitational wave signal.

The gravitational wave energy spectra (eq. \ref{eq:dEdnu}) for a characteristic subset of the
s15A1000 models are shown in Fig. \ref{fig:s15A1000_spect}. Since the gravitational
wave luminosity goes with the slope of the waveform, most of the energy is
radiated for all models in the spike connected to the first bounce. 
We associate the dominant frequencies of the
gravitational wave energy spectra with this first spike of the waveform.
In Table \ref{table:resultss15}, we summarize
the results from all our calculations involving the s15 progenitor model.
Tables \ref{table:resultss11}, \ref{table:resultss20} and \ref{table:resultsheger}
contain the quantitative results of calculations using the s11 and s20 models of
\cite{ww:95} and the models of \cite{heger:00} and \cite{spruit:03}.

The spectrum of s15A1000$\beta$0.2, which shows type I behavior, 
is dominated by frequencies between 300 Hz and 600 Hz and peaks at 460~Hz. Most 
of the smaller peaks are connected to the first spike in the waveform during
which 94\% of the total gravitational wave energy of this model is radiated
(Table \ref{table:resultss15}). There is, however, a contribution by the 
radial and non-radial ring-down pulsations that have 
characteristic periods of 2 - 2.5 ms in this model, translating to frequencies of 400-500~Hz.
The peak at 700 Hz and the one-order-of-magnitude-suppressed peak at about 1400~Hz
are higher harmonics of lower frequency contributions, as well as higher
frequency modes of the ring-down oscillation. With increasing $\beta_i$
the spectrum shifts to lower frequencies and lower absolute values, 
peaking at 152 Hz ($\beta_i$=0.40\%),
91 Hz ($\beta_i$=0.60\%), and 38 Hz ($\beta_i$=0.80\%). Furthermore, a prominent peak 
at low frequencies (in this series beginning with model 
s15A1000$\beta$0.4) can be directly associated with the oscillation 
frequency of the post bounce expansion-collapse-bounce cycles seen in type II models.

As shown in Fig. \ref{fig:s15A1000_spect}, the energy spectra provide an excellent way to learn about 
overall collapse dynamics and clearly exhibit the great changes 
brought about by even moderate rotation.  We now summarize the most
important overall characteristics of the dynamical types: Type \bf I \rm is
characterized by bounce at supranuclear densities and exhibits a clear subdivision into
infall (1), bounce (2) and ring-down phases (3). The gravitational wave energy 
spectra of type I models peak above 350 Hz and the integrated radiated
energy achieves its maximum, reaching 10$^{-8}$ \modot $c^2$ for the fastest rotating
model that still bounces at supranuclear density.  On the other hand, type \bf II \rm models
bounce at subnuclear densities due to the influence of centrifugal forces, 
exhibit much longer timescales, and manifest several
harmonic oscillator-like expansion-collapse-bounce cycles.
Their spectra peak at lower frequencies and the total energies
radiated are, for the fastest rotators, one to two orders of magnitude smaller
than those for type I (see Tables 3-6). 

We must point out that the specific value of $\beta_i$ needed
to produce a given type (I or II) varies with
the degree of differential rotation  
and (more strongly) with the effective $\Gamma$ 
(\S\ref{section:rotation}), determined by the equation of state in 
combination with the microphysical processes
occuring during collapse (e.g. electron capture, leading to  
a lower effective $\Gamma$).  Moreover, the maximum value
of $\beta_i$ needed to maintain type I behavior, depends on the effective $\Gamma$ 
through eq. (\ref{eq:betacrit}). Note that the final $\beta$s achieved in our
calculations are all below 10\%.

Since we have not included radiative transfer, 
our results should be considered preliminary.   
Furthermore, the collapse will depend on the rotation law used and
on general relativistic effects.
Hence, we cannot say, though the results of this study suggest, that a $\beta$ high enough
for dynamical or secular instability might not occur during realistic stellar core
collapse.

\subsection{Variation with Degree of Differential Rotation}
\label{section:intramodel}

We have performed calculations with three different values of the parameter A
of the rotation law described in \S \ref{section:rotation}. 
To achieve rigid rotation throughout the entire core, we set A 
to 50000 km in eq. \ref{eq:rotlaw} and to explore moderate
differential rotation, we use A=1000 km and A=500 km. Figure \ref{fig:initialomega}
shows a sample of the rotation laws for different values of A.

\begin{figure}
\plotone{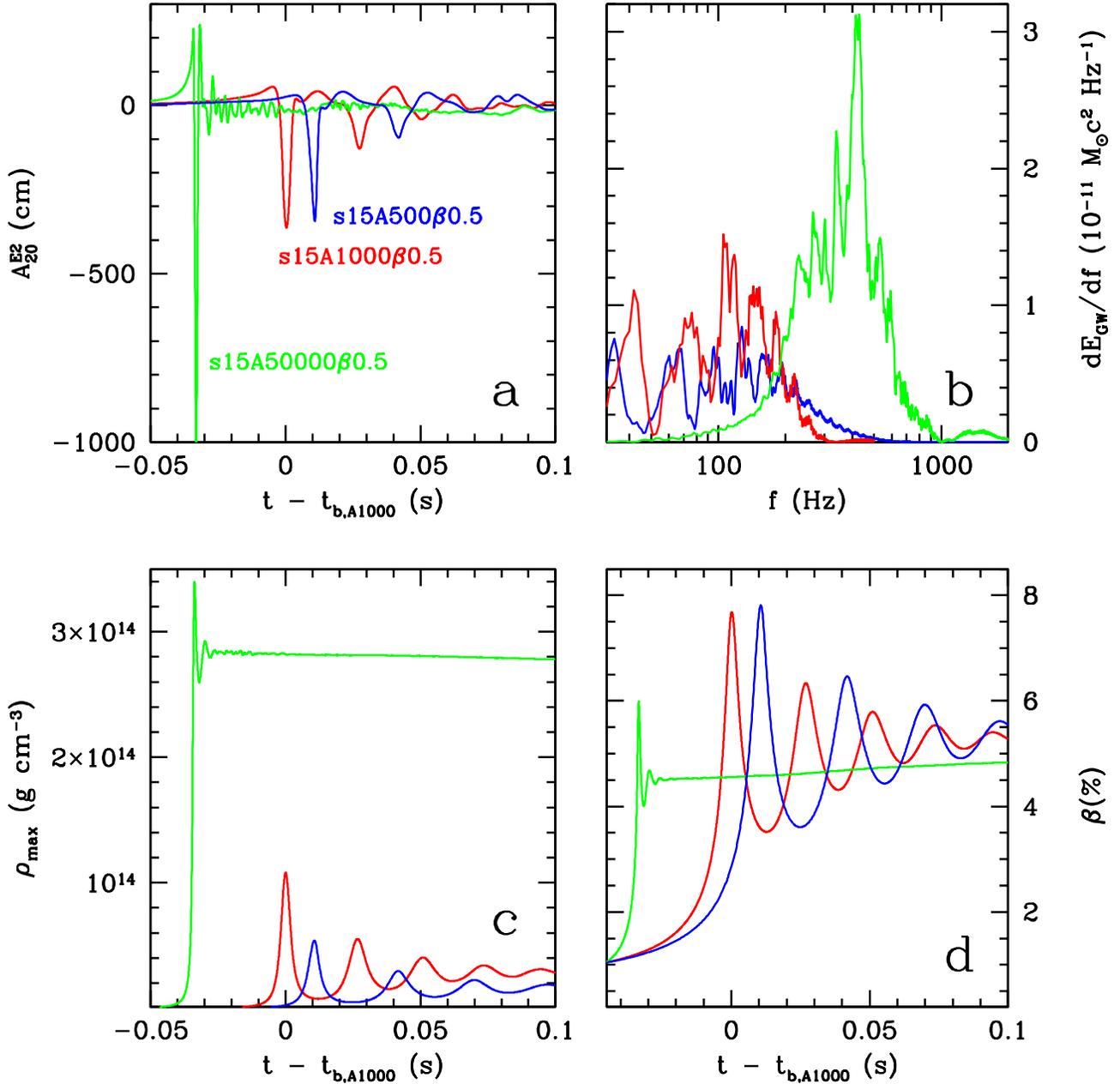}
\caption{Comparison between different initial distributions of angular momentum
at fixed $\beta$=0.50\% for the s15 model from \cite{ww:95}.  Models A50000, 
A1000, and A500 are nearly rigidly rotating interior to 50000, 1000, and 500
kilometers, respectively.
\undertext{{\bf a (upper left):}}
Time evolution of the gravitational wave amplitude for the models
s15A50000$\beta$0.5, s15A1000$\beta$0.5, and s15A500$\beta$0.5. The time axis is
relative to the time of bounce of the s15A1000$\beta$0.5 model. For a given model
the transition from a bounce dominated by nuclear repulsive forces to one
in which centrifugal forces play a significant role occurs at progressively lower $\beta_i$ 
with decreasing A.  \undertext{{\bf b (upper right):}} Energy spectra of the three models. Note
the distinct peak of the rigidly rotating s15A50000$\beta$0.5 model (green) at about
400 Hz that is directly associated with the post-bounce ringing of the compact 
remnant, also seen in the waveform. The spectra of the models with smaller A exhibit
a local maximum at low frequencies that also directly correspond to the frequencies of their
post-bounce expansion-collapse-bounce cycles.
\undertext{{\bf c (lower left):}} 
Evolution of the maximum density. Model s15A50000$\beta$0.5 still bounces at 
supranuclear densities, while the two models with smaller A have already made
the transition to type II behavior. 
\undertext{{\bf d (lower right):}} 
Evolution of the rotation parameter $\beta$
with the time given relative to the time of bounce of model s15A1000$\beta$0.5. The models
with A=1000 and A=500 have more angular momentum in their central regions and, hence,
are more strongly influenced by centrifugal forces.  For a given $\beta_i$, they also achieve 
larger final $\beta$s and $\beta$s at bounce. 
\label{fig:intramodel0.5}}
\end{figure}

In Fig. \ref{fig:intramodel0.5}, we present results from our study of the A dependence
of the gravitational wave signature for $\beta_i$=0.5\% and the s15 progenitor.
Generally, as A is lowered,
more rotational energy is (for a fixed $\beta_i$) moved interior to A. 
Hence, the influence of rotation on the collapsing central regions
is larger in those models with smaller A and the transition from type I to type II
behavior occurs for smaller $\beta_i$. This is illustrated in Fig. 
\ref{fig:intramodel0.5}c where model s15A50000$\beta_i$0.5 still shows type I behavior,
even though models s15A1000 and s15A500 have transitioned to type II. The same effect
is seen in Fig. \ref{fig:intramodel0.5}a.
The energy spectra in Fig. \ref{fig:intramodel0.5}b show low-frequency peaks at the frequencies
of the expansion-collapse-bounce cycles for models s15A500 and s15A1000, while the spectrum of 
the s15A50000 model is still type I. 

As indicated, for larger values of A, i.e. for a more rigidly
rotating configuration, the value of $\beta_i$ must be higher for 
type II behavior to occur. This shifts to higher values the $\beta_i$s required for maximum
gravitational wave amplitudes and final $\beta$s. This A dependence 
is generic and translates into corresponding behavior for s11 and s20 progenitors 
(see Tables \ref{table:resultss11}, \ref{table:resultss15}, and \ref{table:resultss20}).

\subsection{Initial Model Comparisons}
\label{section:intermodel}

To investigate the effect of changes in presupernova stellar structure
on the collapse dynamics and, ultimately, on the gravitational wave signature, we have 
performed calculations using the s11, s15 and s20 progenitor models of 
\cite{ww:95} at given values of A and $\beta_i$.  
We choose A=1000 km and $\beta_i$=0.5\% for our progenitor model comparisons.  

Initially, the s11 and s15 models are quite similar in their structures (see Figs.
\ref{fig:initialrho} and \ref{fig:initialye}) and this similarity persists throughout
their evolution. The s20 model, however, has a
smaller initial central density, a larger $Y_e$, and a significantly
larger iron core, encompassing about 1.7 \modot and reaching out to 2200 km (Table
\ref{table:initialmodels}). 

\begin{figure}
\plotone{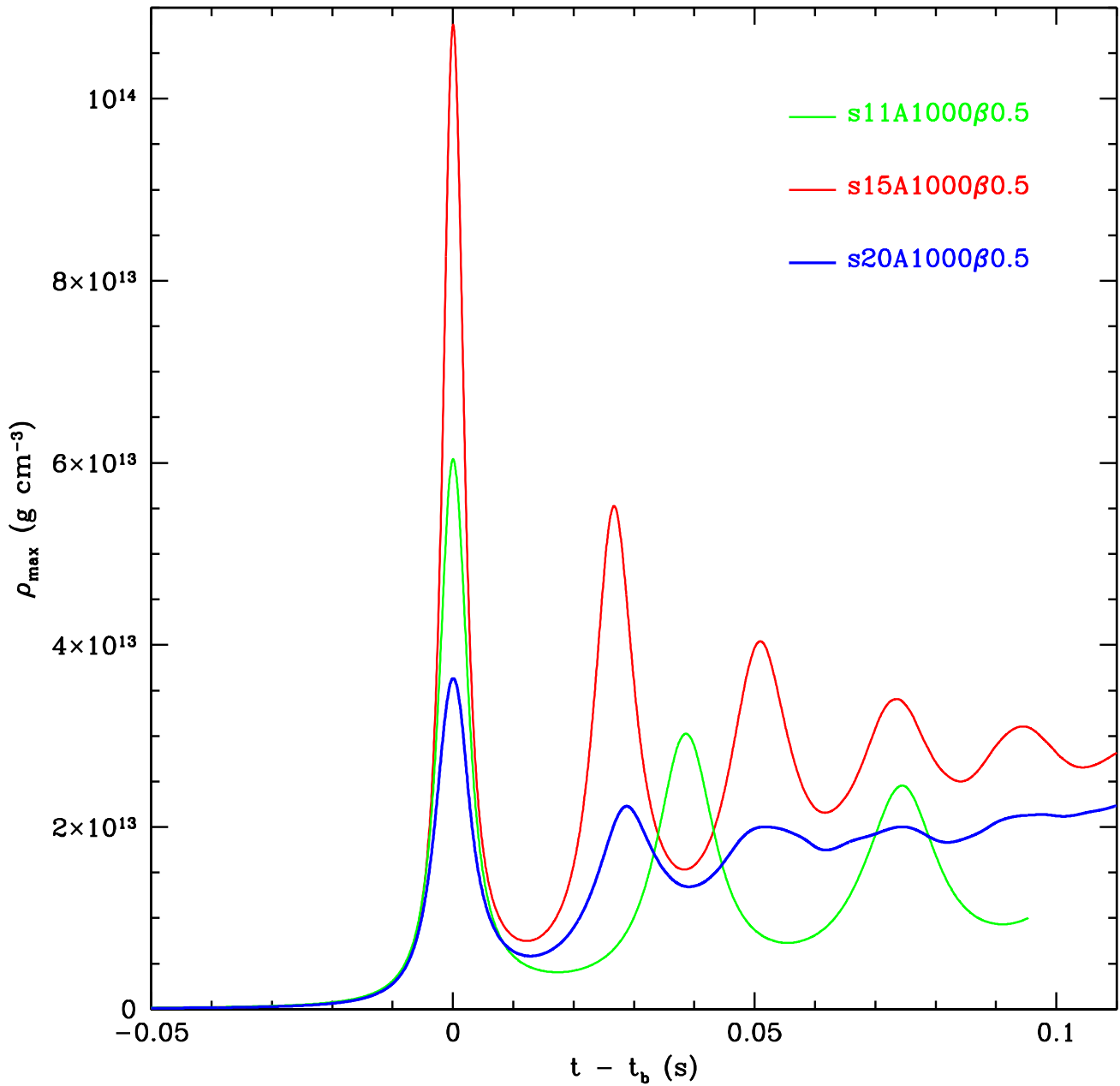}
\caption{Inter-model comparison between s11A1000, s15A1000 and s20A1000 for 
$\beta_i$=0.50\% of the evolution of the peak density. The time is 
given relative to the time of bounce for each individual model.
Model s15 reaches higher maximum densities and has the shortest post-bounce oscillation periods.
Model s11, which has a similar initial density
and angular momentum distribution, reaches lower densities and has longer 
post-bounce oscillation periods. The s20A1000 model, however, with significantly
different initial density, angular momentum, and compositional profiles,
bounces for the same A and $\beta_i$ at even lower densities. It also has shorter
oscillation periods than the s11 model and is more quickly damped by the proximity 
of its stalled shock.
\label{fig:intermodelrho}}
\end{figure}

\begin{figure}
\epsscale{1.0}
\begin{center}
\includegraphics[width=10cm]{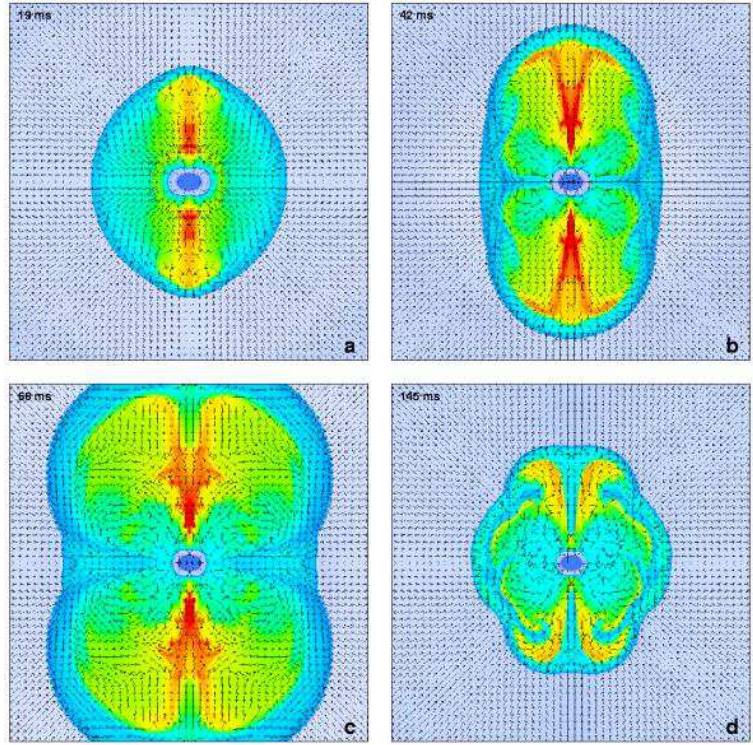}
\end{center}
\caption{2-D plots of the specific entropy 
for model s20A1000$\beta$0.3. Red denotes high
entropy ($\sim$14 k$_B$) and dark blue denotes 
low entropy ($\sim$0.9 k$_B$, seen in the compact
remnant). Shown are the inner 1200$\times$1200 km$^2$ of the
hydrodynamic grid.  As in Fig. \ref{fig:2ds15}, velocity
vectors are superposed and the times after bounce are given in the
top left of each panel.  Since the core is oblate, core bounce happens
first and at smaller radii along the poles. 
After bounce, the shock is able to propagate much faster
along the rotation axis than in the equatorial region. High entropy,
jet-like structures form along the rotation axis.
This is seen in panel (a), which shows the core 19 ms after bounce.
The bounce shock has already reached about 400 km at the poles and
300 km at the equator. Forty-two ms after bounce (panel b) the axis ratio has
increased even more and violent vortical motion
has set in. At about 68 ms after bounce (panel c), the bounce shock has stalled
and most velocities point inward.  The shock recedes.
Panel (d) shows the shocked region at the end of the evolution. 
The shock has receded even further.  Interior to
the shock, vortical motion has lead to the breaking of equatorial
symmetry.
(This figure is available in high-quality format from
http://www.ita.uni-heidelberg.de/\~{}cott/gwpaper~.)
\label{fig:2ds20}}
\end{figure}

Figure \ref{fig:intermodelrho} displays the evolution of the maximum density of 
models s11A1000$\beta$0.5, s15A1000$\beta$0.5, and s20A1000$\beta$0.5. The s15
model reaches the highest densities and bounces due to centrifugal forces. After
bounce, the s15 core executes the typical damped expansion-collapse-bounce cycles, 
as does model s11. However, model s20, 
with a significantly different initial structure,  
reaches lower densities than the s11 and s15 models and exhibits 
shorter periods in its post-bounce oscillations. Furthermore, its
post-bounce oscillations are more quickly damped by the proximity
of its stalled shock.  A shock acts
like a strong absorber of sound waves.  As a result, the characteristic
time for the purely hydrodynamic damping of inner-core oscillations is roughly
the round-trip sound travel time to the shock radius.  The larger
this radius,  the weaker the damping.
The 2-dimensional plots (Fig. \ref{fig:2ds20}) of the specific entropy of model
s20A1000$\beta$0.3 reveal some of the dynamics of the s20 models.
As in s11 and s15 models with rotation, the shock forms first at the
poles and propagates out faster along the rotation axis than at the equator.
This is shown in the transition from Fig. \ref{fig:2ds20}a to
Fig. \ref{fig:2ds20}b, in which the axis ratio 
is almost 2:1 and the shock has a prolate shape.
Despite centrifugal forces, Fig. \ref{fig:2ds20}c shows that matter
can flow in at the equator.  The bounce shock stalls, but still executes
oscillations in radius.  We note that equatorial 
symmetry is clearly broken in Fig. \ref{fig:2ds20}d. This
emphasizes the importance of including all 180$^{\circ}$ in an
axisymmetric simulation.  

\begin{figure}
\epsscale{1.0}
\plotone{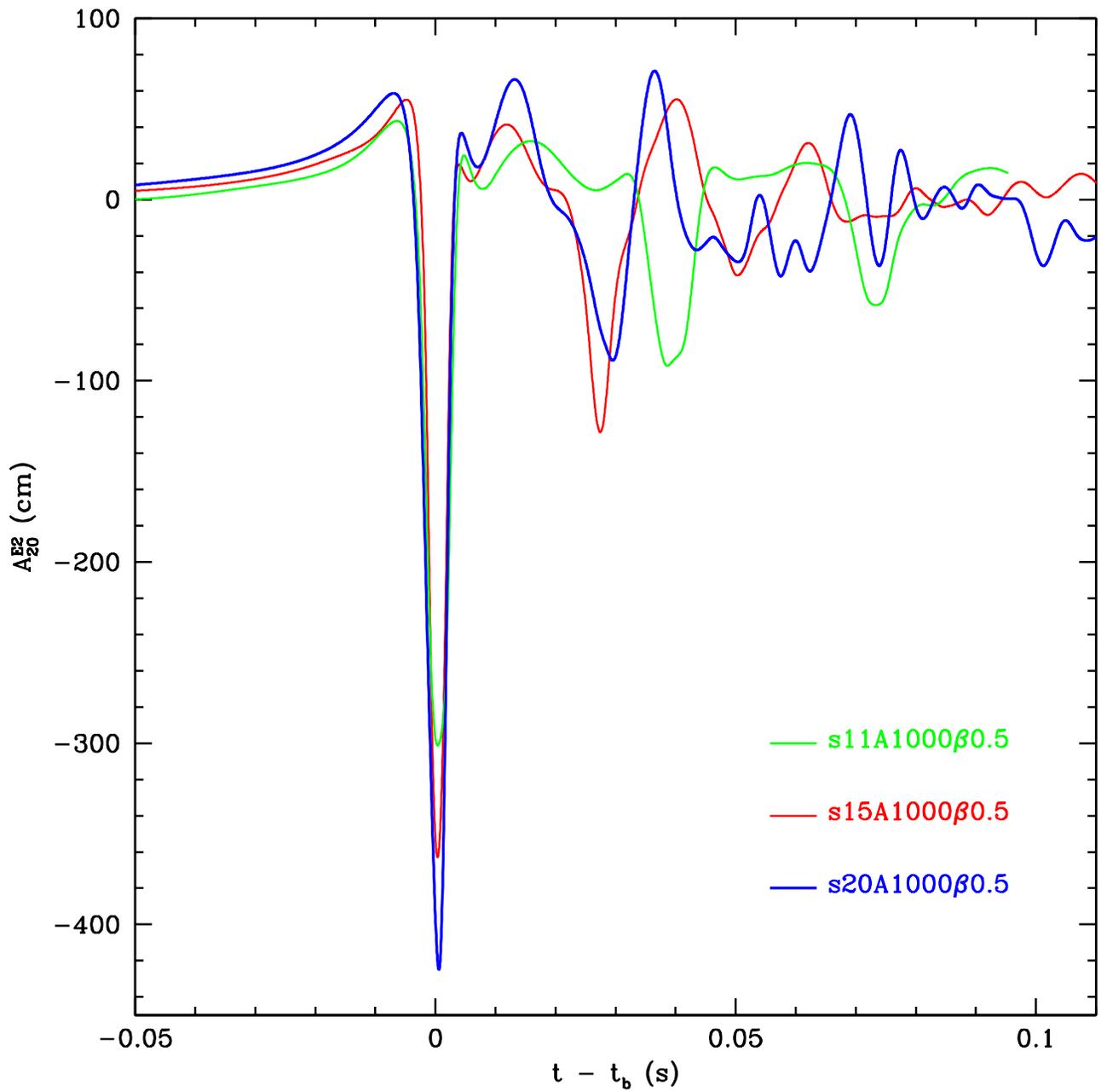}
\caption{Comparison of the evolution of the gravitational wave signal $A^{E2}_{20} (t)$ of 
models s11, s15 and s20 for A=1000 km and $\beta_i$=0.50\%. 
The waveforms reflect the density evolution seen in Fig. \ref{fig:intermodelrho}.
For the s20 signal, additional high-frequency contributions 
appear to be correlated with the vortical motions and aspherical infall
seen behind its stalled shock.
\label{fig:intermodelA}}
\end{figure}

\begin{figure}
\plotone{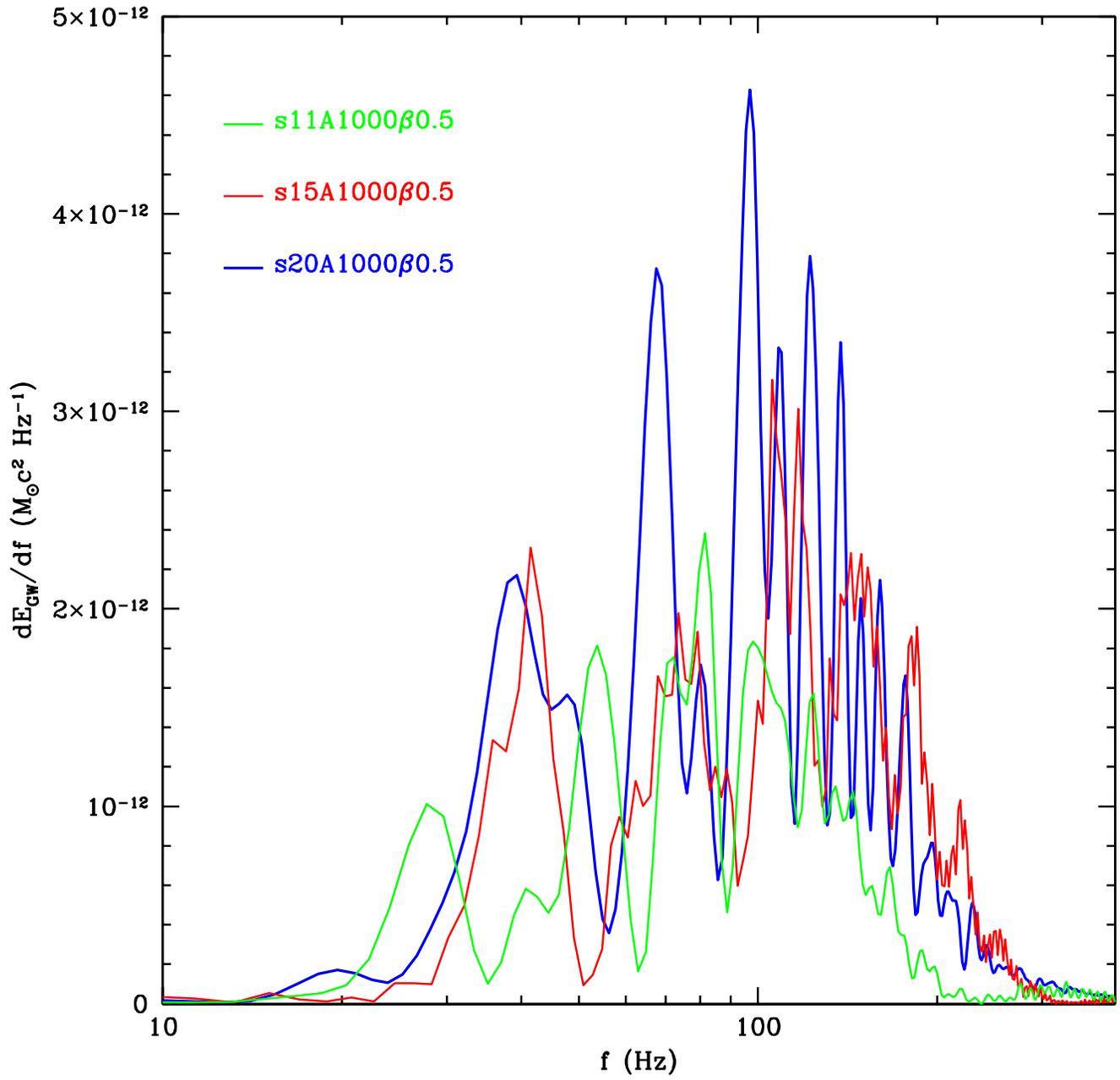}
\caption{Comparison of the gravitational wave energy spectra for s11, s15 and s20
models.  Note the linear scale of the ordinate.
\label{fig:intermodelspect}}
\end{figure}

Figure \ref{fig:intermodelA} shows the waveforms
of models s11A1000, s15A1000, and s20A1000 for $\beta_i$=0.5\%. Models s11 and s15 exhibit clear type II
behavior, with the variation in their wave forms paralleling that of their
maximum densities.   
The wave form of the s20 model, however, exhibits different behavior;  
the stall of its bounce shock and subsequent accumulation 
of outer-core material onto the compact remnant introduce additional higher-frequency
components. As Fig. \ref{fig:intermodelspect} indicates, this is also reflected in its gravitational 
wave energy spectrum, which contains more power at higher frequencies.  

The bounce shock stalls in all our models with the s20 progenitor.  
For larger $\beta_i$, standard type II behavior is altered, since the post-bounce 
expansion-collapse-cycles are strongly damped by the stalled shock.
We point out that to date the effect of a stalled shock 
upon the gravitational waves signature from rotating stellar collapse has not been 
discussed.  However, such effects might 
be common for core-collapse supernovae, since in realistic models energy losses
due to neutrinos - at least initially - lead generically to stalled bounce 
shocks.

\subsection{The Heger Models}
\label{section:hegermodels}

We have chosen the e15 and e20 models of
\cite{heger:00} and the m15b4, m20b4, and m25b4 models of \cite{spruit:03}
for our study of progenitors evolved with rotation and 
magnetic fields (\S\ref{section:rotation}). Models from \cite{heger:00} 
that end with the letter 'b' (with $\beta_i$ $\ge$ 3\%) were not used, since they
undergo expansion and not collapse once mapped onto our two-dimensional grid. We would
have had to artificially force these models to collapse by altering their thermodynamic
structure.  We have chosen not to do so and postpone the   
investigation of the gravitational wave signature from these models 
to future research that will include detailed weak interaction
physics and neutrino transport.  

\begin{figure}
\plotone{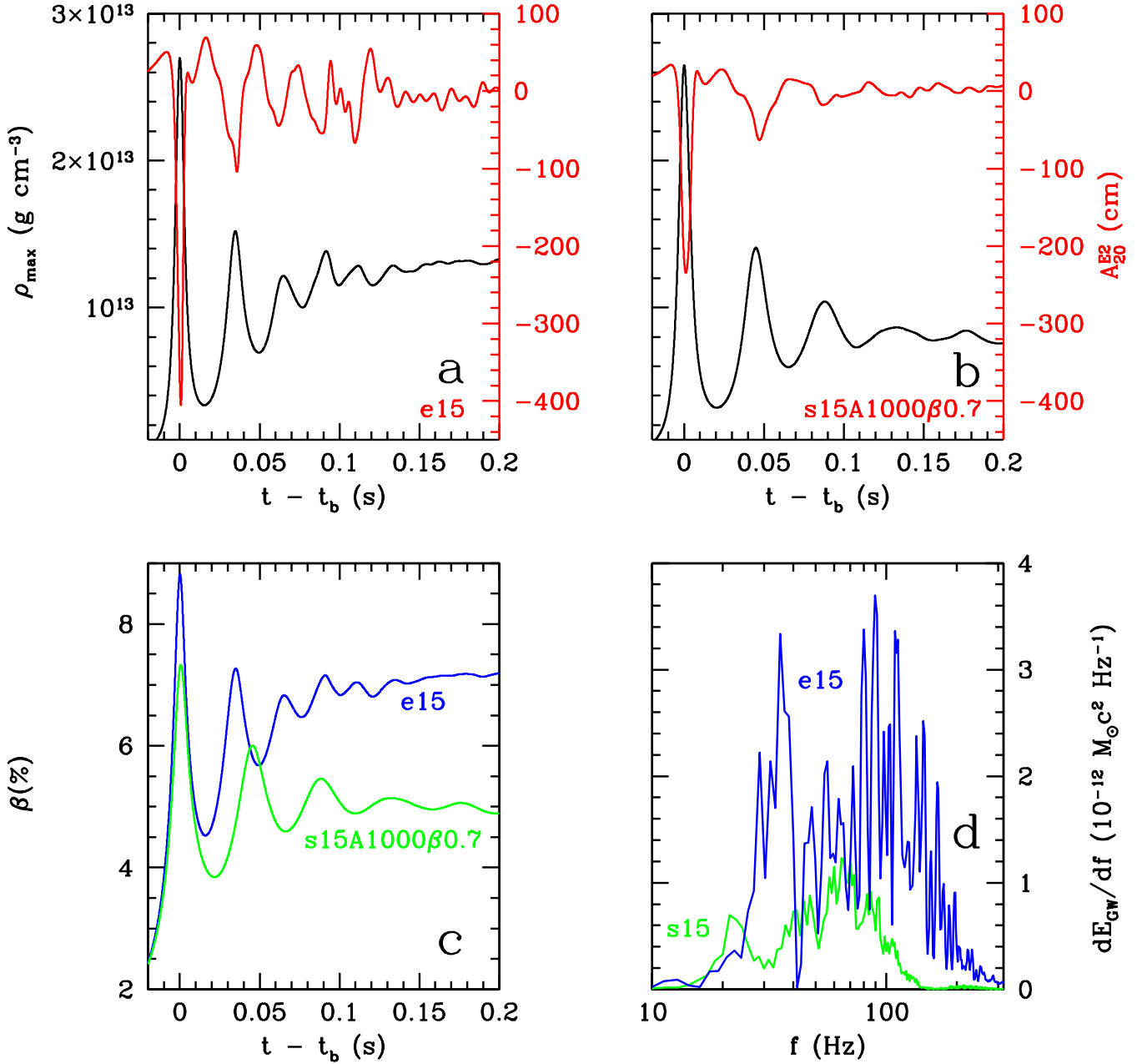}
\caption{Comparison of the e15 model from \cite{heger:00} 
with the s15A1000$\beta$0.7 model from \cite{ww:95}.
Both models have similar initial $\beta_i$ (0.645\% for the e15 model versus 0.70\% for the 
s15A1000$\beta$0.7 model) and angular velocity ($\Omega$) profiles but
differ significantly in their total angular momenta. This is due to differences in 
their initial density profiles. \undertext{{\bf a (upper left)}} and
\undertext{{\bf b (upper right):}}  
Evolution of the maximum density and the gravitational wave amplitude, respectively, for models
e15 and s15A1000$\beta$0.7.  The waveform of e15 exhibts 
additional high-frequency components about 100 ms
after bounce that we associated with the sudden damping of the 
post-bounce expansion-collapse-bounce cycles by infalling matter.  This is also 
reflected in the evolution of the maximum density. 
\undertext{{\bf c (lower left):}}  Evolution of the
rotation parameter $\beta$ for the two models. The larger angular momentum
of model e15 and its specific distribution translate into a larger final $\beta$.
\undertext{{\bf d (lower right):}} Energy spectra for the two
models under consideration. The e15 model radiates significantly more energy 
and has distinctly more fine spectral structure.  
\label{fig:hegere15s15}}
\end{figure}

The collapse, bounce, and post-bounce behavior up to about 100 ms after bounce of the
e15 model ($\beta_i$ = 0.645\%) conforms to type II behavior.   
Figure \ref{fig:hegere15s15}a shows the evolution of the maximum density and the
gravitational waveform of model e15 and Fig. \ref{fig:hegere15s15}b shows
the evolution of these quantities for model s15A1000$\beta$0.7. Even though they experience
core bounce at the same density, the period of the post-bounce oscillations
of model e15 is significantly shorter than that of the s15 model.  Since shorter
periods mean greater accelerations, the individual peaks of the waveform, associated
with the first and subsequent bounces of the e15 model, are greater 
than those of the s15 model. The same holds for the integrated radiated energy.

As with the s20 model, the bounce shock of the
e15 model stalls, damping the type II-like post-bounce
expansion-collapse-bounce cycles. This process introduces high frequency components
into the waveform and, thus, into the energy spectrum of the gravitational wave 
signal (Fig. \ref{fig:hegere15s15}d). The e20 model has a lower initial
$\beta_i$ than e15 and similar total angular momentum. Hence, it reaches a larger 
density at bounce and its gravitational wave energy spectrum peaks at about
193 Hz (for e15, it is 90 Hz).  However, the overall behavior 
is similar and the bounce shock of the e20 model
also stalls.  The subsequent infall of matter onto the inner core
is non-spherical (see, for example, Fig. \ref{fig:2ds20} and Table \ref{table:resultsheger}). 

\begin{figure}
\epsscale{1.0}
\plotone{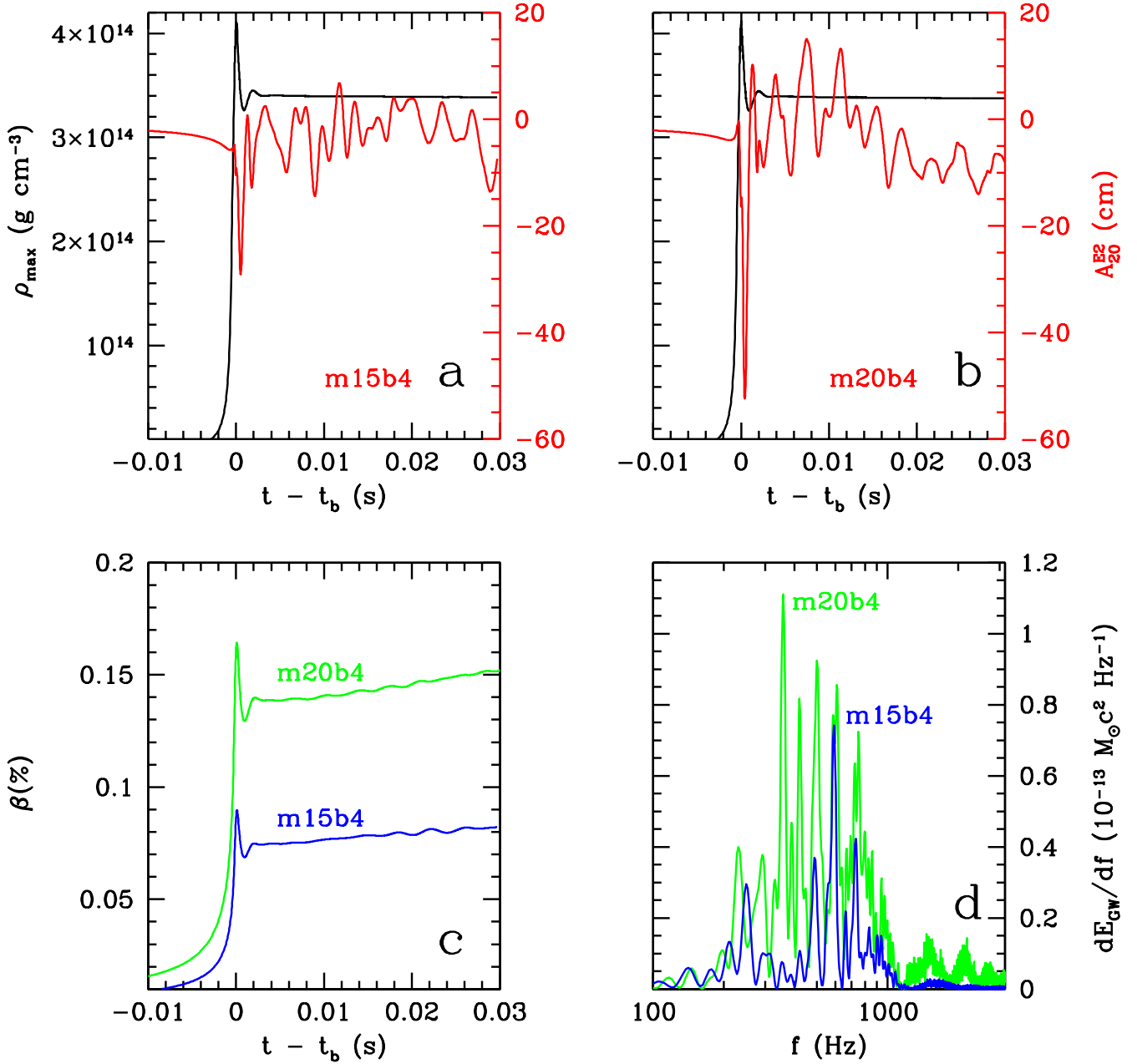}
\caption{The 15 and 20 \modot models from \cite{spruit:03}. 
\undertext{{\bf a (upper left):}} Evolution of the maximum density (black) 
and the gravitational wave amplitude (red) of the m15b4 model. 
\undertext{{\bf b (upper right):}} 
The same for the m20b4 model. Both models rotate very slowly and
rigidly and show only small deviations from spherical symmetry. Hence, the  
gravitational wave amplitudes are small. \undertext{{\bf c (lower left):}}
This figure depicts the evolution of the rotation parameter $\beta$ for these
models. Both models bounce due to the stiffening of the equation of state at
nuclear density, with little contribution due to centrifugal forces. The model with the
greater $\beta_i$, m20b4, reaches a larger final $\beta$.  
\undertext{{\bf d (lower right):}} Energy spectra. The spectrum of the m20b4 model
peaks at lower frequencies and contains significantly more energy (4 times more; 
see Table 6) than the specrum of the m15b4 model. This shows the sensitivity of the
gravitational wave signature to small differences in the initial rotation profile and 
stellar structure.
\label{fig:hegerm}
}
\end{figure}

The models m15b4, m20b4 and m25b4 have initial rotation parameter 
$\beta_i = E_{rot}/|E_{grav}|$ between 0.002\% and 0.005\%. These values of $\beta_i$ are
much smaller than the ones we investigated with the models of \cite{ww:95}. The effects
of rotation on the m-models are minimal and, as can be seen in Fig. 
\ref{fig:hegerm} for models m15b4 and m20b4 (m25b4 is
similar), the gravitational wave amplitudes from the
bounce and ring-down phases are one to two orders of magnitude smaller than those of any
s11/s15/s20 model with $\beta_i$=0.1\% (see Fig. \ref{fig:s15A1000_1}). In
fact, as in the nonrotating case discussed below (\S \ref{section:nonrot}), 
most of the gravitational wave energy is being emitted by aspherical convective 
overturn after bounce (Table \ref{table:resultsheger}) and the spectra are dominated
by the characteristic frequencies of the convective 
motions (Burrows and Hayes 1996; M\"uller and Janka 1997).

\subsection{Nonrotating Models}
\label{section:nonrot}

\begin{figure}
\plotone{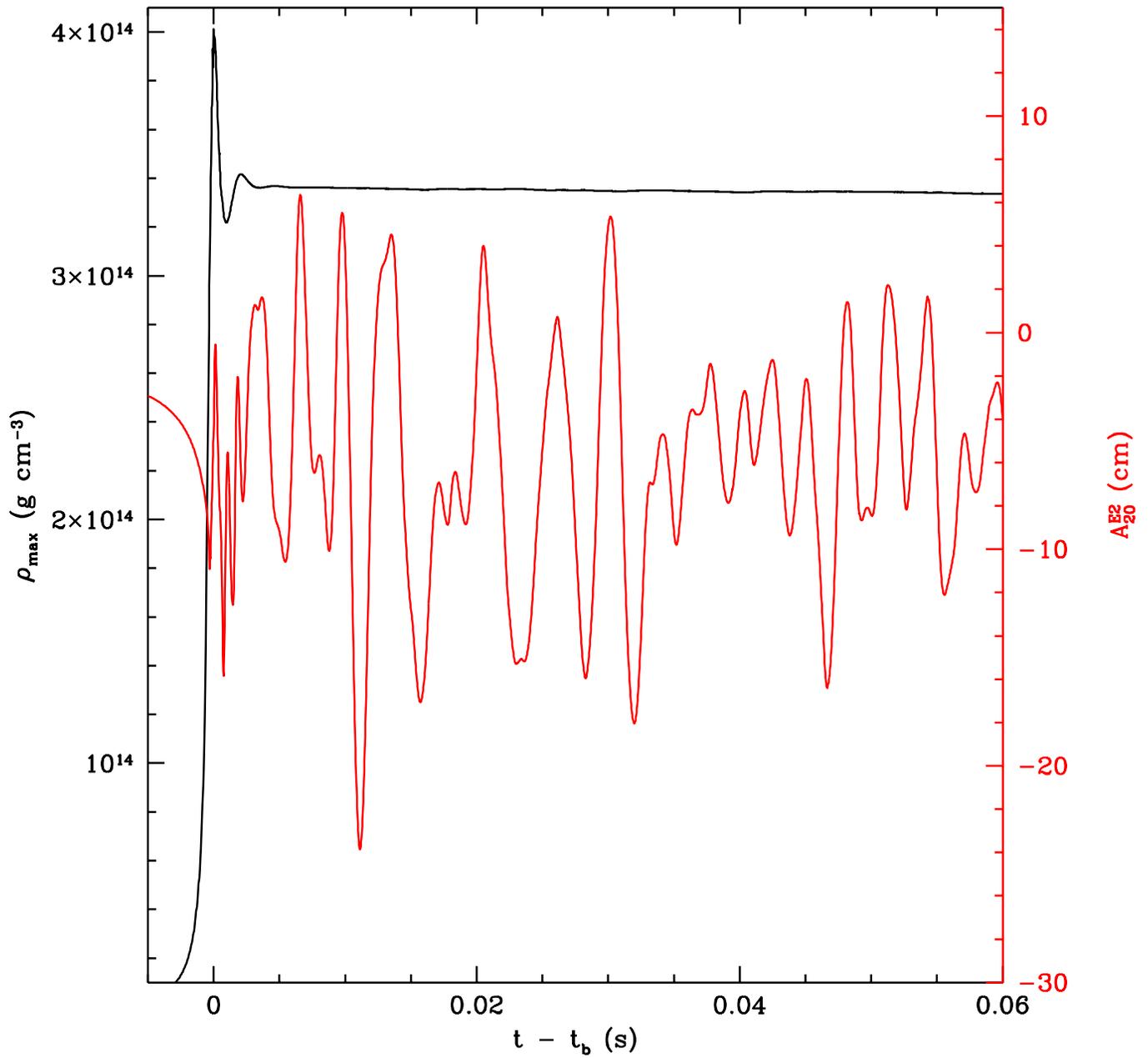}
\caption{Evolution of the maximum density (black) and the gravitational wave amplitude (red)
of a s15 model evolved without rotation. The time is given relative to the time of core
bounce.  Small scale perturbations, introduced by the finite-difference approximation
and post-bounce convective instability, lead to continuous gravitational wave emission with
amplitudes that are one to two orders of magnitude smaller then those observed from
the collapse of a rotating model. 
\label{fig:nonrot}}
\end{figure}

As calibration, we performed calculations without rotation with the s11, s15 and s20 progenitor
models from \cite{ww:95}. Figure \ref{fig:nonrot} shows the evolution of the maximum
density and the gravitational wave amplitude for a nonrotating
s15 model. The negative spike in the gravitational wave amplitude
associated with core bounce exhibited by this model is about two orders of magnitude 
weaker than the one seen for models with slow rotation (Table 
\ref{table:resultss15}). This is a consequence of the much smaller reduced mass-quadrupole moment of 
this nonrotating model,  not deformed by centrifugal forces. Even though the 
collapse proceeds spherically, its reduced mass-quadrupole moment is initially 
not exactly zero, since small perturbations introduced by the numerical scheme and by the
mapping of the one-dimensional progenitor model onto the two-dimensional computational grid
give rise to some initial asphericity which grows during collaps.  

After core bounce convective instability leads to aspherical bulk mass motion 
in the central regions, emitting small amplitude gravitational waves with 
frequencies corresponding to the characteristic turnover periods. Since this type
of gravitational wave emission is not connected to the dynamical event of 
core bounce, it lasts for a prolonged interval, and can eventually radiate more
energy than the dynamical event itself.   The characteristic frequencies
of the gravitational radiation from the overturning motions range from 225 Hz
to 960 Hz.

\section{Evolution of the Rotation Rate}
\label{section:omegaev}

The evolution of the rotation parameter $\beta$
and the angular velocity is of particular interest, since they are connected
to still unanswered questions in core-collapse supernovae physics: What is
the period of newborn neutron stars? Can protoneutron
stars become dynamically or secularily unstable to triaxial rotational
modes? 

Our study is Newtonian and lacks a full treatment of the microphysics involved.
Even though we have included realistic presupernova models and a realistic
equation of state, we cannot claim to provide final answers
to the above questions. Moreover, our calculations have been done
in 2D, not 3D, and, thus, are not free of symmetry constraints. 
General relativistic gravity would lead to more compact configurations
with higher $\beta$s.  Hence, the results presented in this section
should be seen only as indications of the systematic behavior of the
rotation rate evolution and of the changes in the distribution of 
angular velocity with total angular momentum.

\begin{figure}
\epsscale{1.0}
\plotone{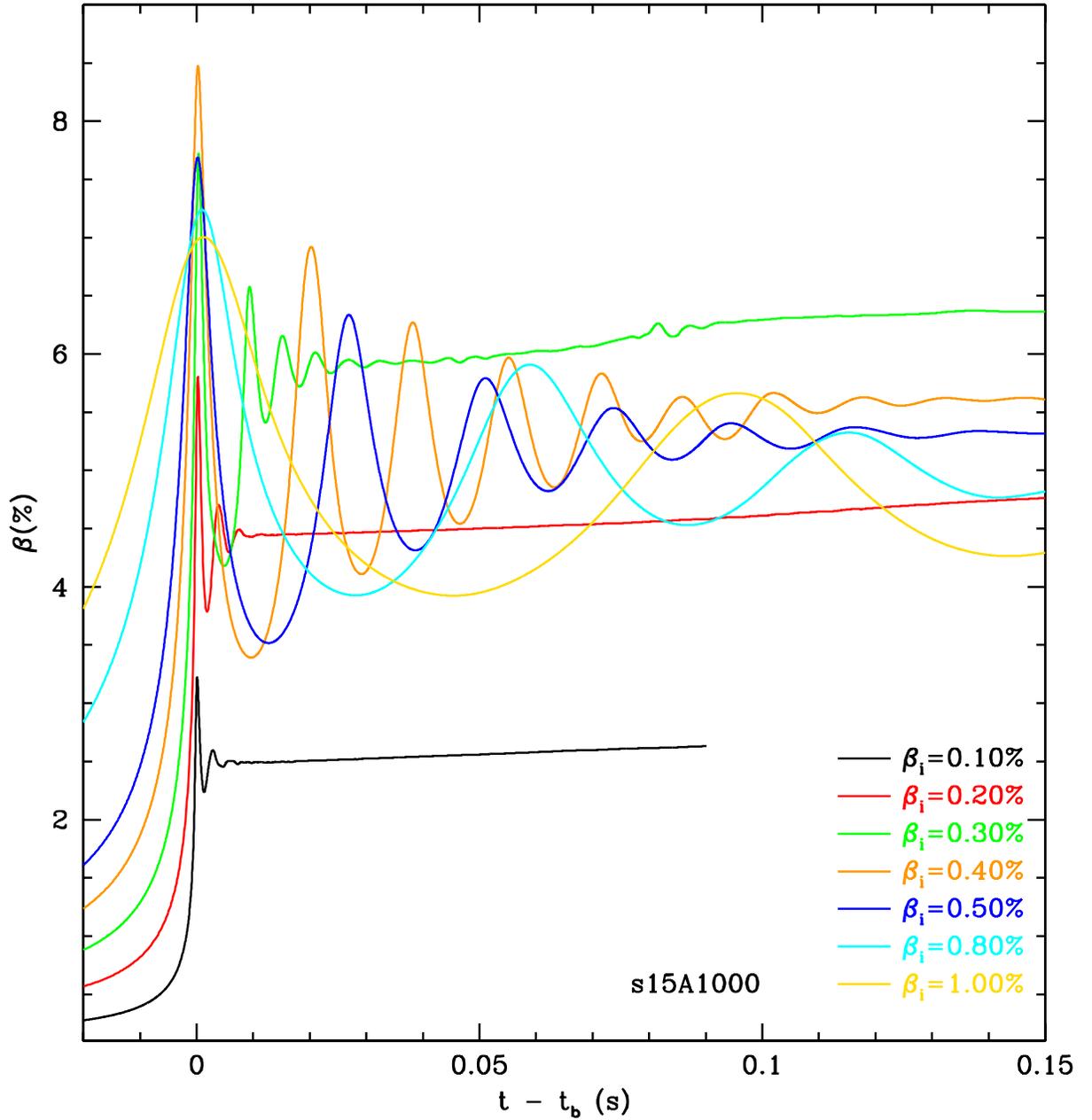}
\caption{Evolution of the ratio of rotational to gravitational energy (rotation
parameter $\beta$) for the s15A1000 model series. The zero of the time
axis is set to the time of bounce for each model. The $\beta_i$=0.40\% model, which is the
first to bounce by centrifugal forces, reaches the maximum $\beta$
at bounce for all models. The $\beta_i$=0.30\% model has a smaller $\beta$ at bounce but
shows, due to the greater compactness of its remnant, the largest final $\beta$. The models
with $\beta_i >$ 0.40\% collapse so slowly that centrifugal forces are able to 
halt collapse before greater compression can lead 
to very large $\beta$s. None of our models exceeds a $\beta$ of 10\%. 
\label{fig:s15A1000_beta}
}
\end{figure}

A look at Fig. \ref{fig:s15A1000_beta} and at Tables \ref{table:resultss11},
\ref{table:resultss15}, and \ref{table:resultss20} discloses that there exists a maximum 
value of $\beta$ at bounce for a given progenitor model and value of A. Interestingly,
the maximum $\beta$ is not reached by the model with the maximum $\beta_i$, but
by a model with some intermediate value of $\beta_i$ (in Fig. \ref{fig:s15A1000_beta}
this is 0.40\%). 
$\beta$ at bounce is determined by the subtle interplay between initial angular 
momentum distribution, the equation of state, 
centrifugal forces and gravity. The ``optimal'' configuration
leads to the overall maximum $\beta$ at bounce for a given $\beta_i$.  
Moreover, the model yielding the maximum final $\beta$ is not
necessarily the model yielding the maximum $\beta$ at bounce (see Fig. 
\ref{fig:s15A1000_beta} and Tables 3-6).

\begin{figure}
\epsscale{1.0}
\plotone{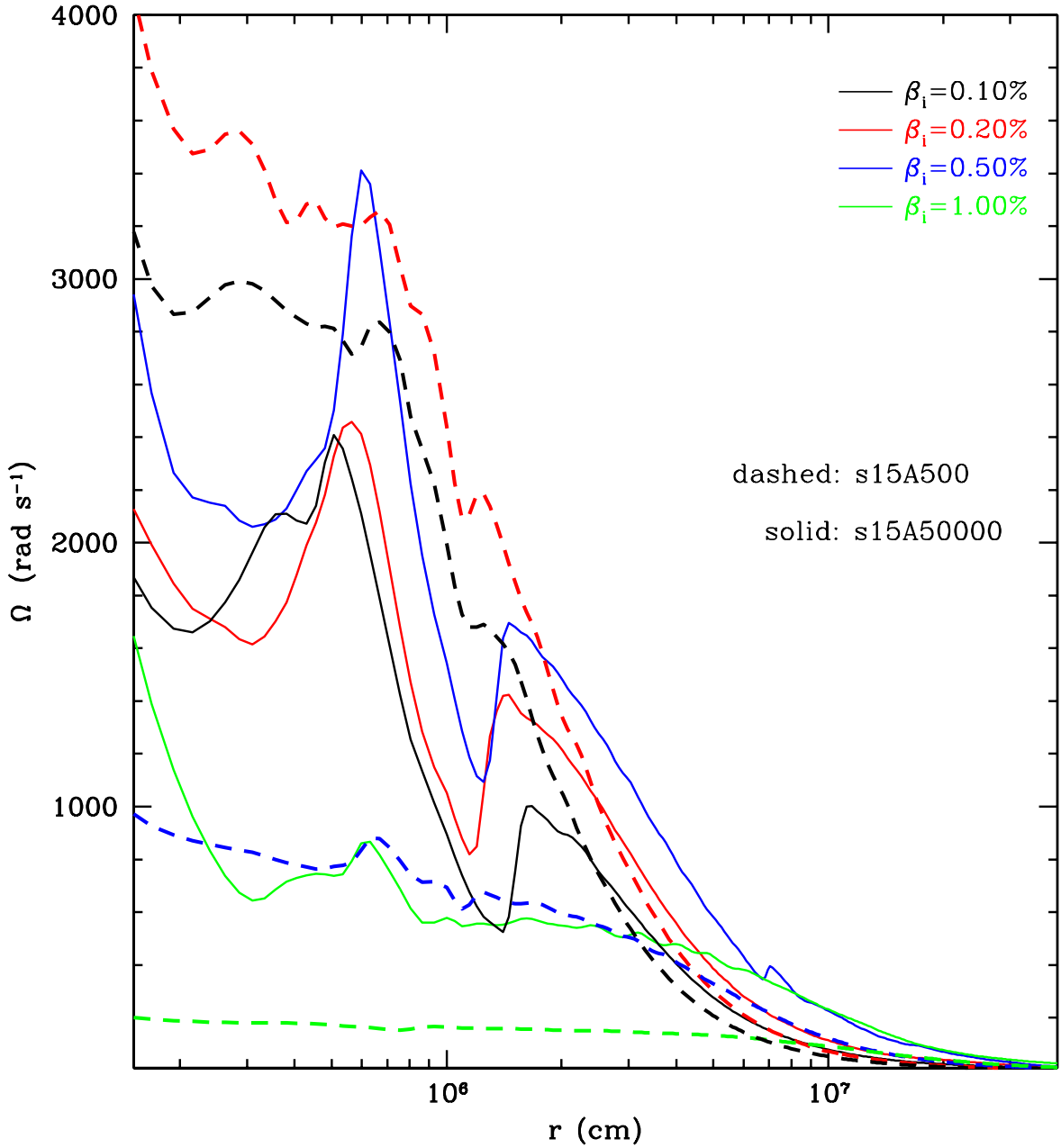}
\caption{Angular velocity profiles in the equatorial region at the end of the
evolution of s15 models with A=500 km (dashed) and A=50000 km (solid)
and for $\beta_i$s of 0.10, 0.20, 0.50, and 1.00\%. Interestingly, the initially
rigidly rotating models (A=50000 km) exhibit a more differentially
rotating central region than the initially more differentially rotating
models do. The local peak in the angular velocity at $\sim$6-8 km
and the strong  $\Omega$-gradients associated with it
have been considered possible drivers of the
magneto-rotational instability (MRI)
(\citealt{wheeler:03}).
\label{fig:s15_omega}}
\end{figure}

Figure \ref{fig:s15_omega} shows the final angular velocity distribution versus radius at the equator of the
s15 model for A=500 km and A=50000 km and for a variety of $\beta_i$s. As with 
$\beta$, overall the angular velocity increases with 
increasing $\beta_i$ until a maximum is reached. It subsequently decreases
with the further increase of $\beta_i$. Both initial settings of A lead to
strongly differential rotation in the central regions, while the initially more 
rigidly rotating model (A=50000 km, 
solid lines in Fig. \ref{fig:s15_omega}) actually yields larger  
post-bounce angular velocity gradients inside 30 km. 
Its equatorial velocity profile peaks off center for moderate
$\beta_i$ at radii between 6 and 8 km. The initially more differentially rotating
model (A=500 km, dashed lines in Fig. \ref{fig:s15_omega}) leads to the highest central values of the
angular velocity while its angular velocity profile quickly drops to low 
values and practically rigid rotation for $\beta_i \ge$ 0.3\%. Model s15A500$\beta$0.2 results in the shortest rotation period near the center ($\sim$1.5 ms).
Model s15A50000$\beta$0.5 yields the shortest period of the A=50000 km model series ($\sim$1.85 ms). 
The angular velocity shear exterior to the peak at 6-8 km 
exhibited by these models has also been identified in the one-dimensional 
study of \cite{Akiyama:03}. These authors 
consider such shear a possible driver for the magneto-rotational
instability (MRI), which could be a generator of strong magnetic fields. 

None of our models develop off-center density maxima (i.e.\ become toroidal) which could lead to 
dynamical growth of an azimuthal m=1 mode at $\beta$ \sgreater\ 14\% as suggested by \cite{cent:01}.
\cite{shibata:03} have shown that strongly differentially rotating polytropes with centrally 
peaked density profiles and EOS-$\Gamma$ above $\sim$ 4/3 can become dynamically unstable 
to triaxial m=2 deformations (bar modes) 
for $\beta$ on the order of 1\%. As seen in Fig. \ref{fig:s15_omega}
and discussed in the previous paragraph, our models exhibit strong differential rotation inside
the protoneutron star. However, it is impossible to judge whether they are stable or unstable
to triaxial rotational instabilities in context of \cite{shibata:03}, since these protoneutron
stars differ greatly from the simple polytropic equilibrium models employed in that study.
Three-dimensional simulations employing a finite-temperature EOS and realistic 
post-collapse models are needed to address this issue conclusively.  

\section{Prospects for Detection}
\label{section:detection}

To assess the detectability of the gravitational waves radiated by our
models we follow the discussions in \cite{abra:92} and \cite{flanhughes:98}. For
a given frequency, $f$, \cite{flanhughes:98} define the characteristic gravitational 
wave strain 

\begin{equation}
\label{eq:charstrain}
h_{char} (f) = \sqrt{\frac{2}{\pi^2} \frac{G}{c^3} \frac{1}{D^2} \frac{dE(f)}{df}},
\end{equation}
where D is the distance of the source from the detector (for galactic sources we
set this equal to 10 kpc) and $dE(f)/df$ is the spectral energy density of the gravitational
radiation defined by eq. (\ref{eq:dEdnu}). 

The matched-filter signal-to-noise ratio for a
source that emits at optimal orientation and polarization is given by

\begin{equation}
(SNR_{optimal})^2 = \int d (ln f) \frac{h_{char}(f)^2}{h_{rms}(f)^2}
\end{equation}
with the (single) detector rms noise strain $h_{rms}$ being defined as the square root of
frequency times the detector noise power spectral density
\begin{equation}
h_{rms}(f) = \sqrt{f\,S(f)}\, .
\label{hrms}
\end{equation}

For two observatories, averaging over all angles and both polarizations, and assuming
a SNR of 5, the detector burst sensitivity is considered to be  
$h_{SB} \simeq 11 h_{rms}$ (\citealt{abra:92}; \citealt{gust:99}).

Figure \ref{fig:ligo} shows the detector $h_{rms}$ noise strains of the initial and 
advanced LIGO designs (\citealt{gust:99}). The solid squares mark the peaks of 
the optimal characteristic strain $h_{char} (f_{max})$ for each 
of our models at an assumed distance of 10 kpc. 
The most important parameters governing the position of a given
model in Fig. \ref{fig:ligo} are $\beta_i$ and A.
Generally, the models with moderate initial rotation ($\beta_i$ below 0.5\%)
peak at frequencies well above the LIGO peak sensitivity (near 100 Hz), but 
also give the largest overall gravitational wave amplitudes. 
With increasing $\beta_i$, the models shift towards lower frequencies 
and smaller amplitudes, but at $D$=10 kpc remain for the most part above the initial
LIGO sensitivity limit. Only for the strongest rotators, which peak at low
frequencies, will detectability at 10 kpc by the initial 
LIGO interferometers be problematic.  For a given $\beta_i$, the model spectra shift upwards
and to the right with increasing A. Overall, our s20 models peak
at the largest $h_{char}$. The peaks of the s11 and s15 models are
very similar for slow rotation, but the fast rotating s11 models peak at
lower frequencies and higher strains than the corresponding s15 models.
If we were to draw an imaginary line through the field of points on Fig. \ref{fig:ligo}, 
we would find that peak $h_{char}$ is very roughly proportional to $f_{max}$
to the 0.8 power.

The models evolved without rotation are marked with little stars in Fig. \ref{fig:ligo}. 
Models marked with a diamond correspond to rotating 
progenitors from the studies of \cite{heger:00} and \cite{spruit:03}. The three models
at higher frequencies below the sensitivity of the first-generation 
(initial) LIGO are the three ``magnetic'' progenitors that
rotate very slowly ($\beta_i$ < 0.01\%). 
In sum, approximately 80\% of our models are within the optimal sensitivity limit of 
the first-generation LIGO and about 10\% should be detectable even
using the $h_{SB}$ condition. Almost all of our models should be detectable
with the 2nd-generation (advanced) LIGO.

\section{Summary and Discussion}
\label{summary}

In this parameter study, we have investigated
the emission of gravitational radiation from rotational stellar 
core collapse using Newtonian gravity, realistic initial models, 
and a realistic equation of state. 
We have performed a total of \numberofmodels simulations, investigating
the dependence upon the degree of differential rotation, the initial ratio of
rotational energy to gravitational potential energy ($\beta_i$), and progenitor. 
For this, we employed the 11, 15 and 20 \modot (s11, s15, s20) presupernova models from the stellar evolution
study of \cite{ww:95} and put them into rotation via a rotation law that
assumes rotation on cylinders. In addition, we have performed
simulations with the progenitor models of \cite{heger:00} and \cite{spruit:03}
that include a one-dimensional prescription for rotational evolution. All of our models
encompass the full 180$^{\circ}$ of the symmetry domain.
Nothing artificial was done to initiate collapse.

\begin{figure}
\plotone{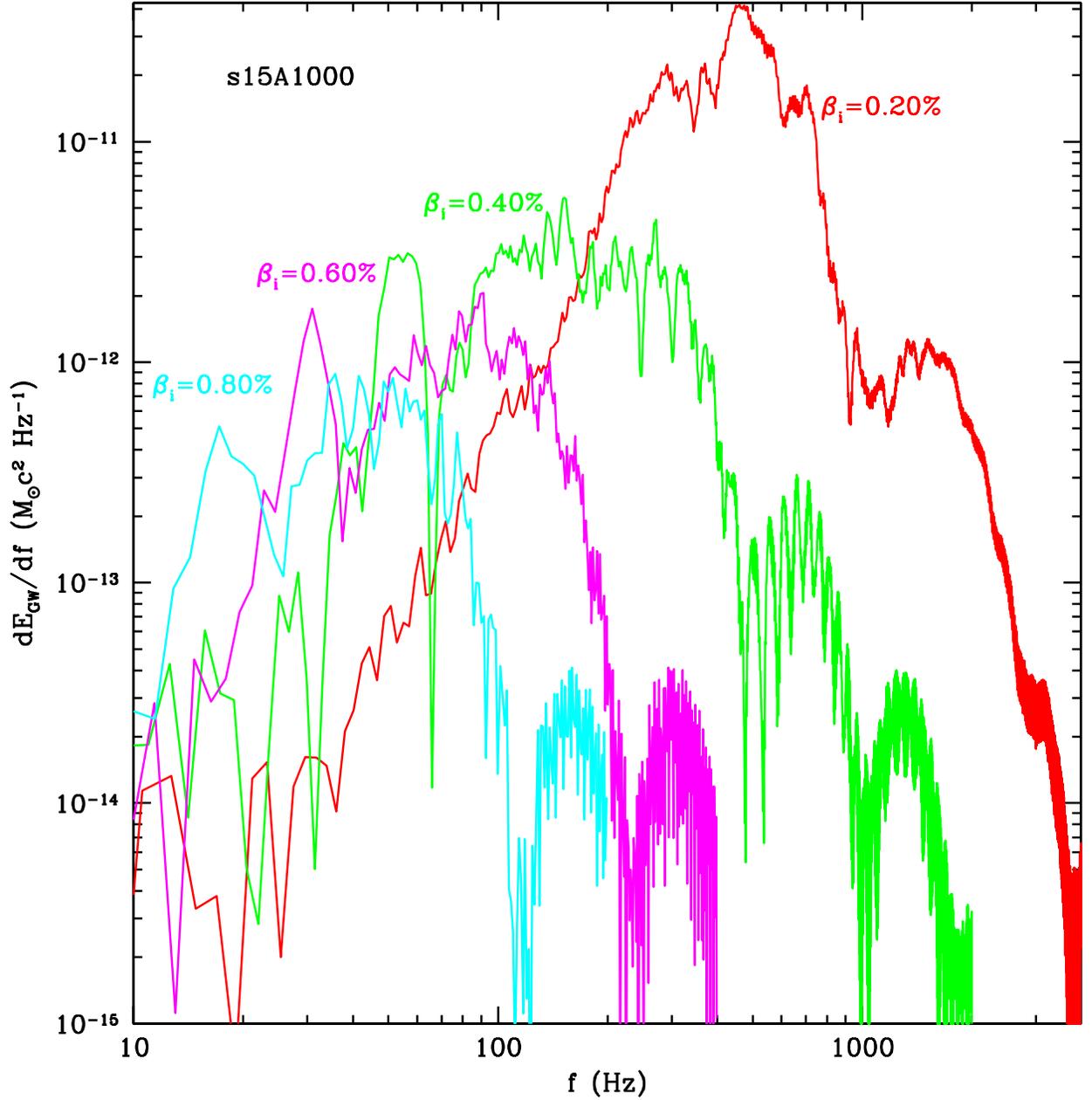}
\caption{Energy spectra of the gravitational radiation emitted from representative
models of the s15A1000 model series.  Note the shift of the spectra to lower
frequencies with increasing $\beta_i$ and the logarithmic scale of the ordinate. The spectra have been
cut off at a frequency beyond which generic high frequency noise at constant
magnitude sets in. The first pronounced peak in the spectra of the more strongly 
rotating models can be identified with the frequency of the 
expansion-collapse-bounce cycles exhibited by these models (type II). The spectra of the slower
rotators peak at the dominant frequencies of the post-bounce ringing of the compact 
remnant.  The higher harmonics are clearly seen in these models. 
\label{fig:s15A1000_spect}}
\end{figure}

\begin{figure}
\epsscale{1.0}
\plotone{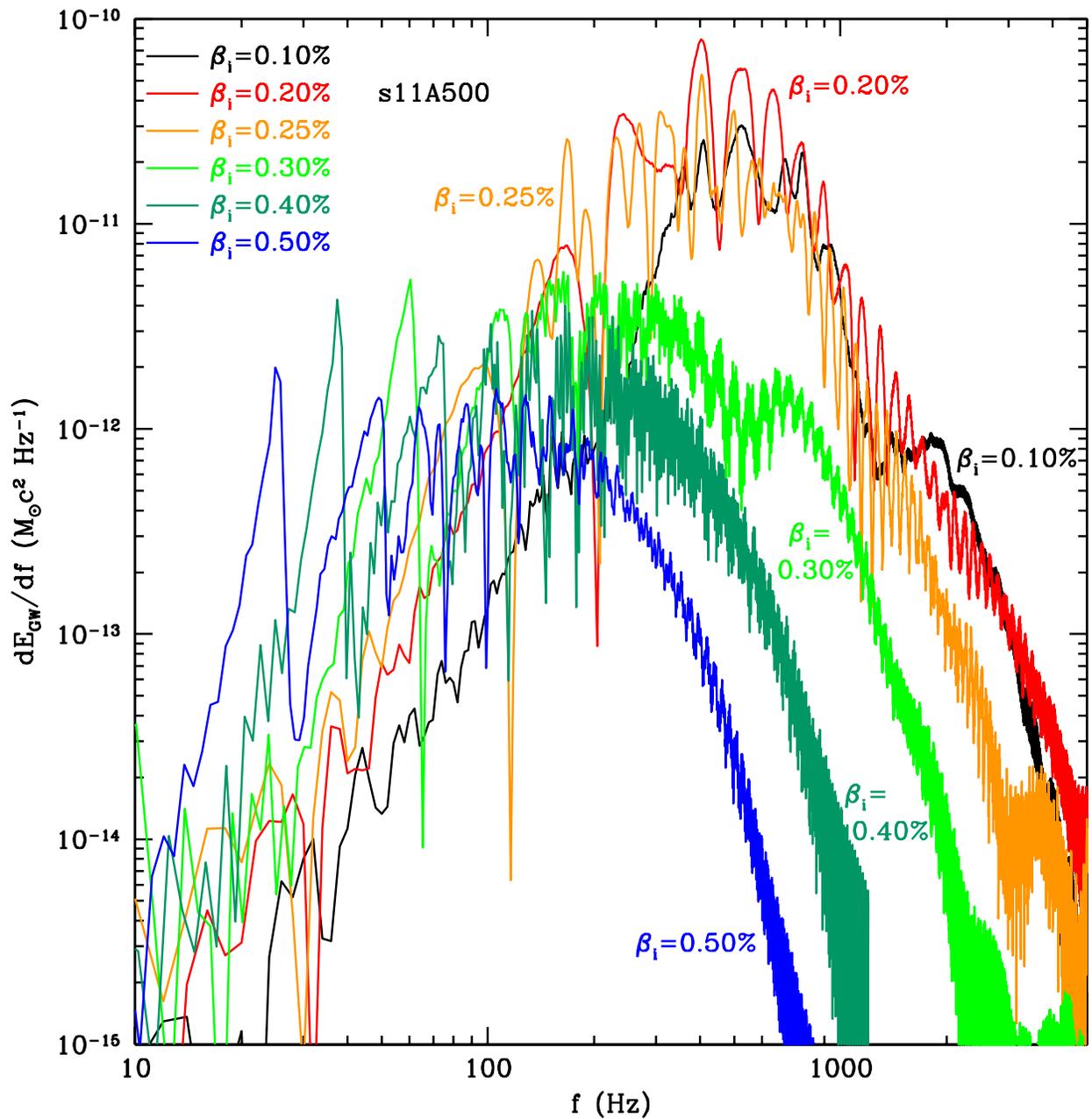}
\caption{Same as Fig. \ref{fig:s15A1000_spect}, but for the s11A500 sequence
and for $\beta_i$s from 0.1\% to 0.5\%.
\label{s11A3_spect}}
\end{figure}

Our results indicate that there are two types of characteristic behavior
for the collapse dynamics and the resulting gravitational wave signature. Type I occurs
for slow initial rotation and is characterized by core bounce at supranuclear
densities. The wave signal of a type I model exhibits a sharp spike and 
high frequency oscillations as the compact remnant rings down after core bounce.
Type I models have the largest gravitational wave amplitudes, their
energy spectra peak at the highest frequencies and they radiate the largest amount
of energy. On the other hand, type II models bounce at subnuclear densities
due to the influence of centrifugal forces. They exhibit several 
damped harmonic oscillator-like post-bounce expansion-collapse-bounce 
cycles whose periods grow with increasing $\beta_i$.   
The gravitational wave amplitudes,
frequencies, and total energies of type II models are smaller than for
those that exhibit type I behavior. The frequency at which the coherent post-bounce
oscillations occur is clearly discernible in the energy spectra of type II models. 
Figure \ref{s11A3_spect} portrays the gravitational wave energy spectrum
for the s11A500 sequence at various $\beta_i$s from 0.1\% to 0.5\%.  This
figure is similar to Fig. \ref{fig:s15A1000_spect} for the s151000 series.  The systematic
shift with increasing $\beta_i$ above the transition $\beta_i$ ($\sim$0.25\%)  
from higher frequencies and strength to lower values of both is clearly seen.  
Transitional models that exhibit features of both
type I and type II behavior are also present. For a given initial model, 
we find that the transition from type I to type II happens for
smaller $\beta_i$ if the core is initially more differentially rotating and
for larger $\beta_i$ if its core has rigid initial rotation.  The s11 and
s15 progenitor models give very similar results, the s11 model giving
slightly larger gravitational wave amplitudes, frequencies, and energies than
the s15 model for a given set of parameters. The s20 model, however, having a
shallower initial density distribution, leads to different
post-bounce behavior, since its hydrodynamic bounce shock stalls. 
Subsequent to stall, matter infalls 
aspherically through and onto the central core.  This leads to
near-critical damping of the post-bounce oscillations of the compact remnant
and introduces high-frequency components into the wave form. Since 
sophisticated one-dimensional studies that include weak interaction
physics and neutrino transport indicate that the bounce shock
does indeed initially stall (Rampp and Janka 2002; Liebend\"orfer et al. 2001a,b;
Thompson, Burrows, and Pinto 2003), the distinctive character of the s20 model may in fact
be more generic than that of our s11 and s15 models.  

The rotating progenitor models e15 and e20 from \cite{heger:00} 
behave similarly to their s15 and s20 counterparts 
with similar $\beta_i$s and As. 
However, due to differences in presupernova structure,
the bounce shock of the e15 model stalls,  
leading to more infall into
the central core and strong damping of post-bounce
oscillations. The same is true for the e20 model.
Hence, the gravitational wave signatures for both models are 
similar to those for the s20 models. The models of \cite{spruit:03} 
include the effects of magnetic fields on the distribution 
of angular momentum.  This leads to very rigid rotation throughout the star
and, thus, to little angular momentum in the iron core itself. Our simulations
show that these extremely slowly rotating cores ($\beta_i < 0.005\%$)
yield gravitational wave amplitudes from core bounce that are two
orders of magnitude smaller than for the s11, s15 and s20 models
with the smallest $\beta_i$.  Hence, the 
quantitative and qualitative behavior of these slowly rotating 
models is very similar to the models that were evolved without any
rotation at all.

Our models yield absolute values of the dimensionless 
maximum gravitational wave strain
in the interval 2.0 $\times$ 10$^{-23}$ $\le$ $h^{TT}_{max}$ $\le$ 
1.25 $\times$ 10$^{-20}$ at a distance of 10 kpc. 
The total energy radiated ($E_{GW}$) lies 
in the range 1.4 $\times$ 10$^{-11}$ \modot c$^2$ $\le$ $E_{GW}$ $\le$ 
2.21 $\times$ 10$^{-8}$ \modot c$^2$ and the energy spectra peak (with the exception
of very few models) in the frequency interval 20 Hz \sles$\,$ f$_{peak}$ \sles$\,$~600~Hz.  
Furthermore, our hydrodynamic results indicate (if the canonical critical conditions
apply), that none of our collapsed cores will 
undergo triaxial rotational instabilities, since the maximum ratio of rotational to
gravitational energy ($\beta$) reached is 9.16\% at bounce and 8.23\% at the end
of the evolution.  Both these values are below the putative critical $\beta$ for
dynamical ($\beta_{dyn}\simeq 27\%$) or secular 
($\beta_{sec}\simeq 14\%$) triaxial instability (in the simplified case of 
a MacLaurin spheroid).  However, given the extreme density profiles
in realistic progenitor cores, and the canonical bifurcation of the flow
into outer supersonic and inner subsonic collapse, the actual critical
$\beta$s for triaxial deformation of realistic cores have yet to be determined. 
Recent studies by \cite{cent:01} and \cite{shibata:03} which employed polytropic 
neutron star models suggest that dynamical rotational instabilities could occur 
at much lower $\beta$. These results have to be verified in three-dimensional
simulations including a finite-temperature EOS and realistic protoneutron star
models.

Since including weak-interaction physics and neutrino transport will change the
overall evolution, our specific results for the $\beta_i$ dependence should be taken with care. 
However, the qualitative behavior and trends we have identified should be robust.

In comparison with the  
studies of \cite{zm:97} (Newtonian gravity) and \cite{harry:02b}  
(relativistic gravity), who used polytropic progenitors and a 
simplified equation of state, our  
models radiate on average about an order of magnitude 
less energy, exhibit a factor of 4 to 10 smaller maximum wave amplitudes, 
and peak at lower frequencies. \cite{zm:97} categorized
the collapse, bounce and post-bounce behavior of their models in three types
of which their types I and II match our types I and II. 
However, we have not encountered their type III behavior (``rapid collapse'') 
in any of our calculations. 

Aside from this study, there have been only two major efforts to calculate
the gravitational wave signature of core collapse with a sophisticated 
realistic equation of state and realistic progenitor models.  These were \cite{mm:91}
and the recent study by \cite{kotake:03}. Both studies included
electron capture and a leakage scheme for approximate treatment 
of neutrino transport, but calculated only a very small set of models and used only
one pre-supernova progenitor. In accord with our results, none of them 
found \cite{zm:97} type III behavior. Their gravitational wave 
amplitudes and total radiated energies are qualitatively and quantitatively similar to ours.

\begin{figure}
\plotone{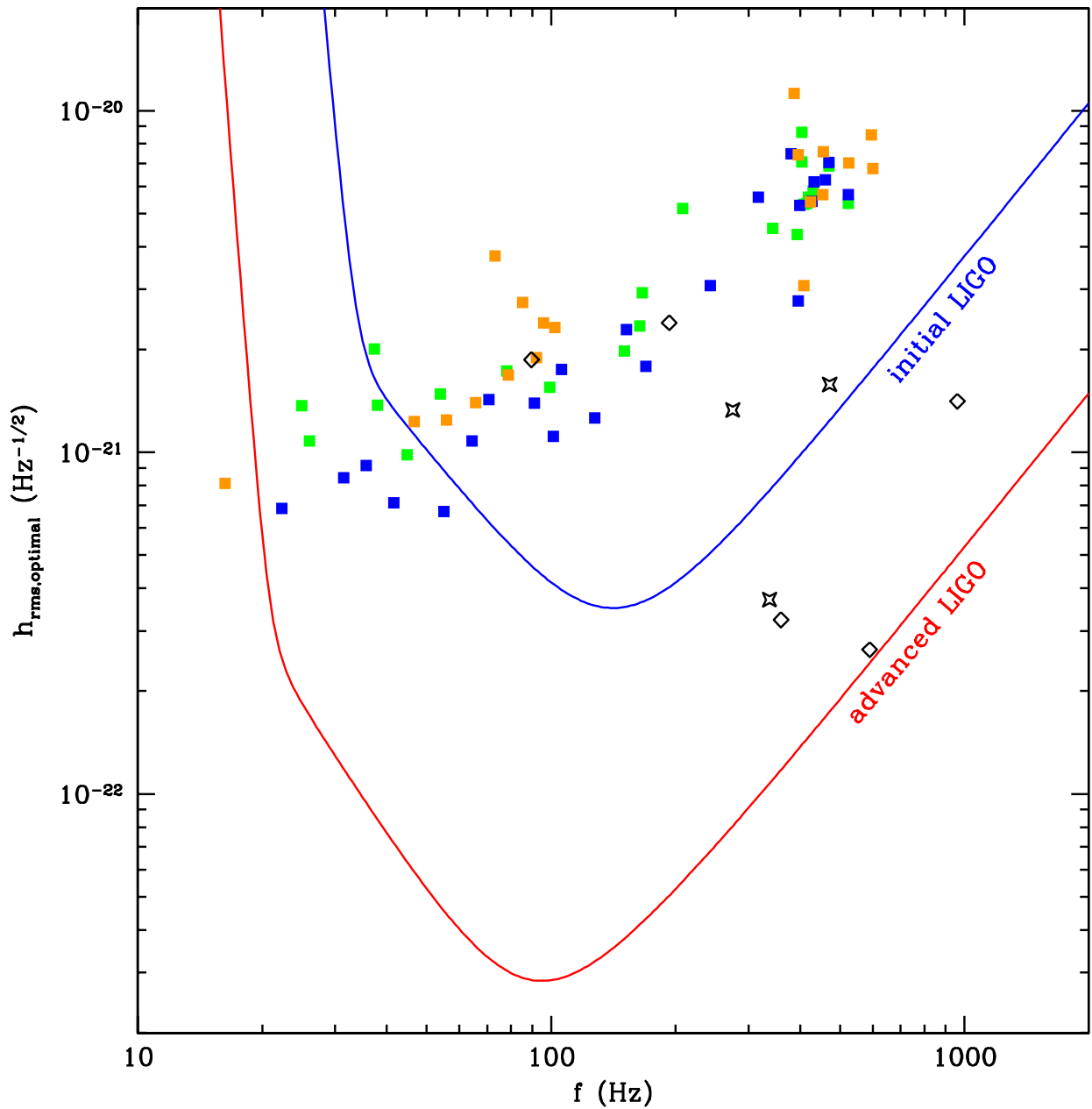}
\caption{LIGO sensitivity plot. Plotted are the optimal root-mean-square noise
strain amplitudes $h_{rms} = \sqrt{f S(f)}$ of the initial and advanced LIGO
interferometer designs. Optimal means that the gravitational waves are
incident at an optimal angle and optimal polarization for detection and that there
are coincident measurements of gravitational waves by
multiple detectors. For gravitational waves from
burst sources incident at random times from a random direction and a signal-to-noise
ratio (SNR) of 5, the rms noise level $h_{rms}$ is approximately a factor of 11 above
the one plotted here (\citealt{abra:92}; \citealt{flanhughes:98}).
We have plotted solid squares at the maxima of the characteristic gravitational wave
strain spectrum ($h_{char} (f)$; \S\ref{section:detection}) of
our s11 (green), s15 (blue), and s20 (orange) models from \cite{ww:95} that
were artificially put into
rotation. Our nonrotating models are marked with stars; diamonds stand for models
from \cite{heger:00} and \cite{spruit:03}. The distance to Earth was set to 10 kpc for
all models. Most of our models lie above the optimal design sensitivity limit of LIGO I.
Hence, the prospects for detection are good.
Those models that are not detectable by the 1st-generation LIGO
are those that rotate most slowly (the \citealt{spruit:03} models) 
and those which are the fastest rotators. \label{fig:ligo}}
\end{figure}

\begin{figure}
\epsscale{1.0}
\plotone{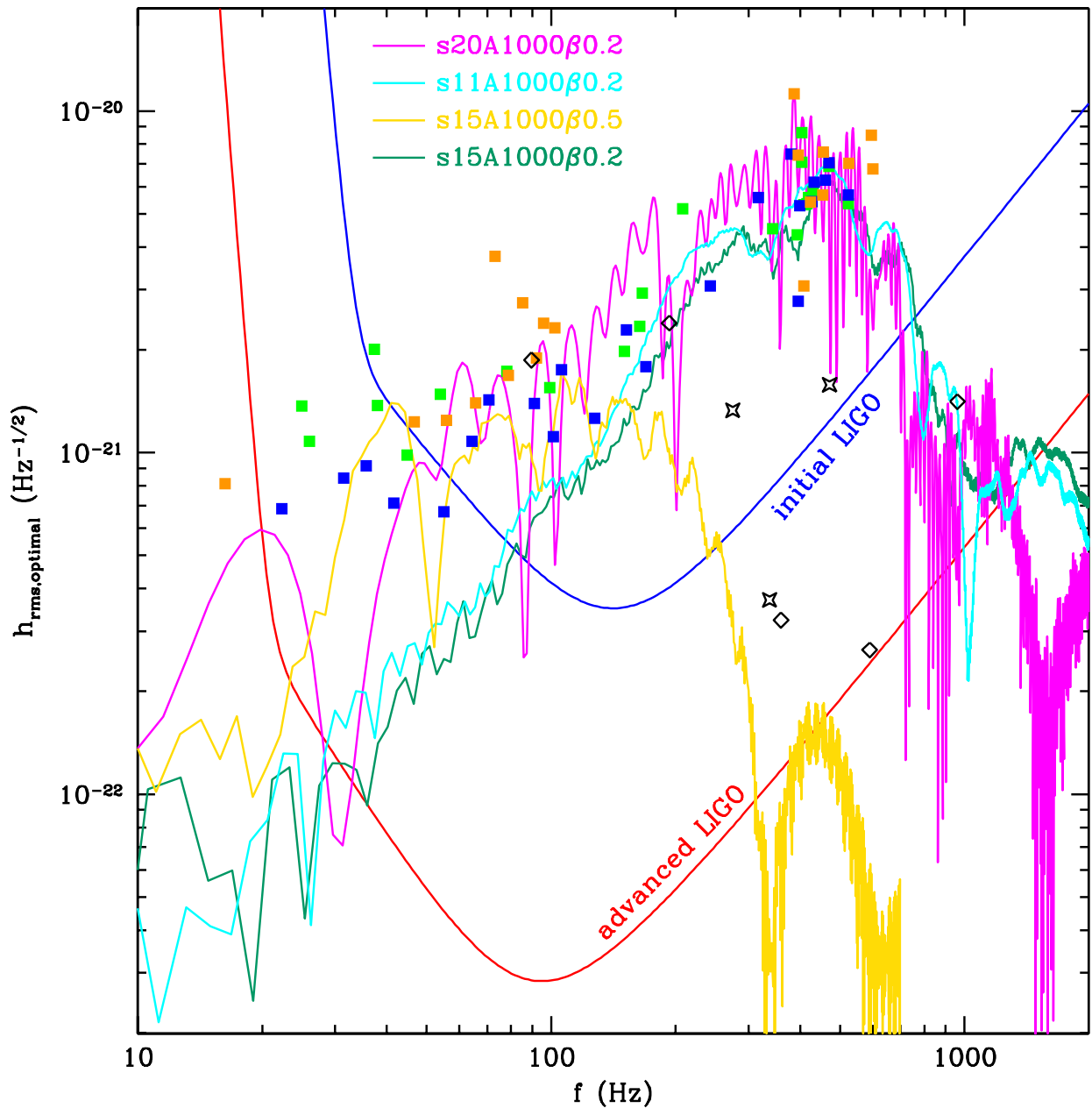}
\caption{Same as Fig. \ref{fig:ligo}, but with various full $h_{char}$ spectra
(using eq. \ref{eq:charstrain}) superposed.  This plot makes clear the large width of actual spectra and
the deviation from even quasi-periodic behavior of rotating collapse wave signatures.
\label{ligo.h}}
\end{figure}

To assess the detectability of gravitational waves from core collapse, we applied the 
method proposed by \cite{flanhughes:98}.  We find that 
at a distance of 10 kpc, i.e. for galactic
distances, the 1st-generation LIGO, once it has reached its design sensitivity level,
will be able to detect more than 80\% of our core collapse
models under optimal conditions and orientations.  
Assuming random polarizations and angles of incidence, this reduces to 10\%.  
Advanced LIGO, however, should be able to detect virtually all models at galactic distances. 
Figure \ref{ligo.h} is similar to Fig. \ref{fig:ligo} in that it presents peak $h_{char}$,
but it also includes the actual $h_{char}$ spectra of selected models (eq. \ref{eq:charstrain}).
These spectra are complementary to the energy spectra of previous figures,
and serve to put the issues of detectability in a noisy detector into sharper relief.  

To conclude, we point out that even though this
study has advanced our knowledge of gravitational waves from
core collapse a bit further, many important questions remain unanswered.  
Detailed microphysics and neutrino transport are apt to change the dynamics 
of collapse, bounce and post-bounce phases. The coherent post-bounce oscillations of 
type II models will most likely be strongly damped by the stalled 
bounce shock in simulations that account for neutrino energy losses. Furthermore,
gravitational radiation is to be expected from anisotropic neutrino emission (Burrows and Hayes 1996), 
and neutrino-driven convection behind the stalled shock and in the protoneutron
star. Both sources are potentially significant emitters of gravitational wave energy.   

The detection of gravitational radiation from collapse supernovae 
would open a new window into the violent dynamics at the core of the
supernova and neutron-star birth phenomena.  Furthermore, features
seen in gravitational radiation may have their counterparts in the neutrino
signal.  Seeing a correlation in these two disparate channels 
could break the study of supernovae wide open.  As we have shown, 
measurements of wave frequency, wave form, and power can
in principle reveal the rotational structure of the massive star interior that,
though of central importance across a broad spectrum of astrophysics,
has to date remained almost completely out of reach.

\acknowledgments

We acknowledge helpful conversations with Todd Thompson, 
Itamar Lichtenstadt, Harald Dimmelmeier, Ian Hawke, Chris Fryer, Ed Seidel,
Sam Finn, Joan Centrella, Kim New, Wolfgang Duschl, 
Casey Meakin, Jeremiah Murphy and Sven Marnach.  In addition, we thank
Jeff Fookson and Neal Lauver of the Steward Computer Support Group
for their invaluable help with the local Beowulf cluster. 
Support for this work is provided in part by the Scientific Discovery 
through Advanced Computing (SciDAC) program of the DOE, grant number 
DE-FC02-01ER41184.


\clearpage

\begin{deluxetable}{lrrccrlc}

\tablecaption{Initial Model Data\label{table:initialmodels}}
\tablehead{
\colhead{Model Name}&
\colhead{$R_c$}&
\colhead{$M_c$}&
\colhead{ZAMS Mass }&
\colhead{ZAMS $v_{eq}$}&
\colhead{$\beta_i$}&
\colhead{}&
\colhead{Reference}\\
\colhead{}&
\colhead{($10^8\,$ cm)}&
\colhead{(\modot)}&
\colhead{(\modot)}&
\colhead{(km s$^{-1}$)}&
\colhead{(\%)}&
\colhead{}&
\colhead{}
}
\startdata
n=3 polytrope&$1.55$&$1.46$&&&&&\cite{zm:97}\\
s11WW&$1.33$&$1.31$&$11$&&&&\cite{ww:95}\\
s15WW&$1.16$&$1.28$&$15$&&&&\cite{ww:95}\\
s20WW&$2.21$&$1.73$&$20$&&&&\cite{ww:95}\\
s25WW&$2.28$&$1.78$&$25$&&&&\cite{ww:95}\\
e15&$1.21$ &$1.33$ &15&200&0.650&&\cite{heger:00}\\
e20&$2.19$&$1.68$ &20&200&0.420&&\cite{heger:00}\\
m15b4&$1.23$&$1.37$&15&200&0.002&&\cite{spruit:03}\\
m20b4&$1.53$&$1.49$&20&200&0.003&&\cite{spruit:03}\\
m25b4&$1.93$&$1.64$&25&200&0.005&&\cite{spruit:03}\\ 
\enddata
\tablecomments{
List of progenitor models used in this paper. $R_c$ is the radius of 
the iron core (determined by the discontinuity in $Y_e$ at the 
outer edge of the iron core), $M_c$ is the mass of the core, $v_{eq}$ the 
equatorial velocity of the model at ZAMS, and $\beta_i$
is the initial rotation parameter. 
All progenitor models already have an initial infall 
velocity profile when they are mapped onto our 2-dimensional
grid. Nothing artificial is done to initiate collapse.}
\label{initial.tab}
\end{deluxetable}

\begin{deluxetable}{lccccrcrrcrr}

\tablecaption{Results: Polytropes\label{table:polytropes}}
\tablehead{
\colhead{Model}&
\colhead{$\beta_i$}&
\colhead{J}&
\colhead{t$_b$}&
\colhead{$\Delta$t}&
\colhead{$\beta_b$}&
\colhead{$\rho_{b}$}&
\colhead{$A_{20}^{E2}\,_{max}$}&
\colhead{$h^{TT}_{max}$}&
\colhead{$f_{max}$}&
\colhead{$E_{GW,b}$}&
\colhead{$E_{GW,f}$}\\
\colhead{}&
\colhead{(\%)}&
\colhead{($10^{49}$}&
\colhead{(ms)}&
\colhead{(ms)}&
\colhead{(\%)}&
\colhead{($10^{14}$\hspace*{0.3cm}}&
\colhead{(cm)}&
\colhead{@ 10 kpc}&
\colhead{(Hz)}&
\colhead{($10^{-9}$\hspace*{0.3cm}}&
\colhead{($10^{-9}$\hspace*{0.3cm}}\\
\colhead{}&
\colhead{}&
\colhead{erg s)}&
\colhead{}&
\colhead{}&
\colhead{}&
\colhead{g cm$^{-3}$)}&
\colhead{}&
\colhead{(10$^{-21}$)}&
\colhead{}&
\colhead{\modot c$^2$)}&
\colhead{\modot c$^2$)}
}
\startdata
A50000$\beta$0.25&0.25&1.273&66.78&83.22&4.03&3.52&-876.19&-7.76&540&14.50&18.28\\
A50000$\beta$0.50&0.50&1.746&67.51&82.49&6.75&3.26&-1617.13&-14.31&453&41.43&56.31\\
A50000$\beta$0.90&0.90&2.370&68.90&81.11&9.97&2.83&-2125.30&-18.81&460&49.24&57.08\\
A1000$\beta$0.25&0.25&1.227&67.05&82.95&5.28&3.43&-1211.62&-10.72&467&25.85&36.04\\
A1000$\beta$0.50&0.50&1.646&67.97&82.03&8.40&3.10&-2151.21&-19.04&693&65.61&71.57\\
A1000$\beta$0.90&0.90&2.314&70.11&79.89&12.48&2.36&-1674.60&-14.82&313&17.71&19.58\\
A1000$\beta$1.80&1.80&3.281&74.34&46.26&13.35&0.40&-800.78&-7.09&100&14.35&1.60\\
A500$\beta$0.25&0.25&1.130&67.43&79.57&6.80&3.30&-1782.64&-15.78&509&50.22&64.22\\
A500$\beta$0.90&0.90&2.177&71.81&78.19&13.58&1.19&-1317.80&-11.66&173&7.41&8.08\\
NONROT&-&-&65.96&84.04&-&3.78&-14.82&-0.01&500&-&0.01
\enddata
\tablecomments{Overview of core collapse simulations performed with the polytropic progenitor
model and the hybrid equation of state (\S\ref{section:hybrideos}) for 
comparison with \cite{zm:97} and \cite{harry:02b}. $\beta_i$ is the initial rotation parameter, 
J is the total angular momentum, and t$_b$, $\beta_b$ and $\rho_b$ are the time, central 
density, and rotation parameter at core bounce. $\Delta$t=t$_{f}$-t$_b$ is the time each 
calculation was carried out beyond core bounce. $|A_{20}^{E2}|_{max}$ and
$h_{max}$ are the absolute maximum of the gravitational quadrupole wave amplitude and the
maximum gravitational wave strain, as defined in \S\ref{section:extract}. $f_{max}$ is
the peak frequency of the gravitational wave spectrum, $E_{GW,b}$ is the gravitational
wave energy radiated up to the first post-bounce minimum in the maximum density, and
$E_{GW,f}$ is the total energy radiated in gravitational waves during the entire simulation.}
\label{table:results}
\end{deluxetable}
\newpage

\begin{deluxetable}{lccrrrcrrrrr}

\tablecaption{Results: Realistic Progenitor Model s11WW \label{table:resultss11}}
\tablehead{
\colhead{Model}&
\colhead{$\beta_i$}&
\colhead{J}&
\colhead{$\Delta$t}&
\colhead{$\beta_b$}&
\colhead{$\beta_f$}&
\colhead{$\rho_{b}$}&
\colhead{$A_{20}^{E2}\,_{max}$}&
\colhead{$h^{TT}_{max}$}&
\colhead{$f_{max}$}&
\colhead{$E_{GW,b}$}&
\colhead{$E_{GW,f}$}\\
\colhead{}&
\colhead{(\%)}&
\colhead{(10$^{49}$}&
\colhead{(ms)}&
\colhead{(\%)}&
\colhead{(\%)}&
\colhead{($10^{14}$\hspace*{0.3cm}}&
\colhead{(cm)}&
\colhead{@ 10 kpc}&
\colhead{(Hz)}&
\colhead{($10^{-9}$\hspace*{0.3cm}}&
\colhead{($10^{-9}$\hspace*{0.3cm}}\\
\colhead{}&
\colhead{}&
\colhead{ erg s)}&
\colhead{}&
\colhead{}&
\colhead{}&
\colhead{g cm$^{-3}$)}&
\colhead{}&
\colhead{(10$^{-21}$)}&
\colhead{}&
\colhead{\modot c$^2$)}&
\colhead{\modot c$^2$)}
}
\startdata
s11A500$\beta$0.1&0.10&0.708&167&4.00&3.62&3.70&-958.99&-8.49&524&10.99&11.88\\
s11A500$\beta$0.2&0.20&1.002&145&7.05&5.86&3.30&-1408.17&-12.46&402&19.15&22.11\\
s11A500$\beta$0.25&0.25&1.097&135&7.87&6.43&3.10&-1203.06&-10.65&404&11.21&12.78\\
s11A500$\beta$0.3&0.30&1.227&413&8.80&5.19&2.74&-661.27&-5.85&164&1.70&1.81\\
s11A500$\beta$0.4&0.40&1.417&468&8.00&4.81&1.11&-419.65&-3.71&37&0.46&0.55\\
s11A500$\beta$0.5&0.50&1.584&707&7.71&4.80&0.46&-325.27&-2.88&25&0.18&0.22\\
\hline
\\
s11A1000$\beta$0.1&0.10&0.803&179&3.33&3.03&3.73&-878.86&-7.78&430&8.96&10.04\\
s11A1000$\beta$0.2&0.20&1.135&182&6.06&4.03&3.41&-1304.82&-11.55&470&14.10&15.53\\
s11A1000$\beta$0.3&0.30&1.391&91&8.00&6.21&3.00&-861.14&-7.62&208&3.82&4.26\\
s11A1000$\beta$0.4&0.40&1.606&304&7.98&4.77&2.06&-378.58&-3.35&150&0.35&0.42\\
s11A1000$\beta$0.5&0.50&1.795&284&7.40&4.65&0.60&-301.88&-2.66&78&0.16&0.19\\
s11A1000$\beta$0.6&0.60&1.967&503&7.15&4.49&0.28&-229.35&-2.02&38&0.06&0.07\\
s11A1000$\beta$0.7&0.70&2.124&424&6.95&4.41&0.15&-183.83&-1.63&26&0.03&0.03\\
s11A1000$\beta$0.8&0.80&2.271&510&6.66&4.38&0.09&-139.37&-1.23&44&0.01&0.02\\
\hline
\\
s11A50000$\beta$0.1&0.10&0.856&172&2.17&1.97&3.83&-582.09&-5.15&394&3.93&4.86\\
s11A50000$\beta$0.2&0.20&1.211&156&3.88&3.32&3.64&-979.41&-8.67&416&9.03&9.72\\
s11A50000$\beta$0.25&0.25&1.354&69&4.73&3.83&3.55&-1111.64&-9.84&409&10.00&10.57\\
s11A50000$\beta$0.3&0.30&1.483&229&5.47&4.51&3.45&-1188.38&-10.52&420&10.11&10.61\\
s11A50000$\beta$0.4&0.40&1.713&163&6.47&5.42&3.23&-1075.10&-9.52&344&6.20&6.48\\
s11A50000$\beta$0.5&0.50&1.914&90&7.38&4.11&2.97&-613.39&-5.43&166&1.36&1.52\\
s11A50000$\beta$0.6&0.60&2.097&249&7.34&4.30&2.37&-266.17&-2.36&99&0.16&0.19\\
s11A50000$\beta$0.7&0.70&2.266&404&6.78&4.36&0.86&-238.21&-2.11&54&0.08&0.10\\
\hline
\\
s11nonrot&-&-&88&-&-&4.04&-26.09&-0.02&275&-&0.06
\enddata
\tablecomments{Numerical results of core collapse simulations performed with the
Woosley and Weaver (1995) s11 progenitor model in conjunction with the Lattimer-Swesty
equation of state (\citealt{lseos:91}; \S\ref{section:lseos}).
$\beta_i$ is the initial rotation parameter, J is the total angular momentum, and 
$\beta_b$ and $\rho_b$ are the central density and rotation parameter 
at core bounce. $\Delta$t = t$_{f}$-t$_b$ is the time each 
individual calculation was carried out beyond core bounce. $\beta_f$ is the final 
rotation parameter for the whole grid. $|A_{20}^{E2}|_{max}$ and
$h_{max}$ are the absolute maximum of the gravitational quadrupole wave amplitude and the
maximum gravitational wave strain, as defined in \S\ref{section:extract}. $f_{max}$ is
the peak frequency of the gravitational wave spectrum, $E_{GW,b}$ gives the gravitational
wave energy radiated before the first post-bounce minimum in the maximum density, and
$E_{GW,f}$ is the total energy radiated in gravitational waves during the entire simulation.}
\end{deluxetable}
\newpage

\begin{deluxetable}{lccrrrcrrrrr}

\tablecaption{Results: Realistic Progenitor Model s15WW \label{table:resultss15}}

\tablehead{
\colhead{Model}&
\colhead{$\beta_i$}&
\colhead{J}&
\colhead{$\Delta$t}&
\colhead{$\beta_b$}&
\colhead{$\beta_f$}&
\colhead{$\rho_{b}$}&
\colhead{$A_{20}^{E2}\,_{max}$}&
\colhead{$h^{TT}_{max}$}&
\colhead{$f_{max}$}&
\colhead{$E_{GW,b}$}&
\colhead{$E_{GW,f}$}\\
\colhead{}&
\colhead{(\%)}&
\colhead{(10$^{49}$}&
\colhead{(ms)}&
\colhead{(\%)}&
\colhead{(\%)}&
\colhead{($10^{14}$\hspace*{0.3cm}}&
\colhead{(cm)}&
\colhead{@ 10 kpc}&
\colhead{(Hz)}&
\colhead{($10^{-9}$\hspace*{0.3cm}}&
\colhead{($10^{-9}$\hspace*{0.3cm}}\\
\colhead{}&
\colhead{}&
\colhead{ erg s)}&
\colhead{}&
\colhead{}&
\colhead{}&
\colhead{g cm$^{-3}$)}&
\colhead{}&
\colhead{(10$^{-21}$)}&
\colhead{}&
\colhead{\modot c$^2$)}&
\colhead{\modot c$^2$)}
}

\startdata
s15A500$\beta$0.1&0.10&0.811&215&3.87&3.60&3.69&-862.97&-7.64&524&9.06&10.14\\
s15A500$\beta$0.2&0.20&1.147&181&6.88&5.79&3.31&-1283.00&-11.36&469&16.76&17.88\\
s15A500$\beta$0.25&0.25&1.257&151&7.83&6.55&3.12&-1205.50&-10.67&381&11.87&13.32\\
s15A500$\beta$0.3&0.30&1.409&138&8.71&5.99&2.80&-758.40&-6.71&242&2.69&2.88\\
s15A500$\beta$0.4&0.40&1.557&251&8.10&5.35&1.31&-426.44&-3.77&169&0.51&0.52\\
s15A500$\beta$0.5&0.50&1.812&641&7.81&5.19&0.54&-343.78&-3.04&128&0.23&0.25\\
s15A500$\beta$0.6&0.60&1.987&373&7.71&5.10&0.29&-278.79&-2.47&101&0.11&0.12\\
s15A500$\beta$0.9&0.90&2.421&524&7.52&5.05&0.07&-171.30&-1.52&55&0.02&0.03\\
s15A500$\beta$1.0&1.00&2.564&661&7.40&5.03&0.05&-138.20&-1.22&42&0.01&0.01\\
\hline
\\
s15A1000$\beta$0.1&0.10&0.958&90&3.15&2.63&3.75&-742.56&-6.57&431&6.56&7.90\\
s15A1000$\beta$0.2&0.20&1.355&178&5.43&4.82&3.49&-1164.47&-10.31&461&12.32&13.10\\
s15A1000$\beta$0.3&0.30&1.660&167&7.11&6.37&3.13&-1040.73&-9.21&317&7.10&7.58\\
s15A1000$\beta$0.4&0.40&1.916&312&8.37&5.48&2.66&-491.54&-4.35&152&0.83&0.89\\
s15A1000$\beta$0.5&0.50&2.142&208&7.67&5.22&1.08&-362.96&-3.21&106&0.27&0.29\\
s15A1000$\beta$0.6&0.60&2.292&251&7.43&5.03&0.49&-293.17&-2.59&91&0.13&0.13\\
s15A1000$\beta$0.7&0.70&2.535&313&7.29&4.80&0.26&-234.48&-2.08&64&0.06&0.07\\
s15A1000$\beta$0.8&0.80&2.710&284&7.19&4.77&0.16&-190.31&-1.68&38&0.03&0.03\\
s15A1000$\beta$0.9&0.90&2.633&437&7.11&4.74&0.10&-152.94&-1.35&31&0.02&0.02\\
s15A1000$\beta$1.0&0.80&2.906&441&6.97&4.72&0.06&-124.39&-1.10&22&0.01&0.01\\
\hline
\\
s15A50000$\beta$0.1&0.10&1.081&207&1.40&1.41&3.90&-382.21&-3.38&396&1.65&2.16\\
s15A50000$\beta$0.2&0.20&1.528&108&2.79&2.28&3.77&-654.29&-5.79&400&4.90&5.95\\
s15A50000$\beta$0.5&0.50&2.416&243&5.49&4.94&3.41&-1034.23&-9.15&428&7.46&7.48\\
s15A50000$\beta$1.0&1.00&3.417&494&7.42&4.65&2.16&-299.88&-2.65&71&0.18&0.20\\
\hline
\\
s15nonrot&-&-&60&-&-&4.04&-23.87&-0.02&337&-&0.02
\enddata
\tablecomments{Same as Table \ref{table:resultss11} but for all simulations
using the Woosley and Weaver (1995) s15 progenitor model in conjunction with the Lattimer-Swesty
equation of state.}
\end{deluxetable}
\newpage

\begin{deluxetable}{lccrrrcrrrrr}

\tablecaption{Results: Realistic Progenitor Model s20WW/s25WW \label{table:resultss20}}

\tablehead{
\colhead{Model}&
\colhead{$\beta_i$}&
\colhead{J}&
\colhead{$\Delta$t}&
\colhead{$\beta_b$}&
\colhead{$\beta_f$}&
\colhead{$\rho_{b}$}&
\colhead{$A_{20}^{E2}\,_{max}$}&
\colhead{$h^{TT}_{max}$}&
\colhead{$f_{max}$}&
\colhead{$E_{GW,b}$}&
\colhead{$E_{GW,f}$}\\
\colhead{}&
\colhead{(\%)}&
\colhead{(10$^{49}$}&
\colhead{(ms)}&
\colhead{(\%)}&
\colhead{(\%)}&
\colhead{($10^{14}$\hspace*{0.3cm}}&
\colhead{(cm)}&
\colhead{@ 10 kpc}&
\colhead{(Hz)}&
\colhead{($10^{-9}$\hspace*{0.3cm}}&
\colhead{($10^{-9}$\hspace*{0.3cm}}\\
\colhead{}&
\colhead{}&
\colhead{ erg s)}&
\colhead{}&
\colhead{}&
\colhead{}&
\colhead{g cm$^{-3}$)}&
\colhead{}&
\colhead{(10$^{-21}$)}&
\colhead{}&
\colhead{\modot c$^2$)}&
\colhead{\modot c$^2$)}
}

\startdata
s20A500$\beta$0.1&0.10&1.326&103&5.50&4.97&3.45&-1139.52&-10.09&594&16.03&19.03\\
s20A500$\beta$0.2&0.20&1.882&142&8.96&7.64&2.78&-1077.88&-9.54&396&6.55&8.85\\
s20A500$\beta$0.5&0.50&2.976&416&8.65&8.00&0.22&-355.85&-3.15&92&0.16&0.22\\
\hline
\\
s25A500$\beta$0.2&0.20&2.217&56&9.16&7.55&2.72&-1043.73&-9.24&453&5.90&8.07\\
\hline
\\
s20A1000$\beta$0.1&0.10&1.941&52&4.51&4.01&3.61&-954.61&-8.45&525&10.90&12.67\\
s20A1000$\beta$0.2&0.20&2.745&94&7.71&7.00&3.14&-1247.85&-11.04&387&14.90&18.76\\
s20A1000$\beta$0.3&0.30&3.304&146&9.12&7.52&2.44&-678.39&-6.00&73&1.43&1.74\\
s20A1000$\beta$0.4&0.40&2.858&189&8.49&7.71&0.82&-536.00&-4.74&85&0.66&0.78\\
s20A1000$\beta$0.5&0.50&3.984&113&8.37&7.45&0.36&-424.96&-3.76&97&0.30&0.33\\
s20A1000$\beta$0.6&0.60&4.254&291&8.37&8.10&0.20&-345.78&-3.06&79&0.15&0.17\\
s20A1000$\beta$0.7&0.70&4.549&290&8.31&8.17&0.12&-269.03&-2.38&66&0.07&0.08\\
s20A1000$\beta$0.8&0.80&4.839&300&8.33&8.23&0.08&-220.16&-1.95&56&0.02&0.03\\
s20A1000$\beta$0.9&0.90&5.040&282&8.32&8.15&0.05&-185.81&-1.64&47&0.02&0.03\\
s20A1000$\beta$1.0&1.00&5.226&316&8.36&8.17&0.04&-154.43&-1.37&16&0.01&0.01\\
\hline
\\
s20A50000$\beta$0.1&0.10&1.950&40&1.53&1.26&3.89&-351.07&-3.11&409&1.25&1.79\\
s20A50000$\beta$0.2&0.20&2.758&36&2.89&2.34&3.78&-619.26&-5.48&424&4.48&5.49\\
s20A50000$\beta$0.5&0.50&4.361&230&5.88&5.96&3.39&-1175.30&-10.40&456&10.14&11.06\\
s20A50000$\beta$1.0&1.00&6.168&303&8.38&7.02&2.36&-449.58&-3.98&102&0.61&0.68\\
\hline
\\
s20nonrot&-&-&83&-&-&4.08&-165.02&-1.46&472&-&1.07\\
\enddata
\tablecomments{Same as Table \ref{table:resultss11}, but for all simulations
using the Woosley and Weaver (1995) s20 and s25 progenitor models. 
The s25 run justifies our assumption in \S\ref{section:prog} that 
the s20 and s25 progenitors lead to similar results. 
Note that for the s25 run a larger grid  
was used and the evolution was stopped before the post-bounce 
oscillations had faded. Hence, the differences seen in the total angular momentum 
and maximum frequency.}
\end{deluxetable}
\newpage

\begin{deluxetable}{lccrrrcrrrrr}

\tablecaption{Results: Heger Models \label{table:resultsheger}}

\tablehead{
\colhead{Model}&
\colhead{$\beta_i$}&
\colhead{J}&
\colhead{$\Delta$t}&
\colhead{$\beta_b$}&
\colhead{$\beta_f$}&
\colhead{$\rho_{b}$}&
\colhead{$A_{20}^{E2}\,_{max}$}&
\colhead{$h^{TT}_{max}$}&
\colhead{$f_{max}$}&
\colhead{$E_{GW,b}$}&
\colhead{$E_{GW,f}$}\\
\colhead{}&
\colhead{(\%)}&
\colhead{(10$^{49}$}&
\colhead{(ms)}&
\colhead{(\%)}&
\colhead{(\%)}&
\colhead{($10^{14}$\hspace*{0.3cm}}&
\colhead{(cm)}&
\colhead{@ 10 kpc}&
\colhead{(Hz)}&
\colhead{($10^{-9}$\hspace*{0.3cm}}&
\colhead{($10^{-9}$\hspace*{0.3cm}}\\
\colhead{}&
\colhead{}&
\colhead{ erg s)}&
\colhead{}&
\colhead{}&
\colhead{}&
\colhead{g cm$^{-3}$)}&
\colhead{}&
\colhead{(10$^{-21}$)}&
\colhead{}&
\colhead{\modot c$^2$)}&
\colhead{\modot c$^2$)}
}

\startdata
m15b4&0.0021&0.1345&30&0.087&0.082&4.14&-29.15&-0.26&590&0.004&0.014\\
m20b4&0.0032&0.3034&37&0.161&0.156&4.15&-52.46&-0.46&360&0.026&0.041\\
m25b4&0.0053&0.3561&94&0.324&0.360&4.14&-101.06&-0.89&960&0.110&0.550\\
\hline
\\
e15&0.6454&3.392&311&8.82&7.49&0.27&-405.57&-3.59&90&0.235&0.275\\
e20&0.4176&3.442&294&8.52&7.87&0.55&-436.77&-3.87&193&0.465&0.631\\
\enddata
\tablecomments{Same as Table \ref{table:resultss11}, but for all simulations
using the \cite{heger:00} and \cite{spruit:03} progenitor models with rotation.
Model m25b4, evolved long enough for convective instability behind the shock to grow, shows
considerable energy radiated by the post-bounce convective
bulk mass motions.}
\end{deluxetable}

\newpage

\clearpage
\newpage

\clearpage 

\begin{thebibliography}{83}
\expandafter\ifx\csname natexlab\endcsname\relax\def\natexlab#1{#1}\fi

\bibitem[{Abramovici et al.}(1992)]{abra:92}
Abramovici, A. et al. 1992, Science, 256, 325

\bibitem[{Akiyama et al.}(2003)]{wheeler:03}
Akiyama, S. et al. 2003, \apj, 584, 954

\bibitem[{Ando et al.}(2001)]{ando:01}
Ando, M. et al. 2001, \prl, 86, 3950

\bibitem[{Akiyama {et~al.}(2003){Akiyama} et al.}]{Akiyama:03}
Akiyama, S., Wheeler, J.C., Meier, D., and Lichtenstadt, I.
2003, \apj, 584, 954

\bibitem[{{Bethe} and {Wilson}(1985)}]{bethewilson:85}
{Bethe}, H.~A. and {Wilson}, J.~R. 1985, \apj, 295, 14

\bibitem[{Blanchet, Damour, and Sch\"afer (1990)}] 
{blanchet:90}
Blanchet, L., Damour, T., and Sch{\"a}fer, G. 1990, MNRAS, 242, 289

\bibitem[{{Bonazzola} and {Marck}(1993)}]{bonazzola:93}
{Bonazzola}, S. and {Marck}, J.~A. 1993, \aap, 267, 623

\bibitem[{{Brown}(2001)}]{brown:01}
{Brown}, J.~D. 2001, in AIP Conf. Proc. 575: Astrophysical Sources for
  Ground-Based Gravitational Wave Detectors, 234

\bibitem[{Buras {et~al.}(2003)Buras, Rampp, Janka, and Kifonidis}]{buras:03}
Buras, R., Rampp, M., Janka, H.-T., and Kifonidis, K. 2003, astro-ph/0303171

\bibitem[{Burrows and Hayes(1996)}]{bh:96}
Burrows, A. and Hayes, J. 1996, Phys. Rev. Lett, 76, 352

\bibitem[{Burrows, Hayes, and Fryxell (1995)}]{bhf:95}
Burrows, A., Hayes, J., and Fryxell, B.~A. 1995, ApJ, 450, 830

\bibitem[{{Burrows} and {Thompson}(2003)}]{burrowsthomp:03}
{Burrows}, A. and {Thompson}, T.~A. 2003, in From Twilight to Highlight: The
  Physics of Supernovae. Proceedings of the ESO/MPA/MPE Workshop held in
  Garching, Germany, 29-31 July 2002, 53

\bibitem[{Centrella {et~al.}(2001)Centrella, New, Lowe, \& Brown}]{cent:01}
Centrella, J., New, K., Lowe, L., and Brown, J. 2001, ApJ, 550, L193

\bibitem[{Dimmelmeier, Font, and M\"uller(2002{\natexlab{a}})Dimmelmeier, Font, and
  M{\"u}ller}]{harry:02a}
Dimmelmeier, H., Font, J., and M{\"u}ller, E. 2002{\natexlab{a}}, A\&A, 388,
  917

\bibitem[{Dimmelmeier, Font, and M\"uller(2002{\natexlab{b}})Dimmelmeier, Font, and
  M{\"u}ller}]{harry:02b}
---. 2002{\natexlab{b}}, A\&A, 393, 523

\bibitem[{{Epstein}(1978)}]{epstein:78}
{Epstein}, R. 1978, \apj, 223, 1037

\bibitem[{{Eriguchi} and {M\"uller}(1985)}]{eriguchimueller:85}
{Eriguchi}, Y. and {M\"uller}, E. 1985, \aap, 146, 260

\bibitem[{Evans(1986)}]{evans:86}
Evans, C.~R. 1986, in Dynamical Spacetimes and Numerical Relativity,
  ed.~J.~M.~Centrella (Cambridge, U. K.: Cambridge University Press), ~3--39


\bibitem[{Finn and Evans(1990)}]{finnevans:90}
Finn, L.~S. and Evans, C.~R. 1990, ApJ, 351, 588

\bibitem[{Flanagan and Hughes(1998)}]{flanhughes:98}
\'E. \'E. Flanagan and Hughes, S. A. 1998, \prd, 57, 4535


\bibitem[{{Fryer} {et~al.}(1999){Fryer}, {Benz}, {Herant}, and 
  {Colgate}}]{fryeretal:99}
{Fryer}, C., {Benz}, W., {Herant}, M., and {Colgate}, S.~A. 1999, \apj, 516, 892

\bibitem[{Fryer, Holz, and Hughes (2002)Fryer, Holz, and Hughes}]{fhh:02}
Fryer, C., Holz, D.,  and Hughes, S. 2002, \apj, 565, 430

\bibitem[{Fryer and Heger(2000)}]{fryerheger:00}
Fryer, C.~L. and Heger, A. 2000, ApJ, 541, 1033

\bibitem[{{Fukuda}(1982)}]{fukuda:82}
{Fukuda}, I. 1982, \pasp, 94, 271

\bibitem[{Gustafson {et~al.}(1999)}]{gust:99}
Gustafson, E., Shoemaker, D., Strain, K., and Weiss, R. 1999, \\LSC
white paper on detector research and development, Technical Report
LIGO~T990080-00-D

\bibitem[{Hachisu(1986{\natexlab{a}})}]{hach:86}
Hachisu, I. 1986{\natexlab{a}}, ApJS, 61, 479

\bibitem[{Hachisu(1986{\natexlab{b}})}]{hach2:86}
---. 1986{\natexlab{b}}, ApJS, 62, 461

\bibitem[{Heger {et~al.}(2002)Heger, Fryer, Woosley, Langer, and
  Hartmann}]{heger:02}
Heger, A., Fryer, C.~L., Woosley, S.~E., Langer, N., and Hartmann, D.~H. 2002,
  astro-ph/0212469

\bibitem[{{Heger}, {Langer}, and {Woosley}(2000){Heger}, {Langer}, and {Woosley}}]{heger:00}
{Heger}, A., {Langer}, N., and {Woosley}, S.~E. 2000, \apj, 528, 368

\bibitem[{Heger {et~al.}(2003)Heger, Woosley, Langer, and Spruit}]{spruit:03}
Heger, A., Woosley, S.~E., Langer, N., and Spruit, H.~C. 2003, Stellar Rotation.
  Proceedings IAU Symposium No. 215

\bibitem[{Hillebrandt and Wolff(1985)}]{hillewolff:85}
Hillebrandt, W. and Wolff, R.~G. 1985, in Nucleosynthesis: Challenges and New
  Developments, eds. D. Arnett and J. W. Truran (Chicago, U.S.A.: University
  Chicago Press)

\bibitem[{{Ipser} and {Managan}(1984)}]{ipsermanagan:84}
{Ipser}, J.~R. and {Managan}, R.~A. 1984, \apj, 282, 287


\bibitem[{{Janka}, Zwerger, and M\"onchmeyer (1993){Janka}, {Zwerger},
  {M\"onchmeyer}}]{janka:93}
{Janka}, H.-T., {Zwerger}, T., and {M\"onchmeyer}, R. 1993, \aap, 268, 360

\bibitem[{{Kotake}, Yamada, and Sato(2003){Kotake}, {Yamada},
  {Sato}}]{kotake:03}
{Kotake}, K., {Yamada}, S., and {Sato}, K. 2003, \prd, in press

\bibitem[{Lattimer {et~al.}(1985)Lattimer, Pethick, Ravenhall, and
  Lamb}]{lattimeretal:85}
Lattimer, J.~M., Pethick, C., Ravenhall, D., and Lamb, D. 1985, Nucl. Phys. A,
  432, 646

\bibitem[{Lattimer and Swesty(1991)}]{lseos:91}
Lattimer, J.~M. and Swesty, F.~D. 1991, Nucl. Phys. A, 535, 331

\bibitem[{Ledoux (1945)}]{ledoux:45}
Ledoux, P. 1945, \apj, 102, 143

\bibitem[{{Liebend{\" o}rfer} {et~al.}(2001{\natexlab{a}}){Liebend{\" o}rfer},
  {Mezzacappa}, and {Thielemann}}]{liebendoerfer:01}
{Liebend{\" o}rfer}, M., {Mezzacappa}, A., and {Thielemann}, F.
  2001{\natexlab{a}}, \prd, 63, 104003

\bibitem[{{Liebend{\" o}rfer} {et~al.}(2001{\natexlab{b}}){Liebend{\" o}rfer},
  {Mezzacappa}, {Thielemann}, {Messer}, {Hix}, and {Bruenn}}]{liebendoerfer2:01}
{Liebend{\" o}rfer}, M., {Mezzacappa}, A., {Thielemann}, F., {Messer}, O.~E.,
  {Hix}, W.~R., and {Bruenn}, S.~W. 2001{\natexlab{b}}, \prd, 63, 103004

\bibitem[{{Livne}(1993)}]{livne:93}
{Livne}, E. 1993, \apj, 412, 634

\bibitem[{Misner, Thorne, and Wheeler (1973)Misner, Thorne, and Wheeler}]{MTW}
Misner, C.~W., Thorne, K.~S., and Wheeler, J.~A. 1973, Gravitation (San
  Francisco, U. S. A.: Freeman)

\bibitem[{M{\"o}nchmeyer {et~al.}(1991)M{\"o}nchmeyer, Sch{\"a}fer, M{\"u}ller,
  and Kates}]{mm:91}
M{\"o}nchmeyer, R., Sch{\"a}fer, G., M{\"u}ller, E., and Kates, R. 1991, A\&A,
  246, 417

\bibitem[{{Moncrief}(1979)}]{moncrief:79}
{Moncrief}, V. 1979, \apj, 234, 628

\bibitem[{{M\"uller}(1982)}]{mueller:82}
{M\"uller}, E. 1982, \aap, 114, 53

\bibitem[{{M\"uller} and {Hillebrandt}(1981)}]{muellerhille:81}
{M\"uller}, E. and {Hillebrandt}, W. 1981, \aap, 103, 358

\bibitem[{M{\"u}ller and Janka(1997)}]{jm:97}
M{\"u}ller, E. and Janka, H.-T. 1997, A\&A, 317, 140

\bibitem[{Nakamura and Oohara(1989)}]{nakamura:89}
Nakamura, T. and Oohara, K. 1989, in Frontiers in Numerical Relativity, ed.
  C.~R.~Evans, L.~S.~Finn, and D.~W.~Hobill (Cambridge, U. K.: Cambridge
  University Press), 254

\bibitem[{New(2003)}]{new:03}
New, K. C.~B. 2003, Living Rev. Relativity 6, 2. [Online Article]: cited
on~20.~July~2003, http://www.livingreviews.org/lrr-2003-2

\bibitem[{Punturo(2003)}]{punturo:03}
Punturo, M., VIRGO Note, The VIRGO sensitivity curve,\\ VIR-NOT-PER-1390-51

\bibitem[{Press {et~al.}(1992)Press, Teukolsky, Vetterling, and 
  Flannery}]{numrep}
Press, W.~H., Teukolsky, S.~A., Vetterling, W.~T., and Flannery, B.~P. 1992,
  Numerical Recipes in {C}, 2nd. edition (Cambridge, U. K.: Cambridge
  University Press)

\bibitem[{{Rampp} and {Janka}(2000)}]{ramppjanka:00}
{Rampp}, M. and {Janka}, H.-T. 2000, \apjl, 539, L33

\bibitem[{{Rampp} and {Janka}(2002)}]{ramppjanka:02}
---. 2002, \aap, 396, 361 

\bibitem[{Rampp, M\"uller, and Ruffert (1998)Rampp, M{\"u}ller, and Ruffert}]{rmr:98}
Rampp, M., M{\"u}ller, E., and Ruffert, M. 1998, A\&A, 332, 969


\bibitem[{{Saenz} and {Shapiro}(1978)}]{saenzshapiro:78}
{Saenz}, R.~A. and {Shapiro}, S.~L. 1978, \apj, 221, 286

\bibitem[{{Saenz} and {Shapiro}(1979)}]{saenzshapiro:79}
---. 1979, \apj, 229, 1107

\bibitem[{{Saenz} and {Shapiro}(1981)}]{saenzshapiro:81}
---. 1981, \apj, 244, 1033

\bibitem[{Seidel and Moore(1987)}]{seidelmoore:87}
Seidel, E. and Moore, T. 1987, \prd, 35, 2287

\bibitem[{Seidel, Myra, and Moore(1987)Seidel, Myra, and Moore}]{seideletal:88}
Seidel, E., Myra, E.~S., and Moore, T. 1988, \prd, 38, 2349

\bibitem[{Shibata, Karino, and Eriguchi(2002)Shibata, Karino, and Eriguchi}]{shibata:02}
Shibata, M., Karino, S., Eriguchi, Y. 2002, \mnras, 334, L27

\bibitem[{Shibata, Karino, and Eriguchi(2003)Shibata, Karino, and Eriguchi}]{shibata:03}
Shibata, M., Karino, S., Eriguchi, Y. 2003, \mnras, 343, 619

\bibitem[{Shapiro and Teukolsky(1983)}]{shapteu:83}
Shapiro, L.~S. and Teukolsky, S.~A. 1983, Black Holes, White Dwarfs and Neutron
  Stars (New York U. S. A.: John Wiley \& Sons)

\bibitem[{{Shapiro and Lightman}(1976)}]{shapirolightman:76}
{Shapiro}, S.~L. and {Lightman}, A.~P. 1976, \apj, 207, 263

\bibitem[{{Shapiro}(1977)}]{shapiro:77}
{Shapiro}, S.~L. 1977, \apj, 214, 566

\bibitem[{Shen et al. (1998)}]{shen:98}
Shen, H., Toki, H., Oyamatsu, K., and Sumiyoshi, K. 1998, Nucl. Phys. A, 637, 43

\bibitem[{Tassoul(1978)}]{tassoul:78}
Tassoul, J.-L. 1978, Theory of Rotating Stars (Princeton U. S. A.: Princeton
  University Press)

\bibitem[{Thompson, Burrows, and Pinto (2003)Thompson, Burrows, and Pinto}]{thompetal:03}
Thompson, T.~A., Burrows, A., and Pinto, P.~A. 2003, ApJ 592, July 20, in press
  (astro-ph/0211194)

\bibitem[{{Thorne}(1980)}]{thorne:80}
{Thorne}, K.~S. 1980, Reviews of Modern Physics, 52, 299

\bibitem[{{Timmes} and {Arnett}(1999)}]{timmesarnett:99}
{Timmes}, F.~X. and {Arnett}, D. 1999, \apjs, 125, 277

\bibitem[{{Timmes} and {Swesty}(2000)}]{timmesswesty:00}
{Timmes}, F.~X. and {Swesty}, F.~D. 2000, \apjs, 126, 501

\bibitem[{{Tohline}(1984)}]{tohline:84}
{Tohline}, J.~E. 1984, \apj, 361, 394

\bibitem[{Turner and Wagoner(1979)}]{turnerwagoner:79}
Turner, M.~S. and Wagoner, R.~V. 1979, in Sources of Gravitational Radiation,
  ed. L. Smarr (Cambridge, U. K.: Cambridge University Press)

\bibitem[{{Weaver}, Zimmerman, and Woosley(1978){Weaver}, {Zimmerman}, and {Woosley}}]{wwz:78}
{Weaver}, T.~A., {Zimmerman}, G.~B., and {Woosley}, S.~E. 1978, \apj, 225, 1021

\bibitem[{Willke et al. (2002)}]{wilke:02}
{Willke}, B. et al. 2002, Classical and Quantum Gravity, 19, 1377

\bibitem[{{Woosley}, {Heger}, and {Weaver}(2002){Woosley}, {Heger}, and {Weaver}}]{wwh:02}
{Woosley}, S.~E., {Heger}, A., and {Weaver}, T.~A. 2002, Reviews of Modern
  Physics, 74, 1015

\bibitem[{{Woosley} and {Weaver}(1995)}]{ww:95}
{Woosley}, S.~E. and {Weaver}, T.~A. 1995, \apjs, 101, 181


\bibitem[{{Yamada} and {Sato}(1995)}]{yamadasato:95}
---. 1995, \apj, 450, 245

\bibitem[{Zwerger and M{\"u}ller(1997)}]{zm:97}
Zwerger, T. and M{\"u}ller, E. 1997, A\&A, 320, 209

\end{thebibliography}
\end{document}